\documentclass[journal]{IEEEtran}
\usepackage{cite}
\usepackage{hyperref}
\usepackage{amsmath,amssymb,amsfonts}
\usepackage{algorithmic}
\usepackage{graphicx}
\usepackage{textcomp}
\usepackage[utf8]{inputenc}
\usepackage{url}
\usepackage{marvosym}
\usepackage[dvipsnames]{xcolor}
\usepackage[resetlabels,labeled]{multibib}
\usepackage[T1]{fontenc}

\newcommand{\eg}{e.g.}

\ifCLASSINFOpdf
\else
\fi
\begin{document}
	%
	% paper title
	% Titles are generally capitalized except for words such as a, an, and, as,
	% at, but, by, for, in, nor, of, on, or, the, to and up, which are usually
	% not capitalized unless they are the first or last word of the title.
	% Linebreaks \\ can be used within to get better formatting as desired.
	% Do not put math or special symbols in the title.
	\title{All One Needs to Know about Metaverse: A Complete Survey on Technological Singularity, Virtual Ecosystem, and Research Agenda}%
	%
	% author names and IEEE memberships
	% note positions of commas and nonbreaking spaces ( ~ ) LaTeX will not break
	% a structure at a ~ so this keeps an author's name from being broken across
	% two lines.
	% use \thanks{} to gain access to the first footnote area
	% a separate \thanks must be used for each paragraph as LaTeX2e's \thanks
	% was not built to handle multiple paragraphs
	%
	
	\author{Lik-Hang Lee$^1$,
		Tristan Braud$^2$,
		Pengyuan Zhou$^{3,4}$,
		Lin Wang$^1$,
		Dianlei Xu$^6$,
		Zijun Lin$^5$,
		Abhishek Kumar$^6$,
		Carlos Bermejo$^2$,
		and~Pan Hui$^{2,6}$,~\IEEEmembership{Fellow,~IEEE,
		}% <-this % stops a space
		\thanks{Corresponding Authors: Lik-Hang Lee, E-mail: (likhang.lee@kaist.ac.kr)}%, and Pan Hui, E-mail: (panhui@ust.hk)}% <-this % stops a space
		\thanks{$^1$ KAIST, South Korea; $^2$ HKUST, Hong Kong SAR; $^3$ USTC China; $^4$ MCT Key Lab of CCCD; $^5$ UCL, UK; $^6$ Uni. Helsinki, Finland.}%
		%<-this % stops a space
		%\thanks{(1) KAIST, South Korea; (2) HKUST, Hong Kong SAR; (3) USTC China; (4) UCL, UK; (5) Uni. Helsinki, Finland.}%
		\thanks{Manuscript submitted in October 2021.}}
	
	% note the % following the last \IEEEmembership and also \thanks - 
	% these prevent an unwanted space from occurring between the last author name
	% and the end of the author line. i.e., if you had this:
	% 
	% \author{....lastname \thanks{...} \thanks{...} }
	%                     ^------------^------------^----Do not want these spaces!
	%
	% a space would be appended to the last name and could cause every name on that
	% line to be shifted left slightly. This is one of those "LaTeX things". For
	% instance, "\textbf{A} \textbf{B}" will typeset as "A B" not "AB". To get
	% "AB" then you have to do: "\textbf{A}\textbf{B}"
	% \thanks is no different in this regard, so shield the last } of each \thanks
	% that ends a line with a % and do not let a space in before the next \thanks.
	% Spaces after \IEEEmembership other than the last one are OK (and needed) as
	% you are supposed to have spaces between the names. For what it is worth,
	% this is a minor point as most people would not even notice if the said evil
	% space somehow managed to creep in.

	% The paper headers
	\markboth{Journal of \LaTeX\ Class Files,~Vol.~14, No.~8, September~2021}%
	{Shell \MakeLowercase{\textit{et al.}}: Bare Demo of IEEEtran.cls for IEEE Journals}
	% The only time the second header will appear is for the odd numbered pages
	% after the title page when using the twoside option.
	% 
	% *** Note that you probably will NOT want to include the author's ***
	% *** name in the headers of peer review papers.                   ***
	% You can use \ifCLASSOPTIONpeerreview for conditional compilation here if
	% you desire.

	% If you want to put a publisher's ID mark on the page you can do it like
	% this:
	%\IEEEpubid{0000--0000/00\$00.00~\copyright~2015 IEEE}
	% Remember, if you use this you must call \IEEEpubidadjcol in the second
	% column for its text to clear the IEEEpubid mark.

	% use for special paper notices
	%\IEEEspecialpapernotice{(Invited Paper)}

	% make the title area
	\maketitle
	
	% As a general rule, do not put math, special symbols or citations
	% in the abstract or keywords.
	\begin{abstract}
		
		Since the popularisation of the Internet in the 1990s, the cyberspace has kept evolving. We have created various computer-mediated virtual environments including social networks, video conferencing, virtual 3D worlds (e.g., VR Chat), augmented reality applications (e.g., Pokemon Go), and Non-Fungible Token Games (e.g., Upland). Such virtual environments, albeit non-perpetual and unconnected, have bought us various degrees of digital transformation. The term `metaverse' has been coined to further facilitate the digital transformation in every aspect of our physical lives. At the core of the metaverse stands the vision of an immersive Internet as a gigantic, unified, persistent, and shared realm. While the metaverse may seem futuristic, catalysed by emerging technologies such as Extended Reality, 5G, and Artificial Intelligence, the digital `big bang' of our cyberspace is not far away.
		
		This survey paper presents the first effort to offer a comprehensive framework that examines the latest metaverse development under the dimensions of state-of-the-art technologies and metaverse ecosystems, and illustrates the possibility of the digital `big bang'. First, technologies are the enablers that drive the transition from the current Internet to the metaverse. We thus examine eight enabling technologies rigorously - Extended Reality, User Interactivity (Human-Computer Interaction), Artificial Intelligence, Blockchain, Computer Vision, IoT and Robotics, Edge and Cloud computing, and Future Mobile Networks. In terms of applications, the metaverse ecosystem allows human users to live and play within a self-sustaining, persistent, and shared realm. Therefore, we discuss six user-centric factors -- Avatar, Content Creation, Virtual Economy, Social Acceptability, Security and Privacy, and Trust and Accountability. Finally, we propose a concrete research agenda for the development of the metaverse. 
	\end{abstract}
	
	% Note that keywords are not normally used for peerreview papers.

	% For peer review papers, you can put extra information on the cover
	% page as needed:
	% \ifCLASSOPTIONpeerreview
	% \begin{center} \bfseries EDICS Category: 3-BBND \end{center}
	% \fi
	%
	% For peerreview papers, this IEEEtran command inserts a page break and
	% creates the second title. It will be ignored for other modes.
	\IEEEpeerreviewmaketitle
	
	\section{Introduction}
	\label{sec:introduction}
	% \PARstart{M}
	\IEEEPARstart{M}{etaverse}, combination of the prefix ``meta'' (implying transcending) with the word ``universe'', describes a hypothetical synthetic environment linked to the physical world. %facilitated by the 
	The word `metaverse' was first coined in a piece of speculative fiction named \textit{Snow Crash}, written by \textit{Neal Stephenson} in 1992~\cite{joshua_information_2017}.
	In this novel, Stephenson defines the metaverse as a massive virtual environment parallel to the physical world, in which users interact through digital avatars. 
	Since this first appearance, the metaverse as a computer-generated universe has been defined through vastly diversified concepts,
	such as \textit{lifelogging}~\cite{Bruun2019LifeloggingIT}, \textit{collective space in virtuality}~\cite{iii_2018}, \textit{embodied internet/ spatial Internet}~\cite{chayka_2021}, \textit{a mirror world}~\cite{nast}, an omniverse: a venue of simulation and collaboration~\cite{nvidia_developer_2021}.
	In this paper, we consider the metaverse as a virtual environment blending physical and digital, facilitated by the convergence between the Internet and Web technologies, and %the indispensable technology of 
	Extended Reality (XR). 
	%Extended Reality (XR) enables a seamless blend of physical and virtual objects across the physical world and digital environments.
	%is a collection of physical and virtual objects across the physical world and digital environments, 
	According to the \textit{Milgram and Kishino's Reality-Virtuality Continuum}~\cite{continuum}, XR integrates digital and physical to various degrees, e.g., augmented reality (AR), mixed reality (MR), and virtual reality (VR). 
	%we have seen vastly diversified concepts related to the metaverse describing computer-generated universes, 
	% It is important to note that the 
	Similarly, the metaverse scene in \textit{Snow Crash} projects the duality of the real world and a copy of digital environments. In the metaverse, all individual users own their respective avatars, in analogy to the user's physical self, to experience an alternate life in a virtuality that is a metaphor of the user's real worlds.

	\begin{figure}[!t]
		\centering
		\includegraphics[width=\columnwidth]{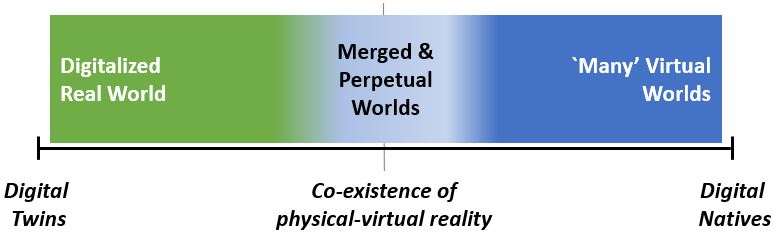}
		\caption{We propose a `\textbf{digital twins-native continuum}', on the basis of duality. This metaverse vision reflects three stages of development. We consider the \textit{digital twins} as a starting point, where our physical environments are digitised and thus own the capability to periodically reflect changes to their virtual counterparts. According to the physical world, digital twins create digital copies of the physical environments as `many' virtual worlds, and human users with their avatars work on new creations in such virtual worlds, as \textit{digital natives}. It is important to note that such virtual worlds will initially suffer from limited connectivity with each other and the physical world, i.e., information silo. They will then gradually connect within a massive landscape. Finally, the digitised physical and virtual worlds will eventually merge, representing the final stage of the \textit{co-existence of physical-virtual reality} similar to the surreality). Such a connected physical-virtual world give rise to the unprecedented demands of perpetual and 3D virtual cyberspace as the metaverse. }
		\label{fig:duality-continuum}
	\end{figure}
	
	\begin{figure*}[!t]
		\centering
		\includegraphics[width=\textwidth]{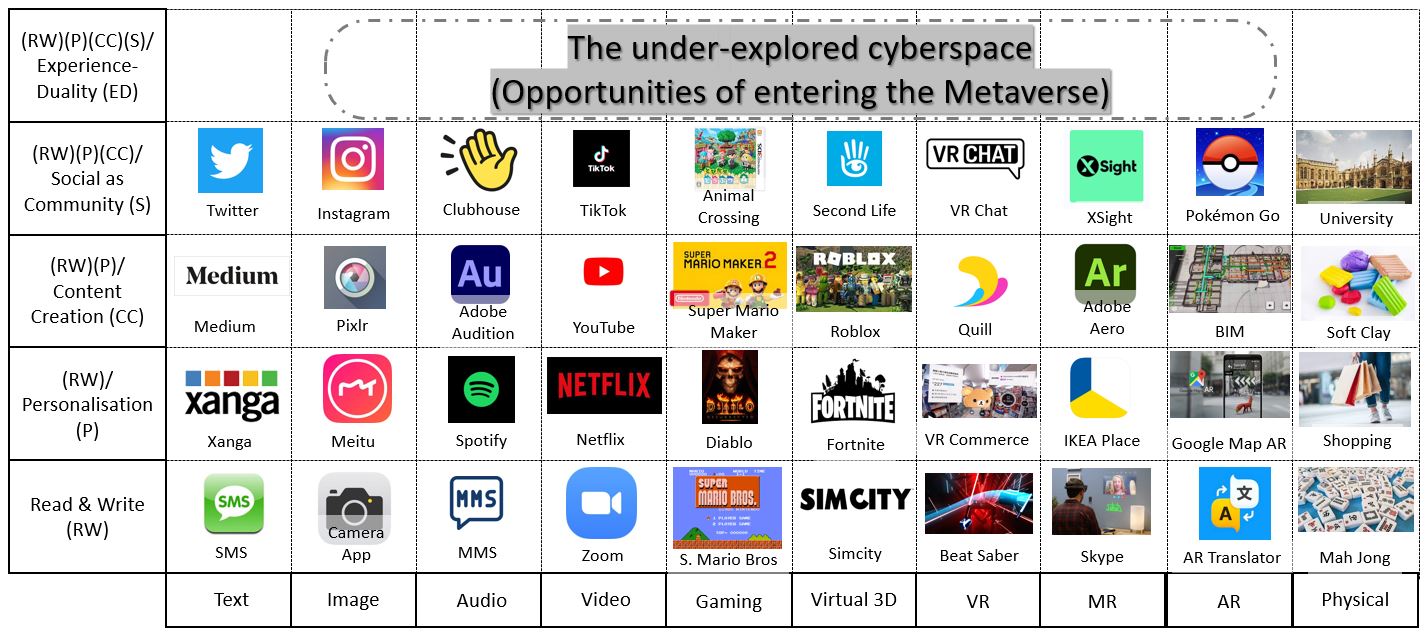}
		\caption{The cyberspace landscape of real-life applications, where superseding relationships exists in the information richness theory (left-to-right) as well as transience-permanence dimension (bottom-to-top). 
		}
		\label{fig:cyberspace}
	\end{figure*}
	
	To achieve such \textbf{duality}, the development of metaverse has to go through three sequential stages, namely (I) \textit{digital twins}, (II) \textit{digital natives}, and eventually (III) \textit{co-existence of physical-virtual reality} or namely the surreality. 
	Figure~\ref{fig:duality-continuum} depicts the relationship among the three stages.
	\textit{Digital twins} refer to large-scale and high-fidelity digital models and entities %instead of single objects (e.g., nut and bolt) 
	duplicated in virtual environments. \textit{Digital twins} reflect the properties of their physical counterparts~\cite{u194}, including the object motions, temperature, and even function. The connection between the virtual and physical twins is tied by their data~\cite{Grieves2017DigitalTM}. The existing applications are multitudinous such as computer-aided design (CAD) for product design and building architectures, smart urban planning, AI-assisted industrial systems, robot-supported risky operations~\cite{10.1109/SESoS/WDES.2019.00018, 10.1145/3387940.3392199, HRI-risk, Cureton2020DigitalTO, Langen2017ConceptFA}. 
	After establishing a digital copy of the physical reality, the second stage focuses on \textit{native content creation}. 
	Content creators, perhaps represented by the avatars, involve in digital creations inside the digital worlds. Such digital creations can be linked to their physical counterparts, or even only exist in the digital world. 
	Meanwhile, connected ecosystems, including culture, economy, laws, and regulations (e.g, data ownership), social norms, can support these digital creation~\cite{bush_2021}. Such ecosystems are analogous to real-world society's existing norms and regulations, supporting the production of physical goods and intangible contents~\cite{Viljoen2020ThePA}. However, research on such applications is still in a nascent stage, focusing on the first-contact point with users, such as input techniques and authoring system for content creation~\cite{Lau-chi21, XR-studio, TutorVR-CHI19, communication-chi16}.
	In the third and final stage, the metaverse could become a self-sustaining and persistent virtual world that \textit{co-exists and interoperates} with the physical world with a high level of independence. As such, the avatars, representing human users in the physical world, can experience heterogeneous activities in real-time characterised by unlimited numbers of concurrent users theoretically in %a number of 
	multiple virtual worlds~\cite{Grieves2017DigitalTM}. 
	Remarkably, the metaverse can afford interoperability between platforms representing different virtual worlds, i.e., enabling users to create contents and widely distribute the contents across virtual worlds. For instance, a user can create contents in a game, e.g., \textit{Minecraft}\footnote{\url{https://www.minecraft.net/en-us}}, and transfer such contents into another platform or game, e.g., \textit{Roblox}\footnote{\url{https://www.roblox.com/}}, with a continued identity and experience. To a further extent, the platform can connect and interact with our physical world through various channels, user's information access through head-mounted wearable displays or mobile headsets (e.g. Microsoft Hololens\footnote{\url{https://www.microsoft.com/en-us/hololens}}), contents, avatars, computer agents in the metaverse interacting with smart devices and robots, to name but a few. 
	
	\begin{figure*}[!t]
		\centering
		\includegraphics[width=\linewidth]{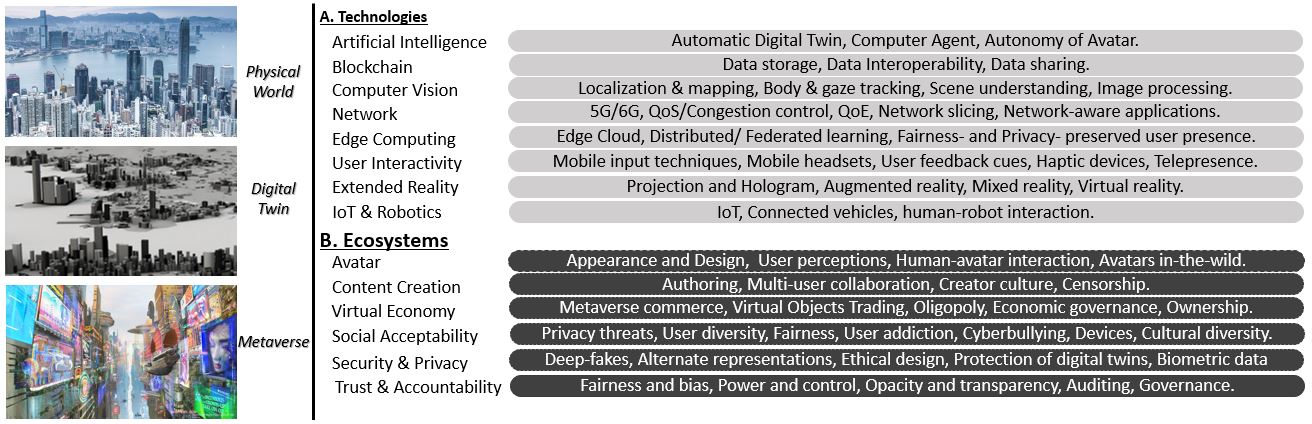}
		\caption{Connecting the physical world with its digital twins, and further shifting towards the metaverse: (A) the key technologies (e.g., blockchain, computer vision, distributed network, pervasive computing, scene understanding, ubiquitous interfaces), and; (B) considerations in ecosystems, in terms of avatar, content creation, data interoperability, social acceptability, security/privacy, as well as trust/accountability.}
		\label{fig:framework}
	\end{figure*}
	
	% existing examples of Games, VRchat, other virtual reality space.
	According to the diversified concepts of computer-mediated universe(s) mentioned above, one may argue that we are already situated in the metaverse. Nonetheless, this is only partially correct, and we examine several examples to justify our statement with the consideration of the three-stage metaverse development roadmap. 
	The \textit{Earth 3D map}\footnote{\url{https://earth3dmap.com/}} offers picture frames of the real-world but lacks physical properties other than GPS information, while social networks allow users to create contents but limited to texts, photos, and videos with limited options of user engagements (e.g., liking a post). 
	Video games are getting more and more realistic and impressive. Users can experience outstanding graphics with in-game physics, e.g., \textit{Call of Duty: Black Ops Cold War}, that deliver a sense of realism that resembles the real world in great details.
	A remarkable example of an 18-year-old virtual world, \textit{Second Life}\footnote{\url{https://id.secondlife.com}}, is regarded as the largest user-created 3D Universe. Users can build and shape their 3D environments and live in such a virtual world extravagantly. However,video games still lack interoperability between each other. 
	The emerging platforms leveraging virtual environments (e.g., VRChat\footnote{\url{https://hello.vrchat.com/}} and Microsoft Mesh\footnote{\url{https://www.microsoft.com/en-us/mesh?activetab=pivot\%3aprimaryr7}}) offer enriched environments that emulate virtual spaces for social gatherings and online meetings. However, these virtual spaces are not perpetual, and vanish after the gatherings and meetings.
	Virtual objects in AR games (e.g., Pokémon Go\footnote{\url{https://pokemongolive.com/en/}}) have also been attached to the physical reality without reflecting any principles of the digital twins.
	%Furthermore, the authors attempt to illustrate the gap between the current cyberspace and the metaverse. 
	
	Figure~\ref{fig:cyberspace} further demonstrates the significant gap that remains between the current cyberspace and the metaverse. Both x- and y-axes demonstrate superseding relationships: Left-to-Right (e.g., Text $<$ Image) and Bottom-to-Top (e.g., Read and Write (RW) $<$ Personalisation). 
	The x-axis depicts various media in order of information richness~\cite{Daft1986OrganizationalIR} from text, image, audio, video, gaming, virtual 3D worlds, virtuality (AR/MR/AR, following \textit{Milgram and Kishino's Reality-Virtuality Continuum}~\cite{continuum}) and eventually, the physical world.
	The y-axis indicates user experience under a spectrum between transience (Read and Write, RW) and permanence (Experience-Duality, ED). We highlight several examples to show this superseding relationship in the y-axis. 
	At the \textit{Read \& Write} level, the user experience does not evolve with the user. Every time a user sends a SMS or has a call on Zoom, their experience is similar to their previous experiences, as well as these of all the other users. 
	With \textit{personalisation}, users can leverage their preference to explore cyberspaces like Spotify and Netflix. Moving upward to the next level, users can proactively participate in \textit{content creation}, e.g., Super Mario Marker allows gamers to create their tailor-made game level(s).
	%With the user interaction enabling personalization and content creation, 
	Once a significant amount of user interaction records remain in the cyberspace, under the contexts of personalisation and content creation, the cyberspace evolves to a social community.
	%Once the users' interaction trace remains personalisation and content creation, they will formulate their digitised community like social networks nowadays (e.g., Twitter). 
	
	However, to the best of our knowledge, we rarely find real-life applications reaching the top levels of experience-duality that involves shared, open, and perpetual virtual worlds (according to the concepts mentioned above in Figure~\ref{fig:duality-continuum}). In brief, the experience-duality emphasises the perpetual virtual worlds that are paired up with the long-lasting physical environments. For instance, a person, namely Paul, can invite his metaverse friends to Paul's physical home, and Paul's friends as avatars can appear at Paul's home physically through technologies such as AR/VR/MR and holograms. Meanwhile, the avatars can stay in a virtual meeting room in the metaverse and talk to Paul in his physical environment (his home) through a Zoom-alike conversation window in a 3D virtual world.
	
	%Once a significant amount of user interaction records remain in the cyberspace, under the contexts of personalisation and content creation, the cyberspace evolves to a social community.

	To realise the metaverse, technologies other than the Internet, social networks, gaming, and virtual environments, should be taken into considerations. The advent of AR and VR, high-speed networks and edge computing , artificial intelligence, and hyperledgers (or blockchain), serve as the building blocks of the metaverse.
	From a technical point of view, we identify the fundamentals of the metaverse and its technological singularity. 
	This article reviews the existing technologies and technological infrastructures to offer a critical lens for building up the metaverse characterised by \textbf{perpetual, shared, concurrent, and 3D virtual spaces concatenating into a perceived virtual universe}. 
	The contribution of the article is threefold. 
	\begin{enumerate}
		\item We propose a technological framework for the metaverse, which paves a way to realise the metaverse. 
		\item By reviewing the state-of-the-art technologies as enablers to the development of the metaverse, such as edge computing, XR, and artificial intelligence, the article reflects the gap between the latest technology and the requirements of reaching the metaverse. 
		\item We propose research challenges and opportunities based on our review, paving a path towards the ultimate stages of the metaverse.
	\end{enumerate}

	This survey serves as the first effort to offer a comprehensive view of the metaverse with both the technology and ecosystem dimensions. Figure~\ref{fig:framework} provides an overview of the survey paper -- among the focused topics under the contexts of technology and ecosystem, the keywords of the corresponding topics reflect the key themes discussed in the survey paper. 
	In the next section, we first state our motivation by examining the existing survey(s) as well as relevant studies, and accordingly position our review article in Section~\ref{sec:related}. Accordingly, we describe our framework for the metaverse considering both technological and ecosystem aspects (Section~\ref{sec:framework}).

	\section{Related Work and Motivation}\label{sec:related}
	
	\begin{figure*}[!t]
		\centering
		\includegraphics[width=\linewidth]{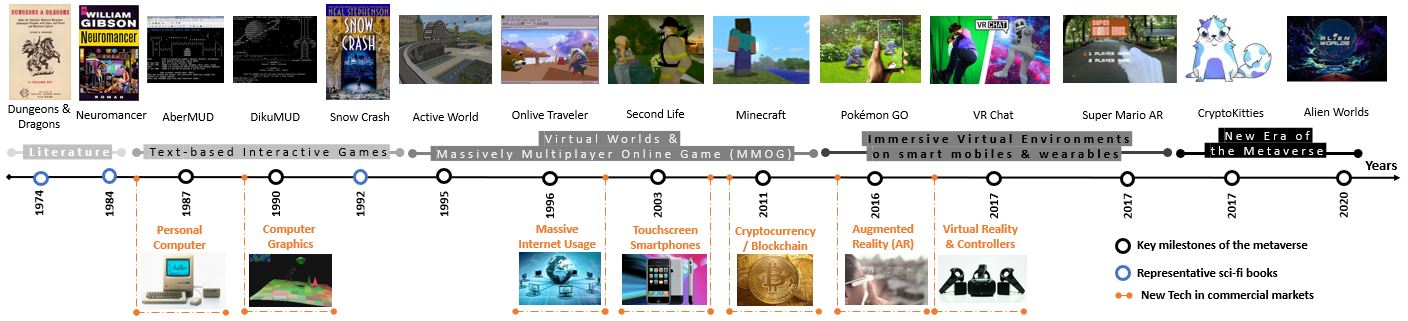}
		%Use of footnotemark, with the footnotetext below
		\caption{A timeline of the Metaverse Development from 1974 to 2020 (information source partially from \textit{T. Frey}\protect\footnotemark
			and~\cite{Duan2021MetaverseFS-ACMMM-2021}), demonstrating the evolving understanding of the metaverse once new technological infrastructures are introduced into the metaverse. With the evolving status of the metaverse, the metaverse has gained more enriched communication media -- text, graphics, 3D virtual worlds. Recently, AR applications demonstrate highly immersive digital overlays in the world, such as Pokémon GO and Super Mario AR, while VR applications (e.g., VR Chat) allow users to be fully immersed in virtual worlds for social gatherings. 
			The landscape of the metaverse is dynamic. For instance, cryptoassets (e.g., CryptoKitties) have appeared as in-game trading, while Alien Worlds encourages the users to earn non-fungible tokens (NFT) that can be converted into currencies in the real world\protect\footnotemark.}
		\label{fig:timeline}
	\end{figure*}
	
	To understand the comprehensive landscape of existing studies related to the metaverse, we decided to conduct a review of the relevant literature from 2012 to 2021 (i.e., ten years). In the first attempt of our search, we used the search keyword ``metaverse'' in the title, the abstract, or the body of the articles. We only focused on several primary sources known for high-quality studies on virtual environments, VR, AR, and XR: 
	(ACM CHI) the ACM CHI Conference on Human Factors in Computing Systems; 
	(IEEE ISMAR) IEEE International Symposium on Mixed and Augmented Reality; 
	(IEEE VR) IEEE Virtual Reality conference; 
	(ACM VRST) ACM Symposium on Virtual Reality Software and Technology. We obtained only two effective results from two primary databases of \textit{ACM Library} and \textit{IEEE Xplorer}, i.e., one full article related to the design of artificial moral agents, appeared in CHI~\cite{Metaverse-lit-search-robots-agents}; and one poster paper related to multi-user collaborative work for scientists in gamified environments, appeared in VRST~\cite{VRST-MR-lit}. 
	As the criteria applied in the first-round literature search made only a few eligible research articles, our second attempt relaxed the search criteria to papers with the identical search keyword of `metaverse', regardless of the publication venues. The two primary databases of \textit{ACM Library} and \textit{IEEE Xplorer} resulted in 43 and 24 entities (Total = 67), respectively. Then, we only included research article written in English, and excluded demonstration, book chapters, short papers, posters, and articles appeared as workshops, courses, lectures, interviews, opinions, columns, and invited talks -- when the title, abstracts, and keywords in the articles did not provide apparent reasons for exclusion, we read the entire article and briefly summarise the remaining 30 papers in the coming paragraphs.
	
	First, we spot a number of system solutions and architectures for resolving scalability issues in the metaverse, such as balancing the workload for reduced response time in Modern massively multiplayer online games (MMOGs)~\cite{marzolla_dynamic_2012}, unsupervised conversion of 3D models between the metaverse and real-world environments~\cite{terrace_unsupervised_2012}, High performance computing clusters for large-scale virtual environments~\cite{goel_research_2015}, analyzing underground forums for criminal acts (e.g., trading stolen items and datasets) in virtual worlds~\cite{portnoff_tools_2017}, exploration of new composition and spatialization methods in
	virtual 3D spaces under the context of multiplayer situations~\cite{webster_empty_2017}, governing user-generated contents in gaming~\cite{kasapakis_user-generated_2017}, 
	strengthening the integration and interoperability of highly disparate virtual environments inside the metaverse~\cite{nevelsteen_ipsme-_2021}, and redistributing network throughput in virtual worlds to improve user experiences through avatars in virtual environments~\cite{oliver_mongoose_2012}. 
	% The next one who uses footnotes in figures, I'll personally kick his ass!
	% Captions are processed twice, so the footnote counter increases twice per footnotes. As I have two footnotes here, I need to remove 2 to the counter
	\addtocounter{footnote}{-2}

	% Need to fuck around with the counter once again to put it at the number of the first footnote
	\addtocounter{footnote}{-1}
	\footnotetext{\url{https://futuristspeaker.com/future-trends/the-history-of-the-metaverse/}}
	% As the counter does nto increase automatically, we need to add another one
	\stepcounter{footnote}\footnotetext{\url{https://coinmarketcap.com/currencies/alien-worlds/}}

	Second, we spot three articles proposing user interaction techniques for user interaction across the physical and virtual environments. Young~\emph{et al.} proposed an interaction technique for users to make high-fiving gestures being synchronised in both physical and virtual environments~\cite{young_dyadic_2015}. Vernaza~\emph{et al.} proposed an interactive system solution for connecting the metaverse and real-world environments through tablets and smart wearables~\cite{vernaza_towards_2012}. Next, Wei~\emph{et al.} made user interfaces for the customisation of virtual characters in virtual worlds~\cite{wei_design_2015}.

	Third, the analysis of user activities in the metaverse also gains some attention from the research community. The well-recognised clustering approaches could serve to understand the avatar behaviours in virtual environments~\cite{orgaz_clustering_2012}, and the text content created in numerous virtual worlds~\cite{bello-orgaz_comparative_2013}. As the metaverse may bridge the users with other non-human animated objects, an interesting study by Barin~\emph{et al.} ~\cite{barin_understanding_2017} focuses on the crash incident of high-performance drone racing through the first-person view on VR headsets. The concluding remark of their study advocates that the physical constraints such as acceleration and air resistance will no longer be the concerns of the user-drone interaction through virtual environments. Instead, the design of user interfaces could limit the users' reaction times and lead to the critical reasons for crash incidents. 
	
	Next, we report the vastly diversified scenes of virtual environments, such as virtual museums~\cite{beer_virtual_2015}, ancient Chinese cities~\cite{wei_exploring_2014}, and virtual laboratories or classrooms~\cite{chishiro_global_2017, sebastien_providing_2018, schaf_3d_2012, tarouco_virtual_2013}. We see that the existing virtual environments are commonly regarded as a collaborative learning space, in which human users can finish some virtual tasks together under various themes such as learning environmental IoT~\cite{chishiro_global_2017}, teaching calculus~\cite{tarouco_virtual_2013}, avatar designs and typographic arts in virtual environments~\cite{ayiter_further_2012, ayiter_azimuth_2015}, fostering Awareness of the Environmental Impact of Agriculture~\cite{prada_agrivillage_2015}, and presenting the Chinese cultures~\cite{wei_exploring_2014}.

	Finally, we present the survey articles found in the collection of research articles. Only one full survey article, two mini-surveys, and three position papers~\cite{falchuk_social_2018, ylipulli_chasing_2016} exist. The long survey written by Dionisio~\emph{et al.}~\cite{dionisio_3d_2013} focuses on the development of the metaverse, and accordingly discusses four aspects of realism, ubiquity, interoperability, scalability. The two mini-surveys focus on the existing applications and headsets for user interaction in virtual environments, as well as various artistic approaches to build artwork in VR~\cite{kelley_artistic_2019, plasencia_one_2015}. Regarding the position papers, Ylipulli~\emph{et al.}~\cite{ylipulli_chasing_2016} advocates design frameworks for future hybrid cities and the intertwined relationship between 3D virtual cities and the tangible counterparts, while another theoretical framework classifies instance types in the metaverse, by leveraging the classical Vitruvian principles of Utilitas, Firmitas, and Venustas~\cite{ibnez_cyberarchitecture_2012}. Additionally, as the metaverse can serve as a collective and shared public space in virtual environments, user privacy concerns in such emerging spaces have been discussed in~\cite{falchuk_social_2018}.
	
	As we find a limited number of existing studies emphasising the metaverse, we view that the metaverse research is still in its infancy. Therefore, additional research efforts should be extended in designing and building the metaverse. 
	Instead of selecting topics in randomised manners, we focus on two critical aspects -- technology and ecosystem, with the following justifications.
	First, the technological aspect serves as the critical factor to shape the metaverse. Figure~\ref{fig:timeline} describes the timeline of the metaverse development. The metaverse has experienced four transitions from text-based interactive games, virtual open worlds, Massively Multiplayer Online Game (MMOG), immersive virtual environments on smart mobiles and wearables, to the current status of the metaverse. Each transition is driven by the appearance of new technology such as the birth of the Internet, 3D graphics, internet usage at-scale, as well as hyperledger. It is obvious that technologies serve as the catalysts to drive such transitions of cyberspaces.

	In fact, the research community is still on the way to exploring the metaverse development. 
	Ideally, new technology could potentially unlock additional features of the metaverse and drive the virtual environments towards a perceived virtual universe. 
	Thus, we attempt to bridge various emerging technologies that could be conducive to the further progress of the metaverse. 
	After discussing the potential of various emerging technologies, the game-based metaverse can open numerous opportunities, and eventually may reach virtual environments that is a society parallel to the existing one in the real world, according to the three-stage metaverse as discussed in Section~\ref{sec:introduction}. 
	Our survey paper, therefore, projects the design of metaverse ecosystems based on the society in our real world. The existing literature only focuses on fragmented issues such as user privacy~\cite{falchuk_social_2018}. It is necessary to offer a holistic view of the metaverse ecosystem, and our article serves this purpose.
	% survey The metaverse - limited on privacy, and user sense of presence, separated fragment aspects
	% game-based now, but we go beyond, as stated in technological supports + 3 stages, all rounded society - virtual environments.
	
	Before we begin the discussion of the technologies and the issues of ecosystems in Section~\ref{sec:framework}, here we pinpoint the interdisciplinary nature of the metaverse. Thus, the survey covers \textcolor{black}{fourteen} diversified topics linked to the metaverse. Technical experts, research engineers, and computer scientists can understand the latest technologies, challenges, and research opportunities for shaping the future of the metaverse. This article connects the relationship between the \textcolor{black}{eight} technological topics, and we did our utmost to demonstrate their relationship. On the other hand, social scientist, economists, avatar and content creators, digital policy makers, and governors can understand the indispensable \textcolor{black}{six} building blocks to construct the ecosystems of the metaverse, and how the emerging technologies can bring impacts to both physical and virtual worlds. In addition, other stakeholders who have already engaged in the metaverse, perhaps focusing on the game-oriented developments, can view our article as a reflection of when technological catalysts further drive the evolution of the metaverse, and perhaps the `\textit{Digital Big Bang}'.

	\section{Framework}\label{sec:framework}
	\begin{figure}[!t]
		\centering
		\includegraphics[width=.95\linewidth]{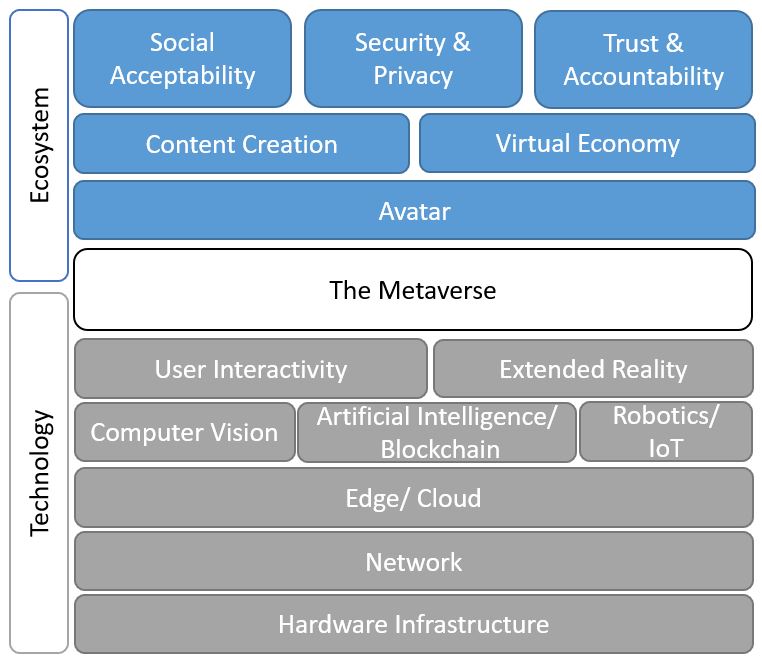}
		\caption{The fourteen focused areas, under two key aspects of technology and ecosystem for the metaverse. The key technologies fuel the `\textit{Digital Big Bang}' from the Internet and XR to the metaverse, which support the metaverse ecosystem. }
		\label{fig:draft-FM}
	\end{figure}
	Due to the interdisciplinary nature of the metaverse, this section aims to explain the relationship between the \textcolor{black}{fourteen} focused areas under two key categories of technologies and ecosystems, before we move on to the discussion on each focused area(s). Figure~\ref{fig:draft-FM} depicts the focused areas under the two categories, where the technology supports the metaverse and its ecosystem as a gigantic application. % views the metaverse as a gigantic application and simultaneously supports the ecosystem. 
	
	Under the technology aspect, i.e., the eight pillars for the metaverse, human users can access the metaverse through extended reality (XR) and  techniques for user interactivity (e.g., manipulating virtual objects). Computer vision (CV), artificial intelligence (AI), blockchain, and robotics/ Internet-of-Things (IoT) can work with the user to handle various activities inside the metaverse through user interactivity and XR. 
	Edge computing aims to improve the performance of applications that are delay-sensitive and bandwidth-hungry, through managing the local data source as pre-processing data available in edge devices, while cloud computing is well-recognised for its highly scalable computational power and storage capacity. Leveraging both cloud-based and edge-based services can achieve a synergy, such as maximising the application performance and hence user experiences. 
	Accordingly, edge devices and cloud services with advanced mobile network can support the CV, AI, robots, and IoT, on top of appropriate hardware infrastructure.
	
	%On the other hand, 
	The ecosystem describes an independent and meta-sized virtual world, mirroring the real world. Human users situated in the physical world  can control their avatars through XR and user interaction technique  for various collective activities such as content creation. Therefore, virtual economy is a spontaneous derivative of such activities in the metaverse. We consider three focused areas of Social acceptability, security and privacy, as well as trust and accountability. Analogue to the society in the physical world, content creation and virtual economy should align with the social norms and regulations. For instance, the production in the virtual economy should be protected by ownership, while such production outcomes should be accepted by other avatars (i.e.,human users) in the metaverse. Also, human users would expect that their activities are not exposed to privacy risks and security threats.

	The structure of the paper is as follows. 
	%Then, we propose our technological framework in Section~\ref{sec:framework}. 
	Based on the proposed framework, we review fourteen key aspects that critically contribute to the metaverse. We first discuss the technological aspect -- XR (Section~\ref{sec:XR}), user interaction in XR and ubiquitous interfaces (Section~\ref{sec:user-interact}), robotics and IoT (Section~\ref{sec:iot-robot}), artificial intelligence (Section~\ref{sec:art-int}), computer vision (Section~\ref{sec:cv}), hyperledger supporting various user activities and the new economy in the metaverse market (Section~\ref{sec:blockchain}), edge computing (Section~\ref{sec:edge}), the future network fulfilling the enormous needs of the metaverse (Section~\ref{sec:network}). 
	Regarding the ecosystem on the basis of the aforementioned technologies, we first discuss the key actors of the metaverse -- avatars representing the human users in Section~\ref{sec:avatar}. Next, we discuss the content creation (Section~\ref{sec:content-create}) and virtual economy (Section~\ref{sec:ntf-econ}), and the corresponding social norms and regulations -- Social Acceptability (Section~\ref{sec:social}), Privacy and Security (Section~\ref{sec:privacy-security}), as well as Trust and Accountability (Section~\ref{sec:trust}).
	Finally, Section~\ref{sec:grand} identifies the grand challenges of building the metaverse, and discusses the key research agenda of driving the `\textit{Digital Big Bang}' and contributing to a unified, shared and collective space virtually.

	\section{Extended Reality (XR)}\label{sec:XR}
	Originated from the \textit{Milgram and Kishino's Reality-Virtuality Continuum}~\cite{continuum}, the most updated continuum has further included new branches of alternated realities, leaning towards the side of physical realities~\cite{10.3389/frvir.2021.647997}, namely MR~\cite{WhatMR-CHI2019} and the futuristic holograms like the digital objects shown in the Star Trek franchise~\cite{Sutherland1965TheUD}. 
	The varied categories inside the continuum allow human users to experience the metaverse through various alternated realities across both the physical and digital worlds~\cite{PAKANEN2022100457}. However, we limited our discussion to four primary types of realities that gain a lot of attention from the academia and industry sectors~\cite{u1, csur-display-2021July, vR-headset-IEEEVR2021}. 
	This section begins with the well-recognised domain of VR, and progressively discusses the emerging fields of AR and its advanced variants, MR and holographic technologies. This section also serves as an introduction to how XR bridging the virtual entities with the physical environments.
	
	\subsection{Virtual Reality (VR)} 
	VR owns the prominent features of totally synthetic views. The commercial VR headsets provide usual way of user interaction techniques, including head tracking or tangible controllers~\cite{vR-headset-IEEEVR2021}. As such, users are situated in fully virtual environments, and interacts with virtual objects through user interaction techniques. In addition, VR is known as `the farthest end from the reality in \textit{Reality-Virtuality Continuum}'~\cite{continuum}. That is, the users with VR headsets have to pay full attention to the virtual environments, and hence separate from the physical reality~\cite{WhatMR-CHI2019}.
	As mentioned, the users in the metaverse will create contents in the digital twins. Nowadays, commercial virtual environments enable users to create contents, e.g., VR painting\footnote{Six artists collaborate to do a VR painting of Star Wars with Tilt Brush:\url{https://www.digitalbodies.net/virtual-reality/six-artists-vr-painting-star-wars/}}. The exploration of user affordance can be achieved by user interaction with virtual entities in a virtual environment, for instance, modifying the shape of a virtual object, and creating new artistic objects. Multiple Users in such virtual environments can collaborate with each other in real-time. This aligns with the well-defined requirements of virtual environments: 
	a shared sense of space, a shared sense of presence, a shared sense of time (real-time interaction), a way to communicate (by gesture, text, voice, etc.), and a way to share information and manipulate objects~\cite{Singhal1999NetworkedVE}. 
	It is important to note that multiple users in a virtual world, i.e., a subset of the metaverse, should receive identical information as seen by other users. Users also can interact with each other in consistent and real-time manners. In other words, how the users should precept the virtual objects and the multi-user collaboration in a virtual shared space would become the critical factors. Considering the ultimate stage of the metaverse, users situated in a virtual shared space should work simultaneously with any additions or interactions from the physical counterpart, such as AR and MR.
	The core of building the metaverse, through composing numerous virtual shared space, has to meld the simultaneous actions, among all the objects, avatars representing their users, and their interactions, e.g., object-avatars, object-object, and avatar-avatar. All the participating processes in virtual environments should synchronise and reflect the dynamic states/events of the virtual spaces~\cite{vR-meld}. However, managing and synchronising the dynamic states/events at scale is a huge challenge, especially when we consider unlimited concurrent users collectively act on virtual objects and interact with each other without sensible latency, where latency could negatively impact the user experiences. 
	
	\subsection{Augmented Reality (AR)}~\label{ssec:xr-ar}
	Going beyond the sole virtual environments, AR delivers alternated experiences to human users in their physical surroundings, which focuses on the enhancement of our physical world. In theory, computer-generated virtual contents can be presented through diversified perceptual information channels, such as audio, visuals, smell, and haptics~\cite{AR-eating-CHI2011, u231, u232}. 
	The first-generation of AR system frameworks only consider visual enhancements, which aim to organise and display digital overlays superimposing on top of our physical surroundings. As shown in very early work in the early 1990s~\cite{u223}, a bulky see-through display did not consider user mobility, which requires users to interact with texts and 2D interfaces with tangible controllers in a sedentary posture. 
	
	Since the very first work, significant research efforts have been made to improve the user interaction with digital entities in AR. It is important to note that the digital entities, perhaps from the metaverse, overlaid in front of the user's physical surroundings, should allow human users to meld the simultaneous actions (analogue to VR). As such, guaranteeing seamless and lightweight user interaction with such digital entities in AR is one of the key challenges, bridging human users in the world physical with the metaverse~\cite{u232}. Freehand interaction techniques, as depicted in most science fiction films like \textit{minority report}\footnote{\url{https://www.imdb.com/title/tt0181689/}}, illustrate intuitive and ready-to-use interfaces for AR user interactions~\cite{u1}. A well-known freehand interaction technique named Voodoo Dolls~\cite{u230} is a system solution, in which users can employ two hands to choose and work on the virtual contents with pinch gestures. HOMER~\cite{u229} is another type of user interaction solution that provides a ray-casting trajectory from a user's virtual hand, indicating the AR objects being selected and subsequently manipulated. 
	
	Moreover, AR will situate everywhere in our living environments, for instance, annotating directions in an unfamiliar place, and pinpointing objects driven by the user contexts~\cite{Lee2020TowardsAR}. As such, we can consider that the metaverse, via AR, will integrate with our urban environment, and digital entities will appear in plain and palpable ways on top of numerous physical objects in urban areas. In other words, users with AR work in the physical environments, and simultaneously communicate with their virtual counterparts in the metaverse. This requires significant efforts in the technologies of detection and tracking to map the virtual contents displayed with the corresponding position in the real environment~\cite{u225,u226,u227,u228}. A more detailed discussion will be available in Section~\ref{sec:cv}.
	Touring Machine is considered as the first research prototype that allows users to experience AR outdoors. The prototype consists of computational hardware and a GPS unit loaded on a backpack, plus a head-worn display that contains map navigation information. The user with Touring Machine can interact with the AR map through a hand-held touch-sensitive surface and a stylus~\cite{u224}.
	In contrast, the recent AR headsets have demonstrated remarkable improvements, especially in user mobility. Users with lightweight AR headsets can receive visual and audio feedback cues indicating AR objects, but other sensory dimensions such as smell and haptics are still neglected~\cite{u1}. 
	It is worth pinpointing that AR headsets are not the only options to access the contents from the metaverse.
	When we look at the current status of AR developments, AR overlays, and even digital entities from the metaverse, can be delivered by various devices, including but not limited to AR headsets~\cite{u1, u2}, hand-held touchscreen devices~\cite{u239}, ceiling projectors~\cite{u235}, and tabletops~\cite{u234}, Pico (wearable) projectors~\cite{Hartmann2020AAR} and so on. 
	Nevertheless, AR headsets own advantages over other approaches, in terms of the switch of user attention and occupying users' hands. First, human users have to switch their attention between physical environments and digital content on other types of AR devices. In contrast, AR headsets enable AR overlays displayed in front of the user's sight~\cite{u7, u144}. Second, the user's hands will not be occupied by the tangible devices as the computational units and displays are mounted on the users' heads. Such advantages enable users with AR headsets to seamlessly experience `\textit{the metaverse through an AR lens}'. More elaboration of the user interactivity is available in Section~\ref{sec:user-interact}. 
	% Although smartphones have become the mainstream testbed for AR applications, with the most remarkable example of Pokémon Go has over 1 billion downloads, two fundamental limitations (dual views and busy hand(s)) exist. First, the user needs to switch his attention between the digital contents on the touchscreen and the physical objects in real-world environments~\cite{u7, u144, u238}. Second, the user's hands are occupied by holding a smartphone, and the user has to lift his hand with the hand-held device to view and interact with the AR objects.
	% In contrast, AR headsets overcome these issues. The headset configuration allows digital overlays to be projected in front of the user's eyes and frees the user's hands from holding the devices. Such advantages facilitate mobile AR, and the users with AR smartglasses are able to experience `the world as the user interfaces'~\cite{u231}. 
	
	\begin{figure}[!t]
		\centering
		\includegraphics[width=\linewidth]{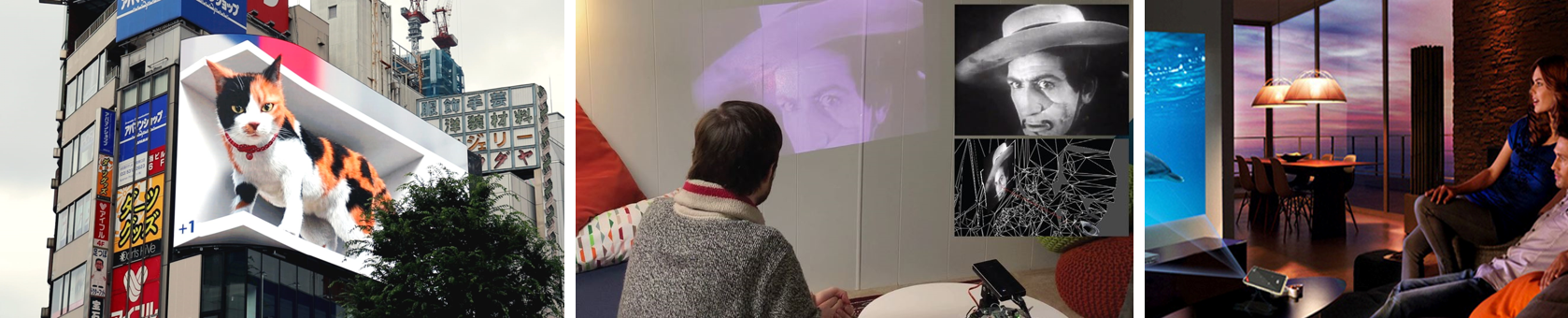}
		\caption{Displaying virtual contents with mature technologies: Public Large Display (Left); Pico-projector attached on top of a wearable computer (Middle), and; mini-projector inside a smartphone (Right).}
		\label{fig:alter-display}
	\end{figure}

	\subsection{Mixed Reality (MR)}
	After explaining the two extremes of the Reality–Virtuality Continuum~\cite{Milgram1994ATO} -- AR and VR, we attempt to discuss the relationship between the metaverse and MR. Unfortunately, there exists no commonly agreed definition for MR, but it is crucial to have a common term that describes the alternated reality situated between two extremes of AR and VR. 
	Nevertheless, the vastly different definitions can be summarised into six working definitions~\cite{WhatMR-CHI2019}, including the ``traditional'' notion of MR in the middle space of the Reality–Virtuality Continuum~\cite{Milgram1994ATO}, MR as a synonym for AR~\cite{36-10.1145/3173574.3174020}, MR as a type of collaboration~\cite{63-10.1145/2702123.2702506}, MR as a combination of AR and VR~\cite{53-10.1109/ISMAR.2015.60}, MR as an alignment of environments~\cite{69-10.1145/3025453.3025743}, a ``stronger'' version of AR~\cite{88-10.1145/3126594.3126601}.

	The above six definitions have commonly appeared in the literature related to MR. The research community views that MR stands between AR and VR, and allows user interaction with the virtual entities in physical environments. It is worthwhile to mention that MR objects, supported by a strong capability of environmental understandings or situational awareness, can work with other tangible objects in various physical environments. For instance, a physical screwdriver can fit turn digital entities of screws with slotted heads in MR, demonstrating an important feature of interoperability between digital and physical entities. In contrast, as observed in the existing applications~\cite{u1}, AR usually simply displays information overlaid on the physical environments, without considering such interoperability. Considering such an additional feature, MR is viewed as a stronger version of AR in a significant number of articles that draw more connected and collaborating relationships among the physical spatial, user interaction, and virtual entities~\cite{world-as-support-CHI-17, u1, gardony2020eye, Lee2020TowardsAR}. 
	
	From the above discussion, albeit we are unable to draw a definitive conclusion to MR, MR is the starting point for the metaverse, and certain properties of the six working definitions are commonly shared between the metaverse and MR. We consider that the metaverse begins with the digital twins that connect to the physical world~\cite{Grieves2017DigitalTM, 10.1109/SESoS/WDES.2019.00018, 10.1145/3387940.3392199, HRI-risk, Cureton2020DigitalTO, Langen2017ConceptFA}. Human users subsequently start content creation in the digital twins~\cite{Viljoen2020ThePA, Lau-chi21, XR-studio, TutorVR-CHI19, communication-chi16}. Accordingly, the digitally created contents can be reflected in physical environments, while human users expect such digital objects to merge with our physical surroundings across space and time~\cite{space-time-MR}. 
	Although we cannot accurately predict how the metaverse will eventually impact our physical surroundings, we see the existing MR prototypes enclose some specific goals such as pursuing scenes of realism~\cite{realism-MR-TVCG}, bringing senses of presence~\cite{presence-sen-haptic}, creating empathetic physical spatial~\cite{empathy-MR}. These goals can be viewed as an alignment with the metaverse advocating that multiple virtual worlds work complementary with each other\cite{Grieves2017DigitalTM}.

	\subsection{Large Display, Pico-Projector, Holography}
	\begin{figure}[!t]
		\centering
		\includegraphics[width=\linewidth]{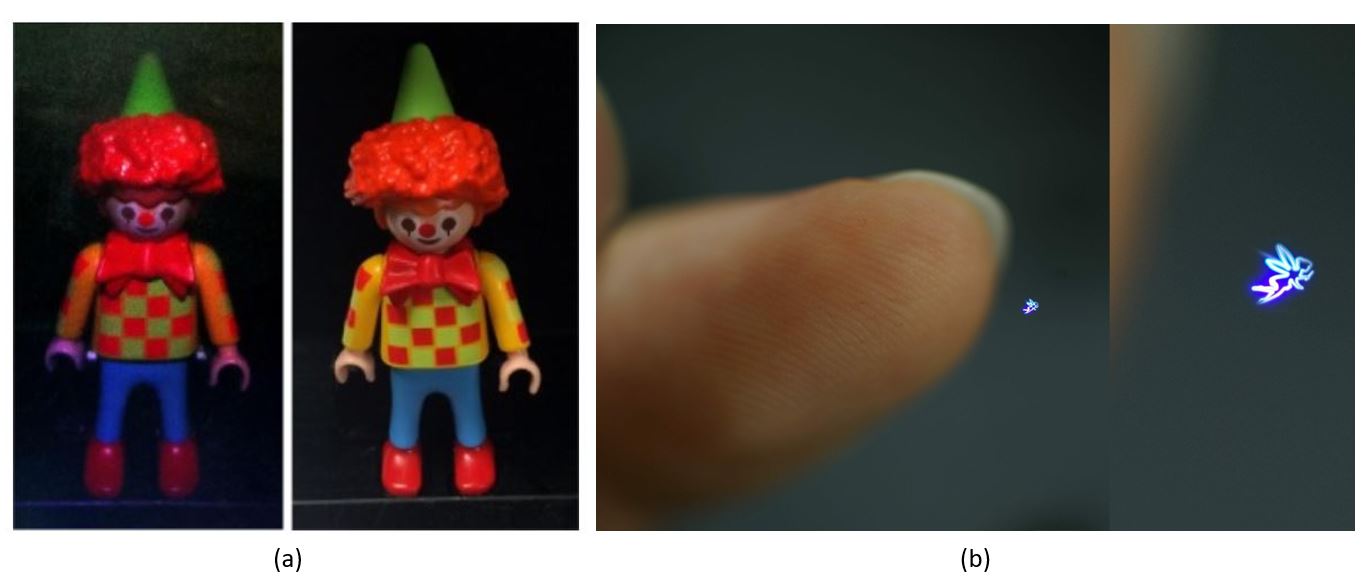}
		\caption{Two holography types: (a) The reflection-based~\cite{VzquezMartn2021FullcolorMR} approach can reproduce colourful holography highly similar to the real object, and; (b) The laser-driven approach can produce a sense of touch to the user's skin surface~\cite{ackerman_2021}. }
		\label{fig:holo-ex}
	\end{figure}
	%https://www.eeweb.com/hologram-types/
	
	Based on the existing literature, this paragraph aims to make speculation for the ways of bringing the uniquely created contents inside the virtual environments (ultimately metaverse) back to the physical counterparts in the shared public space. As the social acceptability of mobile headsets in public spaces is still questionable~\cite{mobile-CHI-social-accept}, we lack evidence that mobile headsets will act as the sole channel for delivering metaverse contents into the public space. Instead, other mature technologies such as large displays and pico-projectors may serve as a channel to project pixels into our real world. 
	Figure~\ref{fig:alter-display} depicts three examples. Large displays\footnote{A giant 3D cat has taken over one of Tokyo's biggest billboards: \url{https://edition.cnn.com/style/article/3d-cat-billboard-tokyo/index.html}}, and pico-projectors~\cite{Hartmann2020AAR} allow users without mobile headsets to view digital entities with a high degree of realism. In addition, miniature projectors embedded inside smartphones, e.g., MOVI Phone\footnote{MOVI-phone: \url{https://moviphones.com/}}, allow content sharing anytime and anywhere. It is also worth noting that smartphones are the most ubiquitous devices nowadays. 
	
	Finally, we discuss the possibility of holographic technology emphasising enriched communication media exceeding the 2D displays~\cite{Kubota1992CreatingAM} and pursuing true volumetric displays (showing images or videos) that show no difference from our everyday objects.
	The current holographic technology can be classified into two primary types: reflection-based and laser-driven holograph\footnote{\url{https://mitmuseum.mit.edu/holography-glossary}}.
	A recent work~\cite{Rogers2021SimulatingVI} demonstrated the feasibility of colourful volumetric display on bulky and sedentary devices, with practical limitations of low resolution that could impact the user perceptions to realism. However, the main advantage of reflection-based holography is to generate the colourful holograms with colour reproduction highly similar to real-life objects~\cite{VzquezMartn2021FullcolorMR} (Figure~\ref{fig:holo-ex}(a)). 
	On the other hand, \textit{Plasma Fairies}~\cite{ackerman_2021} is a 3D aerial hologram that can be sensed by the users' skin surfaces, though the devices can only produce plasmonic emission in a mid-air region no larger than 5 $cm^3$ (Figure~\ref{fig:holo-ex}(b)). 
	We conjecture that if technology breakthrough allows such volumetric 3D objects to appear in the real world ubiquitously, it will come as no surprise that the metaverse can merge with our living urban, as illustrated in Figure~\ref{fig:framework} (top-right corner), and provide a strong sense of presence to the stakeholders in urban areas.
	However, holographic technology suffers from three key weaknesses in the above works, including limited resolution, display size, as well as device mobility. Thus, overcoming such weaknesses becomes the critical turning point of delivering enriched 3D images in the real world. 
	
	\section{User Interactivity}\label{sec:user-interact}
	This section first reviews the latest techniques that enable users to interact with digital entities in physical environments. Then, we pinpoint the existing technologies that display digital entities to human users. We also discuss the user feedback cues as well as the haptic-driven telepresence that connects the human users in physical environments, avatars in the metaverse, and digital entities throughout the advanced continuum of extended reality. 
	
	\subsection{Mobile Input Techniques}\label{ssec:mobile-in}
	
	As the ultimate stage of the metaverse will interconnect both the physical world and its digital twins, all human users in the physical world can work with avatars and virtual objects situated in both the metaverse and the MR in physical environments, i.e., both the physical and virtual worlds constantly impact each other. It is necessary to enable users to interact with digital entities ubiquitously. However, the majority of the existing metaverse only allows user interactions with the keyboards and mice duo, which cannot accurately reflect the body movements of the avatar~\cite{Duan2021MetaverseFS-ACMMM-2021}. Also, such bulky keyboards and mice are not designed for mobile user interaction, and thus enforce users to maintain sedentary postures (e.g., sitting)~\cite{Lee2020TowardsAR, u1}.
	
	Albeit freehand interaction features intuitiveness due to barehanded operations~\cite{u1} and further achieve object pointing and manipulation~\cite{u152}, most freehand interactions rely on computer vision (CV) techniques. Thus, accurate and real-time recognition of freehand interaction is technically demanding, even the most fundamental mid-air pointing needs sufficient computational resources~\cite{u3}. Insufficient computational resources could bring latency to user actions and hence deteriorate the user experience~\cite{Lee2020UbiPointTN}. Apart from CV-based interaction techniques, the research community search vastly diversified input modality to support complicated user interaction, including optical~\cite{u79}, IMU-driven~\cite{u100}, Pyroelectric Infrared~\cite{u78}, electromagnetic~\cite{u147}, capacitive~\cite{u90}, and IMU-driven user interactions~\cite{u100}. Such alternative modalities can capture user activities and hence interact with the digital entities from the metaverse. 
	
	We present several existing works to illustrate the mobile input techniques with alternative input modals, as follows. 
	First, the human users themselves could become the most convenient and ready-to-use interaction surface, named as on-body user interaction~\cite{u1}. For instance, ActiTouch~\cite{u90} owns a capacitive surface attached to the user's forearm. The electrodes in ActiTouch turn the user's body into a spacious input surface, which implies that users can perform taps on their bodies to communicate with other stakeholders across various digital entities in the metaverse. Another similar technique~\cite{u75} enriched the set of input commands, in which users can interact with icons, menus, and other virtual objects as AR overlaid on the user's arm. Additionally, such on-body interaction can be employed as a solution for interpersonal interactions that enable social touch remotely~\cite{u91,u95}. Such on-body user interaction could enrich the communication among human users and avatars.
	The latest technologies of on-body interaction demonstrate the trend of decreasing device size, ranging from a palm area~\cite{u61, u68, u119} to a fingertip~\cite{u5}. The user interaction, therefore, becomes more unnoticeable than the aforementioned finger-to-arm interaction.
	Nevertheless, searching alternative input modalities does not mean that the CV-based techniques are not applicable. The combined use of alternative input modals and CV-based techniques can maintain both intuitiveness and the capability of handling time-sensitive or complicated user inputs~\cite{u1}. For instance, a CV-based solution works complementary to IMU sensors. The CV-based technique determines the relative position between the virtual objects user hands in mid-air, while the IMU sensors enable subtle and accurate manipulation of virtual objects~\cite{u100}. 
	
	Instead of attaching sensors to our body, another alternative is regarded as digital textile. Digital textile integrates novel material and conductive threads inside the usual fabrics, which supports user interactions with 2D and 3D user interfaces (UIs). 
	Research prototypes such as PocketThumb~\cite{u259} and ARCord~\cite{u260} convert our clothes into user interfaces with the digital entities in MR. PocketThumb~\cite{u259} is a smart fabric located at a front trouser pocket. Users can exert taps and touches on the fabrics to perform user interaction, e.g., positioning a cursor during pointing tasks with 3D virtual objects in MR. Also, ARCord~\cite{u260} is a cord-based textile attached to a jacket, and users can rub the cord to perform menu selection and ray-casting on virtual objects in various virtual environments. Remarkably, technology giants have invested in this area to support the next generation of mobile user inputs. For example, Google has launched the Jacquard project~\cite{u258} that attempts to produce smart woven at an affordable price and in a large scale. As a result, the smart woven can merge with our daily outfits such as jackets and trousers, supporting user inputs anywhere and anytime. 
	Although we cannot discuss all types of mobile inputs due to limited space, the research community is searching for more natural, more petite, subtle and unnoticeable interfaces for mobile inputs and alternative input modals in XR, e.g., Electroencephalography (EEG) and Electromyography (EMG)~\cite{kirill-survey, 9156084}.

	\subsection{New Human Visions via Mobile Headsets}
	
	Mobile headsets, as discussed in Section~\ref{ssec:xr-ar}, owns key advantages such as aligned views between physical and virtual realities, and user mobility, which can be regarded as an emerging channel to display virtual content ubiquitously~\cite{mobile-CHI-social-accept}. As VR mobile headsets will isolate human users from the physical realities~\cite{vR-headset-IEEEVR2021} and its potential dangers in public spaces~\cite{dao-breakdown-chi21}, in this section, we discuss the latest AR/MR headsets that are designed for merging virtual contents in physical environments. 
	
	Currently, the user immersiveness in the metaverse can be restricted by limited Field of View (FOV) on AR/MR mobile headsets. Narrowed FOVs can negatively influence the user experience, usability, and task performance~\cite{u7, u26}. The MR/AR mobile headsets usually own FOVs smaller than 60 degrees. The limited FOV available on mobile headsets is far smaller than the typical human vision. For instance, the FOV can be equivalent to a 25-inch display 240 cm away from the user's view on the low-specification headsets such as Google Glass. The first generation of Microsoft Hololens presents a 30 X 17-degree FOV, which is a similar size as a 15-inch 16:9 display located around 60 cm away from the user's egocentric view. We believes that the restricted view will be eventually resolved by the advancement of display technologies, for instance, the second generation of Microsoft Hololens owns an enlarged display having 43 X 29-degree FOV. Moreover, the bulky spectacle frames on MR headsets, such as Microsoft Hololens, can occlude the users' peripheral vision. As such, users can reduce their awareness of incoming dangers as well as critical situations~\cite{u125}. Thus, other form factors such as contact lens can alleviate such drawbacks. A prototypical AR display in the form factor of contact lens~\cite{u159}, albeit offering low-resolution visuals to users, can provide virtual overlays, e.g., top, down, left, right directions in navigation tasks. 
	
	\begin{figure}[!t]
		\centering
		\includegraphics[width=\linewidth]{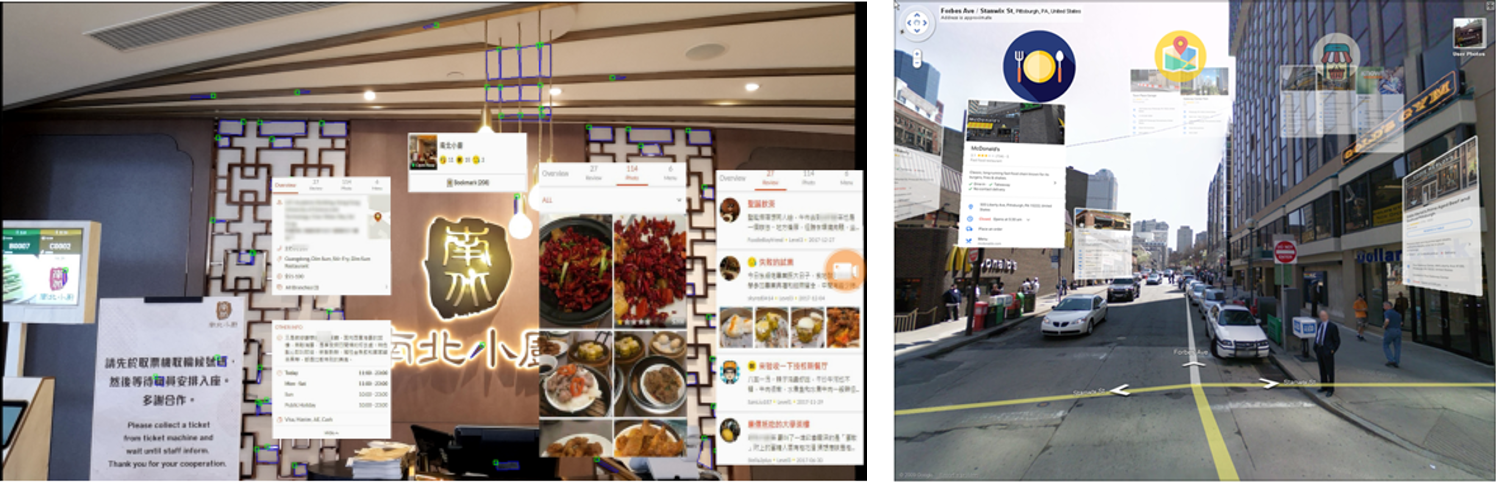}
		\caption{Displaying virtual contents overlaid on top of physical environments: a restaurant (indoor, left)~\cite{8767388}, a street (outdoor, right)~\cite{A2W-LAM2021}.}
		\label{fig:a2w-m2a}
	\end{figure}
	
	The remaining section discusses the design challenges of presenting virtual objects through mobile headsets, how to leverage the human visions in the metaverse. 
	First, one design strategy is to leverage the users' peripheral visual field~\cite{u167} that originally aims to identify obstacles, avoid dangerous incidents, and measure foot placements during a wide range of locomotive activities, e.g., walking, running, driving and other sport activities~\cite{u29}. Combined with other feedback cues such as audio and haptic feedback, users can sense the virtual entities with higher granularity~\cite{u167}. Recent works also present this design strategy by displaying digital overlays at the edge areas of the FOVs on MR/AR mobile headsets~\cite{u2,u7,u16,seen-to-unseen}. The display of virtual overlays at edge areas can result in practical applications such as navigation instructions of straight, left, and right during a navigation task on AR maps~\cite{u7}. 
	A prominent advantage of such designs is that the virtual overlays on the users' peripheral visions highly aligns with the locomotive activities. As such, users can focus on other tasks in the physical world, without significant interruption from the virtual entities from the metaverse. It is important to note that other factors should be considered together when presenting virtual overlays within the users' visual fields, such as colour, illumination~\cite{u15}, content legibility, readability~\cite{u49}, size, style~\cite{u24}, visual fatigue~\cite{u169}, movement-driven shakiness~\cite{u12}. Also, information overflow could ruin the user ability to identify useful information. Therefore, appropriate design of information volume and content placements (Figure~\ref{fig:a2w-m2a}) is crucial to improving the effectiveness of displaying virtual overlays extracted from the metaverse~\cite{8767388, u13, u28,A2W-LAM2021}.
	
	\subsection{The importance of Feedback Cues}\label{ssec:haptic}
	
	After considering the input and output techniques, the user feedback cues is another important dimension for user interactivity with the metaverse. We attempt to explain this concept with the fundamental elements in 3D virtual worlds -- user interaction with virtual buttons~\cite{gordon2019touchscreen-10.1145/3290605, doerrer2002simulating, carlos-eics-button}. Along with the above discussions, virtual environments can provide highly adaptive yet realistic environments~\cite{vr-advantage-10.1145/3359996.3364246}, but the usability and the sense of realism are subject to the proper design of user feedback cues (e.g., visual, audio, haptic feedback)~\cite{faeth2014emergent}. The key difference between touchscreen devices and virtual environments is that touchscreen devices offer haptic feedback cues when a user taps on a touchscreen, thus improving user responsiveness and task performances~\cite{touchscreen1-10.1145/1357054.1357300}. In contrast, the lack of haptic feedback in virtual environments can be compensated in multiple simulated approaches~\cite{tennison2019non-10.1145/3301415}, such as virtual spring~\cite{lecuyer2001boundary}, redirected tool-mediated manipulation~\cite{chi2020-hammer-10.1145/3313831.3376303}, stiffness~\cite{pezent2019tasbi}, object weighting~\cite{weighthaptic-10.1145/3290605.3300550}. With such simulated haptic cues, the users can connect the virtual overlays of the buttons) with the physical metaphors of the buttons~\cite{speicher2019pseudo}. In other words, the haptic feedback not only works with the visual and audio cues, and further acts as an enriched communication signal to the users during the virtual touches (or even the interaction) with virtual overlays in the metaverse~\cite{Ma2015}. More importantly, such feedback cues should follow the principle of user mobility as mentioned in Section~\ref{ssec:mobile-in}. The existing works demonstrate various form factors exoskeletons~\cite{bouzit2002:rutgers, nam2007:smart}, gloves~\cite{in2011:jointless, gloves:cybertouch}, finger addendum~\cite{Gabardi2016, Kim2016}, smart wristbands~\cite{henderson2019leveraging}, by considering numerous mechanisms including air-jets~\cite{sodhi2013:aireal}, ultrasounds~\cite{Carter2013, Arafsha2015, kervegant2017touch}, and laser~\cite{Lee2016, Ochiai2016}. In addition, the full taxonomy of mobile haptic devices is available in~\cite{pacchierotti2017wearable}. 
	
	After compensating the missing haptic feedback in virtual environments, it is important to best utilise various feedback cues and achieve multi-modal feedback cues (e.g., visual, auditory, and haptic)~\cite{lee2008assessing-10.5555/1531514.1531540}, in order to improve the user experiences~\cite{kobayashi2016:towards}, the user's responsiveness~\cite{lecuyer2001boundary}, task accuracy~\cite{cockburn2005multimodal-10.1080/00140130500197260, faeth2014emergent}, the efficiency of virtual object acquisition~\cite{cockburn2005multimodal-10.1080/00140130500197260, gordon2019touchscreen-10.1145/3290605} in various virtual environments. We also consider inclusiveness as an additional benefit of leveraging haptic feedback in virtual environments, i.e., the visually impaired individuals~\cite{kaklanis2008haptic-10.1145/1385569.1385653}. As the prior works on the multi-modal feedback cues do not consider the new enriched instance to appear in varying scenarios inside the metaverse, it is worthwhile to explore the combination of the feedback modals further, and introduce new modals such as smell and taste~\cite{AR-eating-CHI2011}. 
	
	\subsection{Telepresence}\label{ssec:telepresence}
	%carlos
	\begin{figure}[!t]
		\centering
		\includegraphics[width=\linewidth]{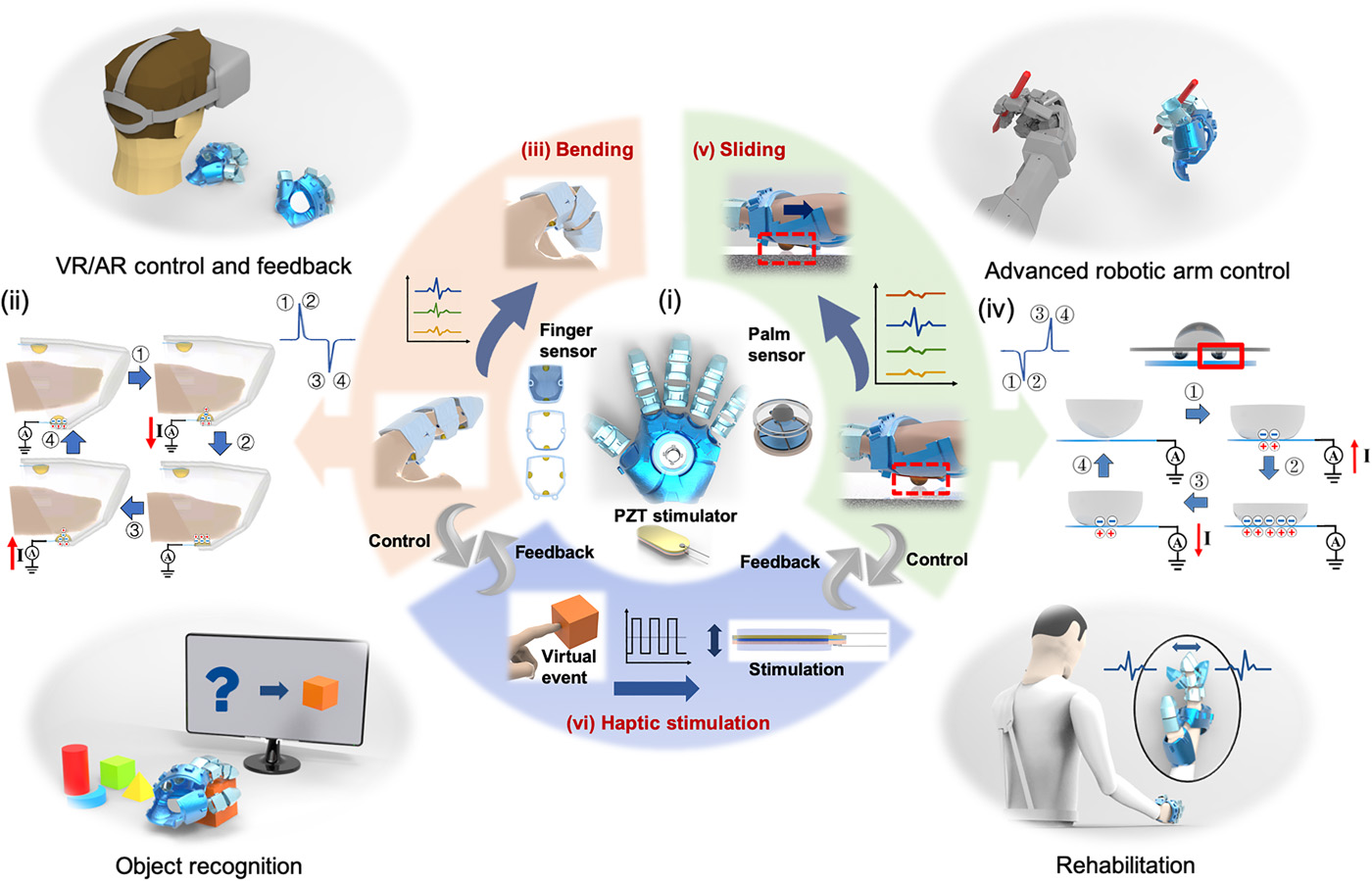}
		\caption{The key principles of haptic devices that support user interaction with various tangible and virtual objects in the metaverse (Image source from~\cite{zhu-haptic}).}
		\label{fig:zhu}
	\end{figure}
	
	The discussion in previous paragraphs can be viewed as the stimuli to achieve seamless user interaction with virtual objects as well as other avatars representing other human users. 
	To this end, we have to consider the possible usage of such stimuli that paves the path towards telepresence through the metaverse. 
	Apart from designing stable haptic devices~\cite{hulin2014:passivity}, the synchronisation of such stimuli is challenging. 
	According to the \textit{Weber-Fechner Law} that describes ``the minimum time gap between two stimuli'' in order to make user feels the two stimuli are distinguishable. 
	Therefore, the research community employs the measures of Just Noticeable Difference (JND) to quantify the necessary minimum time gap~\cite{Lee2015}. Considering the benefits of including haptic feedback in virtual environments, as stated in Section~\ref{ssec:haptic}, the haptic stimuli should be handled separately. As such, transmitting such a new form of haptic data can be effectively resolved by Deadband compression techniques (60\% reduction of the bandwidth)~\cite{Tirmizi2016}. The technique aims to serve cutaneous haptic feedback and further manage the JND, in order to guarantee the user can precept distinguishable haptic feedback.

	\begin{figure*}[!t]
		\centering
		\includegraphics[width=\textwidth]{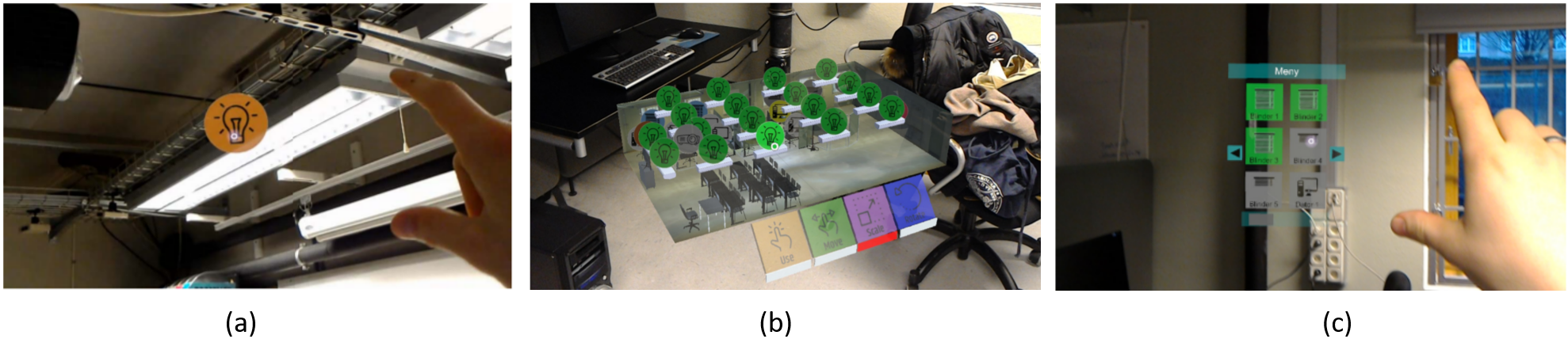}
		\caption{Three basic AR interaction models: (a) The Floating Icons model, with the user gazing at the icon. (b) The WIM model in scale mode, with a hologram being engaged with. (c) The floating menu model, with three active items and three inactive items~\cite{alce2017poster}.}
		\label{fig:ariot}
	\end{figure*}

	Next, the network requirements of delivering haptic stimuli would be another key challenge.
	The existing 4G communication technologies can barely afford AR and VR applications. However, managing and delivering the haptic rendering for user's sensing the realism of virtual environments in a subtle manner are still difficult with the existing 4G network. Although 5G network features with low latency, low jitter, and high bandwidth, haptic mobile devices, considered as a type of machine-type communication, may not be able to adopt in large-scale user interactivity through the current design of 5G network designated for machine-to-machine communication~\cite{siriwardhana2021survey} (More details in Section~\ref{sec:iot-robot}. Additionally, haptic mobile devices is designed for the user's day-long activities anywhere when the network capacity has fulfilled the aforementioned requirements. Thus, the next important issue is to tackle the constraints of energy and computational resources on mobile devices~\cite{Lee2020UbiPointTN}. Apart from reducing the algorithm complexity of haptic rendering, an immediate solution could be offloading such haptic-driven computational tasks to adjacent devices such as cloud servers and edge devices. 
	More detailed information on advanced networks as well as edge and cloud computing are available in Section~\ref{sec:network} and~\ref{sec:edge}, respectively.

	Although we expect that new advances in electronics and future wireless communications will lead to real-time interactions in the metaverse, the network requirements would become extremely demanding if the metaverse will serve unlimited concurrent users.  
	As such, network latency could hurt the effectiveness of such stimuli and hence the sense of realism. To this end, a visionary concept of \textit{Tactile Internet} is coined by Fettweis~\cite{Fettweis2014}, which advocates the redesign of the backbone of the Internet to alleviate the negative impacts from latency and build up ultra-reliable tactile sensory for virtual objects in the metaverse~\cite{Maier2016, Aijaz2016, Simsek2016}. More specifically, \textit{1 ms} is expected as the maximum latency of \textit{Tactile Internet}, which facilitates real-time haptic feedback for the sake of various operations during the telepresence~\cite{Pilz2016}. 
	It is important to note that the network latency is not the only source. Other latency sources could be caused by the devices, i.e., on-device latency~\cite{Steinbach2012,Wildenbeest2013}. 
	For instance, the glass-to-glass latency, representing the round-trip latency from video taken by a smartphone camera to a virtual overlay that appeared in a smartphone screen, is 19.18 ms~\cite{bachhuber2017today}, far exceeding the ideal value of 1 ms for the \textit{Tactile Internet}.
	The aggregation of latency could further deteriorate the user perceptions with virtual environments in the metaverse~\cite{Steinbach2012}. Therefore, we call for additional research attention in this area for building seamless yet realistic user interaction~\cite{zhu-haptic} with various entities linked to the metaverse, as illustrated in Figure~\ref{fig:zhu}.

	\section{Internet-of-Things (IoT) and Robotics}\label{sec:iot-robot}
	
	According to Statista~\cite{statistaiot}, by 2025, the total IoT connected devices worldwide will reach 30.9 billion, with a sharp jump from the 13.8 billion expected in 2021. Meanwhile, the diversity of interaction modalities is expanding. Therefore, many observers believe that integrating IoT and AR/VR/MR may be suitable for multi-modal interaction systems to achieve compelling user experiences, especially for non-expert users. The reason is that it allows interaction systems to combine the real-world context of the agent and immersive AR content~\cite{kim2021multimodal}. To align with our focused discussion on the metaverse, %instead of discussing the full development of IoT, 
	this section focuses on the virtual environments under the spectrum of extended reality, i.e., data management and visualisation, and human-IoT interfacing. Accordingly, we elaborate on the impacts of XR on IoT, autonomous vehicles, and robots/drones, and subsequently pinpoint the emerging issues. 
	%Humans have been exploring to develop machines that can mimic human actions such that they can substitute humans and replicate human actions.   

	\subsection{VR/AR/MR-driven human-IoT interaction}

	The accelerating availability of smart IoT devices in our everyday environments offers opportunities for novel services and applications that can improve our quality of life. However, miniature-sized IoT devices usually cannot accommodate tangible interfaces for proper user interaction~\cite{7613201}. 
	The digital entities under the spectrum of XR can compensate for the missing interaction components. In particular, users with see-through displays can view XR interfaces in mid-air~\cite{Becker2020ConnectingAC}.
	Additionally, some bulky devices like robot arms, due to limitations of form factors, would prefer users to control the devices remotely, in which XR serves as an on-demand controller~\cite{10.1145/3411764.3445398}. Users can get rid of tangible controllers, considering that it is impossible to bring a bundle of controllers for numerous IoT devices.
	Virtual environments (AR/MR/XR) show prominent features of visualising invisible instances and their operations, such as WiFi~\cite{Ar-visual-wifi-Iot} and user personal data~\cite{61a81229ff3e4524beb0120feb07b5ec}.
	Also, AR can visualise the IoT data flow of smart cameras and speakers to the users, thus informing users about their risk in the user-IoT interaction. Accordingly, users can control their IoT data via AR visualisation platforms~\cite{61a81229ff3e4524beb0120feb07b5ec}. 
	
\begin{figure}[t]
		\centering
		\includegraphics[width=\columnwidth]{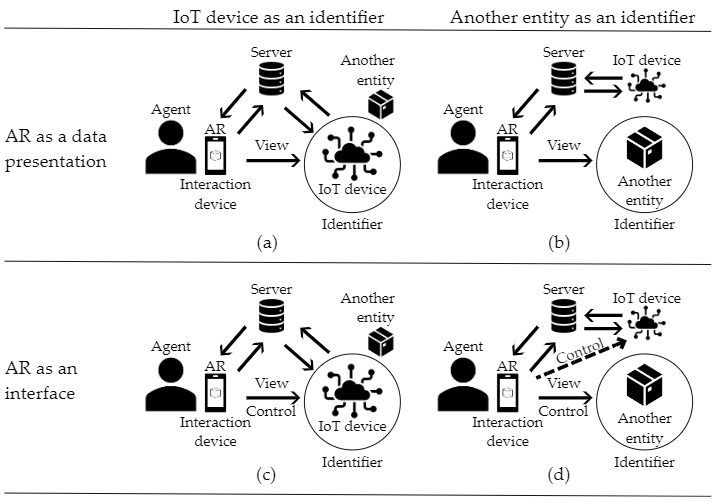}
		\caption{Four interaction models proposed in~\cite{ kim2021multimodal}, categorised by whether an agent can control the IoT device through AR (c,d) or not (a,b), and whether an IoT device (a,c) or another entity (b,d) functions as an AR identifier.}
		\label{fig:armi}
	\end{figure}

	There are several key principles to categorise the AR/VR/MR-directed IoT interaction systems. Figure~\ref{fig:ariot} shows three models defined according to the scale and category of the rendered AR content. 
	Mid-air icons, menus, and virtual 3D objects allow users to control IoT devices with natural gestures~\cite{alce2017poster}.
	Figure~\ref{fig:armi} offers four models depicted according to the controllability of the IoT device and the identifier entity. 
	In short, virtual overlays in AR/MR/XR can facilitate data presentation and interfacing the human-IoT interaction.
	Relatedly, a number of recent works have been proposed in this direction. For example,~\cite{cao2019v} presents V.Ra, a visual and spatial programming system that allows the users to perform task authoring with an AR hand-held interface and attach the AR device onto the mobile robot, which would execute the task plan in a what-you-do-is-what-robot-does (WYDWRD) manner. 
	Moreover, flying drones, a popular IoT device, have been increasingly employed in XR. 
	In~\cite{telepresence-VR-collaborate-drone}, multiple users can control a flying drone remotely and work collaboratively for searching tasks outdoors.
	Pinpointfly~\cite{pinpointfly} presents a hand-held AR application that allows users to edit a flying drone's motions and directions through enhanced AR views.
	Similarly, SlingDrone~\cite{slingdrone} leverages MR user interaction through mobile headsets to plan the flying path of flying drones. 
	
	%MORE to add

	% \begin{figure*}[!t]
	%     \centering
	%     \includegraphics[width=\textwidth]{cv4.jpeg}
	%     \caption{(a) The I2I metaverse of Nissian for assisting driving. (b) The Hyundai Mobility Adventure (HMA) showcasing the future life.}
	%     \label{fig:my_label}
	% \end{figure*}

	\subsection{Connected vehicles}
	As nowadays vehicles are equipped with powerful computational capacity and advanced sensors, connected vehicles with 5G or even more advanced networks could go beyond the vehicle-to-vehicle connections, and eventually connect with the metaverse. Considering vehicles are semi-public spaces with high mobility, drivers and passengers inside vehicles can receive enriched media. 
	With the above incentive, the research community and industry are striving to advance the progress of autonomous driving technologies in the era of AI. %this AI era. 
	Connected vehicles serves as an example of IoT devices as autonomous vehicles could become the most popular scenarios for our daily commute. 
	In recent years, significant progress has been made owing to the recently emerging technologies, such as AR/MR~\cite{riegler2019augmented,riegler2021systematic}. AR/MR play an important role in empowering the innovation of autonomous driving. To date, AR/MR has been applied in three directions for autonomous driving~\cite{in-car-AR-designSpace}. First of all, AR/MR helps the public (bystanders) understand how autonomous vehicles work on the road, by offering visual cues such as the vehicle directions. With such understandings, pedestrian safety has been enhanced~\cite{pedestrain-safety}. 
	To this end, several industrial applications, such as Civil Maps\footnote{\url{https://civilmaps.com/}}, applied AR/MR to provide a guide for people to understand how an autonomous driving vehicle navigates in the outdoor environment. For instance, it shows how the vehicle detects the surroundings, vehicles, traffic lights, pedestrians, and so on. 
	The illustration with AR/MR/XR or even the metaverse can build trust with the users with connected vehicles~\cite{critical-communication-vehicle}. In addition, some AR-supported dynamic maps can also help drivers to make good decisions when driving on the road. Second, AR/MR help to improve road safety. For instance, virtual entities appear in front of the windshield of vehicles, and such entities can augment the information in the physical world to enhance the user awareness to the road conditions. It is important to note such virtual entities are considered as a low-cost and convenient solution, in comparison to largely modified the physical road infrastructure. 
	The latest work also pinpoints the concept of digital twins to enhance road safety, especially for vulnerable road users~\cite{taxonomy-road-users}, instead of inviting the human users to work on risky tasks physically. 
	%As AR/MR tries to add virtual content based on the physical world, it provide a low-cost and convenient platform to test the autonomous driving car. 
	For instance, the Mcity Test Facility at the University of Michigan\footnote{\url{https://record.umich.edu/articles/augmented-reality-u-m-improves\break-driverless-vehicle-testing/}} applies AR to test the driving car. In the platform, the testing and interaction between a real test vehicle and the virtual vehicles are created to test driving safety. In such a MR world, an observer can see a real vehicle passing and stopping at the intersection with the virtual vehicles at the traffic light.
	Last but not least, AR/MR have improved the vehicle navigation and user experience. For example, %Lastly, AR/MR has also been applied in vehicle navigation and user experience. 
	%For instance, 
	WayRay\footnote{\url{https://wayray.com/\#who-we-are}} develops an AR-based navigation system that helps to improve road driving safety. The highlight of this technique is that it alleviates the need for the drivers to rely too much on gauges %too much the gauges 
	when driving. Surprisingly, WayRay provides the driver with highly precise route and environment information in real-time.
	Most recent research also demonstrates the needs of shared views among connected vehicles to enhance user safety, for instance, the view of a front car is shared to the car(s) at the back~\cite{Zhou2021AugmentedIC}. 
	From the above, we see the benefits of introducing virtual entities on connected vehicles and road traffic. Perhaps the metaverse can transform such driving information into interesting animation without compromising road safety.

\begin{figure}[t]
		\centering
		\includegraphics[width=\columnwidth]{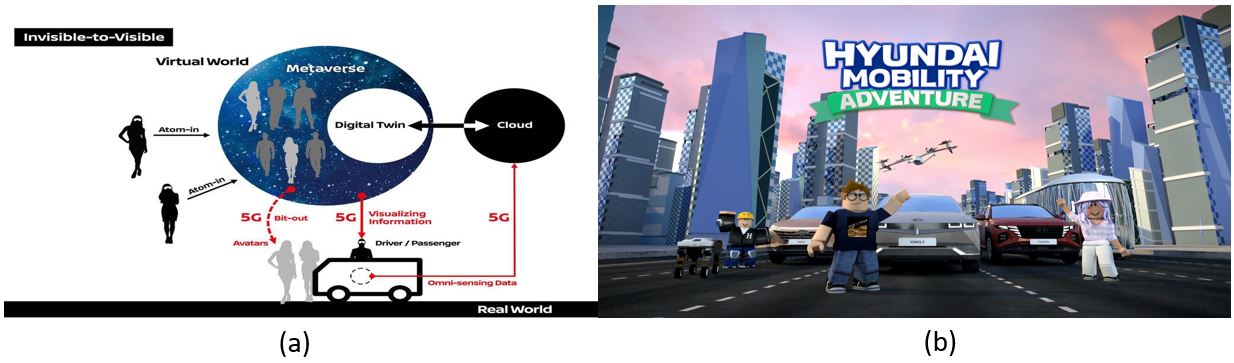}
		\caption{(a) The I2V metaverse of Nissian for assisting driving. I2V can connect drivers, passengers with the people all across the world.(b) The Hyundai Mobility Adventure (HMA) showcasing the future life.}
		\label{fig:meta_i2i}
	\end{figure}

	% Need to mention the potential of metaverse for this field. Will be added 
	% With the perspective of the metaverse, we expect it can bring a brutal force of technology development in autonomous driving. 
	Recent examples also shed lights on the integration between intelligent vehicles and virtual environments. For Invisible-to-Visible (I2V) from Nissian\footnote{\url{https://www.nissan-global.com/EN/TECHNOLOGY/OVERVIEW/i2v.html}} is a representative attempt to build the metaverse platform where an AR interface is designed to connect the physical and virtual worlds together such that the information invisible to the drivers can be visible. As shown in Figure~\ref{fig:meta_i2i}, I2V employs several systems to provide rich information from the inside and outside of vehicle. Specifically, I2V first adopts the omni-sensing technology to gather data in real-time from the traffic and the surrounding vehicles. Meanwhile, the metaverse system seamlessly analyses the road status from the real-time information. Based on the analysis, %analyses, it 
	I2V then identifies the driving conditions around the vehicle immediately. Lastly, the digital twin of the vehicles, drivers, the buildings, and the environment is created via data collected from the omni-sensing system. In such a way, the digital twin can be used to analyse the human-city interaction~\cite{Lee2020TowardsAR} through the perspective of road traffic. The shared information driven by the user activities can further connect to the metaverse. As a result, the metaverse generates the information through the XR interfaces, as discussed in Section~\ref{sec:XR} or the vehicle windshields. To sum up, the digital transformation with the metaverse can deliver human users enriched media during their commutes. % share the information and support AR/VR interface. I2V helps driving in two aspects.
	In addition, I2V helps driving in two aspects.
	The first is visualising the invisible environment for a more comfortable drive. The metaverse system enables displaying the road information and hidden obstacles, traffic congestion, parking guidance, driving in the mountain, driving in poor weather conditions, etc. Meanwhile, I2V metaverse system visualises virtual human communication via MR. For instance, it provides a chance for family members from anywhere in the world to join the metaverse as avatars. It also provides a tourism scenario where a local guide can join the metaverse to guide the driver. 
	
	Furthermore, the Roborace metaverse\footnote{\url{https://roborace.com/}} is another platform blending the physical world with a virtual world where AR generates the virtual obstacles to interact with the track. Hyundai Motor\footnote{\url{https://www.hyundai.news/eu/articles/press-releases/hyundai-vitalizes-future-mobility-in-roblox-metaverse-space.html}} also launched `Hyundai Mobility Adventure (HMA)' to showcase the future lifestyle in the metaverse. The HMA is a shared virtual space where various users/players, which are represented as `avatars', can meet and interact with each other to experience mobility. Through the metaverse platform, the participants can customise their `avatars' and imaginatively interact with each other. %For instance, in the indoor environment of the vehicle, the objected generated AR/VR and other digital entities can pop up in the metaverse, enabling the users to relax in the commute. 
	
	\subsection{Robots with Virtual Environments}\label{ssec:xr-robot}
	
	Virtual environments such as AR/VR/MR are good solution candidates for opening the communication channels between robots and virtual environments, due to their prominent feature of visualising contents~\cite{8956466}. 
	Furthermore, various industrial examples integrate virtual environments to enable human users to understand robot operations, such as task scenario analysis and safety analysis. Therefore, human users build trust and confidence with the robots, leading to the paradigm shift towards human-robot collaboration~\cite{DIANATFAR2021407}. Meanwhile, to date, research studies focus on the user perception with robots and the corresponding interface designs with virtual environments~\cite{10.1145/3411764.3445398, Somaesthetic-drone-2,Somaesthetic}. Also, human users with V.Ra~\cite{cao2019v} can collaboratively develop task plans in AR environments and program mobile robots to interact with stationary IoTs in their physical surroundings.
	
	Nowadays, the emerging MR technology serves as communication interfaces with humanoids in workspace~\cite{10.3389/frobt.2017.00020}, with high acceptance levels to collaborative robots~\cite{10.1145/3399433}. In our daily life, robots can potentially serve as our friends~\cite{10.1145/2701973.2701998} companion devices~\cite{companion-drone}, services drone~\cite{DIS-robot}, caring robots~\cite{caring-robot-1,caring-robot}, 
	an inspector in public spaces~\cite{public-drone}, home guardian (e.g., Amazon Astro\footnote{\url{https://www.aboutamazon.com/news/devices/meet-astro-a-home-robot-unlike-any-other}}), sex partners~\cite{sex-rob-1,sex-rob-2,sex-rob-3}, and even a buddy with dogs~\cite{ACI-robot}, as human users can adapt natural interactions with robots and drones~\cite{10.1145/2750858.2805823}. It is not hard to imagine the robots will proactively serve our society, and engage spontaneously in a wide variety of applications and services. 
	
	The vision of the metaverse with collaborative robots is not only limited to leveraging robots as a physical container for avatars in the real world, and also exploring design opportunities of our alternated spatial with the metaverse. 
	Virtual environments in the metaverse can also become the game changer to the user perception with collaborative robots. It is important to note that the digital twins and the metaverse can serve as a virtual testing ground for new robot designs. 
	The digital twins, i.e., digital copies of our physical environments, allow robot and drone designers to examine the user acceptability of novel robot agents in our physical environments. What are the changes in user perception to our spatial environment augmented by new robot actors, such as alternative humanoids and mechanised everyday objects?
	In~\cite{robot-arhitecture}, designers evaluate the user perceptions to the mechanised walls in digital twins of living spaces, without actual implementation in the real world. The mechanised walls can dynamically orchestrate with user activities of various contexts, e.g., additional walls to separate a user from the crowd, who prefers staying alone at works, or lesser walls for social gatherings. 
	
	% \newpage
	\section{Artificial Intelligence}\label{sec:art-int}
	Artificial intelligence (AI) refers to theories and technologies that enable machines to learn from experience and perform various kinds of tasks, similar to intelligent creatures~\cite{mccarthy1998artificial,russell2002artificial,dick2019artificial}. AI was first proposed in 1956. In recent years, it has achieved state-of-the-art performance in various application scenarios, including natural language processing~\cite{bengio2003neural, collobert2008unified}, computer vision~\cite{kendall2017uncertainties,alhaija2017augmented}, and recommender systems~\cite{zhang2019deep,lu2015recommender}. % AI has been the most popular topic and received significant attention from both academia and industry. 
	AI is a broad concept, including representation, reasoning, and data mining. Machine learning is a widely used AI technique, which enables machines to learn and improve performance with knowledge extracted from experience. There are three categories in machine learning: supervised learning, unsupervised learning, and reinforcement learning. Supervised learning requires training samples to be labelled, while unsupervised learning and reinforcement learning are usually applied on unlabelled data. Typical supervised learning algorithms includes linear regression~\cite{montgomery2021introduction}, random forest~\cite{oshiro2012many}, and decision tree~\cite{myles2004introduction}. K-means~\cite{hamerly2004learning}, principle component analysis (PCA)~\cite{wold1987principal}, and singular value decomposition (SVD)~\cite{paige1981towards} are common unsupervised learning algorithms. Popular reinforcement learning algorithms include Q-learning~\cite{watkins1992q}, Sarsa~\cite{sprague2003multiple}, and policy gradient~\cite{silver2014deterministic}. Machine learning usually requires selecting features manually. Deep learning is involved in machine learning, which is inspired by biological neural networks. In deep neural networks, each layer recieves input from the previous layers, and outputs the processed data to the subsequent layers. %neurons receive input from neurons of the last layer and out-processed data to neurons of the subsequent layers. % next layer. 
	Deep learning is able to automatically extract features from a large amount of data. However, deep learning also requires more data than conventional machine learning algorithms to offer satisfying accuracy. Convolutional neural network (CNN)~\cite{o2015introduction}, recurrent neural network (RNN)~\cite{zaremba2014recurrent} are two typical and widely used deep learning algorithms. 
	
	There is no doubt that the main characteristic of the emerging metaverse is the overlay of unfathomably vast amounts of sophisticated data, which provides opportunities for the application of AI to release operators from boring and tough data analysis tasks, e.g., monitoring, regulating, and planning. In this section, we review and discuss how AI is used in the creation and operation of the metaverse. Specifically, we classify AI applications in the metaverse into three categories: automatic digital twin, computer agent, and the autonomy of avatar.
	%next, the relationship of AI and metaverse, as well as three applications.
	
	\subsection{Automatic Digital Twin}
	%first, talk about three digital: digital model, digital shadow, and digital twin. digital twin not only need to make a digital clone of physical entities, but also feedback to their physical entities. Then talk about something about the application of dt with AI, e.g., prediction, classification, making decision.
	There are three kinds of digitisation, including digital model, digital shadow, and digital twin~\cite{fuller2020digital}. The digital model is the digital replication of a physical entity. There is no interaction between the metaverse and the physical world. The digital shadow is the digital representation of a physical entity. Once the physical entity changes, its digital shadow changes accordingly. In the case of a digital twin, the metaverse and the physical world are able to influence each other. Any change on any of them will lead to a change on the other one. In the metaverse, we focus on this third kind of digitisation.
	
\begin{figure}[t]
		\begin{center}
			\includegraphics[width=\columnwidth]{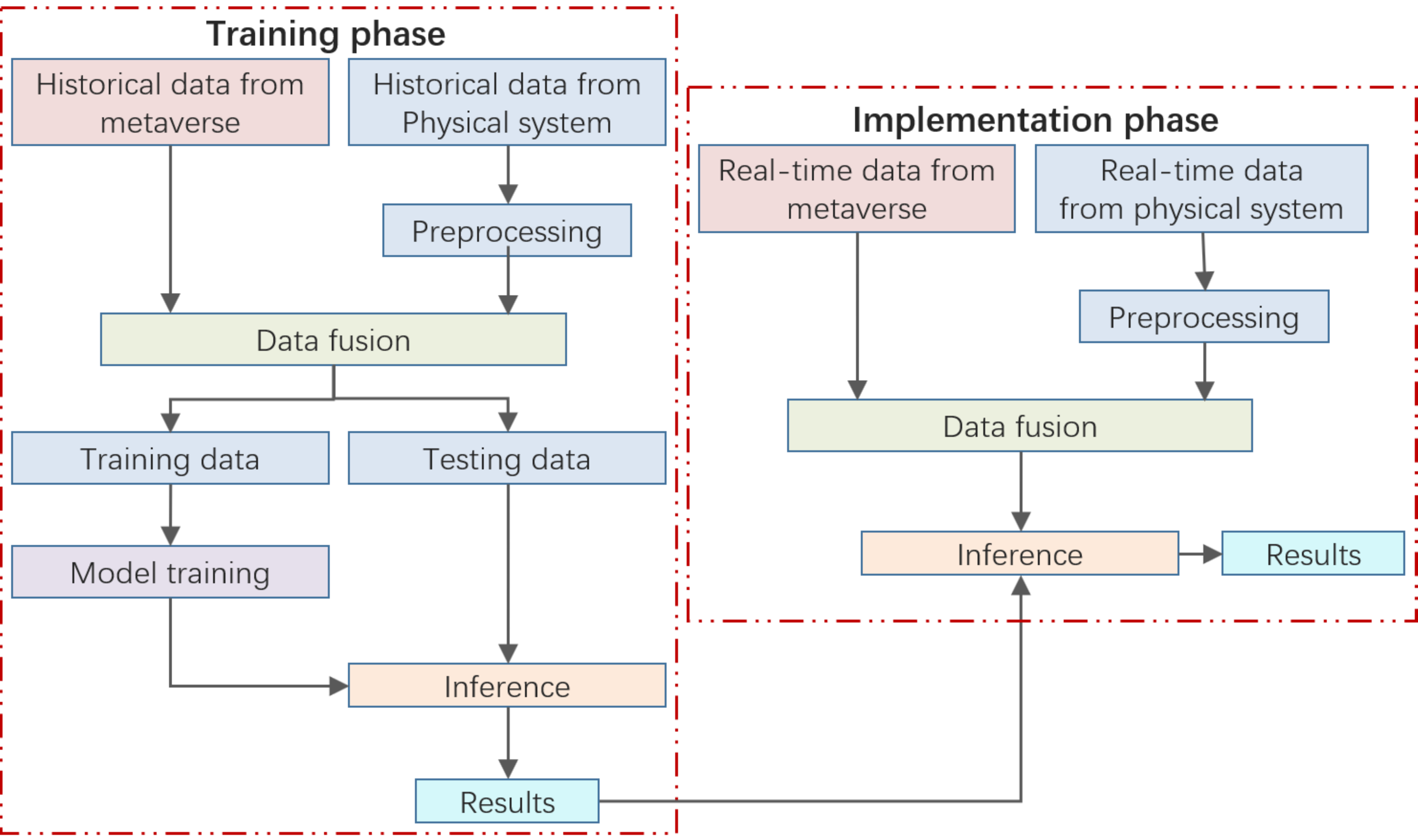}
		\end{center}
		\caption{Illustration of autonomous digital twin with deep learning.}
		\label{autonomous_dt}
	\end{figure}

	Digital twins are digital clones with high integrity and consciousness for physical entities or systems and keeps interacting with the physical world~\cite{fuller2020digital}. These digital clones could be used to provide classification~\cite{farhat2021digital,agnusdei2021classification}, recognition~\cite{piltan2021bearing,yiping2021deep}, prediction~\cite{tuegel2011reengineering,kostenko2018digital}, and determination services~\cite{moi2020digital,hofmann2019implementation} for their physical entities. Human interference and manual feature selection are time-consuming. Therefore, it is necessary to automate the process of data processing, analysis, and training. Deep learning can automatically extract knowledge from a large amount of sophisticated data and represent it in various kinds of applications, without manual feature engineering. Hence, deep learning has great potential to facilitate the implementation of digital twins. Jay \emph{et al.} propose a general autonomous deep learning-enabled digital twin, as shown in Figure~\ref{autonomous_dt}. In the training phase, historical data from both the metaverse and physical systems are fused together for deep learning training and testing. If the testing results meet the requirement, the autonomous system will be implemented. In the implementation phase, real-time data from the metaverse and physical systems are fused for model inference. 
	
	Smart healthcare requires interaction and convergence between physical and information systems to provide patients with quick-response and accurate healthcare services. Hence, the concept of digital twin is naturally applicable to smart healthcare. Laaki et al.\cite{laaki2019prototyping} designs A verification prototype for remote surgery with digital twins. In this prototype, a digital twin is created for a patient. All surgery operations on the digital twin done by doctors will be repeated on the patient with a robotic arm. The prototype is also compatible with deep learning components, e.g., intelligent diagnosis and healthy prediction. Liu \emph{et al.} apply learning algorithms for real-time monitoring and crisis warning for older adults with their digital twins~\cite{liu2019novel}. 
	
	Nowadays, more IoT sensors are implemented in cities to monitor various kinds of information and facilitate city management. Moreover, building information models (BIM) are getting more accurate~\cite{white2021digital}. By combining the IoT big data and BIM, we could create digital twins with high quality for smart cities. Such a smart-city digital twin will make urban planning and managing easier. For example, we could learn about the impact of air pollution and noise level on people's life quality~\cite{wan2019developing} or test how traffic light interval impacts the urban traffic~\cite{wang2020digital}. Ruohomaki \emph{et al.} create a digital twin for an area in urban to monitor and predict the building energy consumption. Such a system could also be used to help to select the optimisation problem of the placement of solar panels~\cite{ruohomaki2018smart}.

	Industrial systems are very complex and include multiple components, e.g., control strategy, workflow, system parameter, which is hard to achieve global optimisation. Moreover, data are heterogeneous, e.g., structured data, unstructured data, and semi-structured data, which makes deep learning-driven digital twin essential~\cite{qi2018digital}. Min \emph{et al.} design a digital twin framework for the petrochemical industry to optimise the production control~\cite{min2019machine}. The framework is constructed based on workflow and expert knowledge. Then they use historical production data to train machine learning algorithms for prediction and optimise the whole system.  
	
	\subsection{Computer Agent}
	%first, talk something about computer agent, what is computer agent, NPC, find a figure on framework, workflow, etc.
	Computer agent, also known as Non-player Character (NPC), refers to the character not controlled by a player. The history for NPCs in games could be traced back to arcade games, in which the mobility patterns of enemies will be more and more complex along with the level increasing~\cite{soni2008bots}. With the increasing requirements for realism in video games, AI is applied for NPCs to mimic the intelligent behaviour of players to meet players' expectations on entertainment with high quality. The intelligence of NPCs is reflected in multiple aspects, including control strategy, realistic character animations, fantastic graphics, voice, etc.
	
	The most straight and widely adopted model for NPC to respond to players' behaviour is finite state machines (FSM)~\cite{gallagher2012no}. FSM assumes there are finite states for an object in its lifecycle. There are four components in FSM: state, condition, action, next state. Once the condition is met, the object will take a new action and change its current state to the next state. Behaviour trees and decision trees are two typical FSM-based algorithms for NPCs to make decisions in games, in which each node denotes a state and each edge represents an action~\cite{isla2008building,zhu2019behavior,kopel2018implementing,agis2020event}. FSM-based strategies are very easy to realise. However, FSM is poor at scalability, especially when the game environment becomes complex. 
	
	Support vector machine is a classifier with the maximum margin between different classes, which is suitable for controlling NPCs in games. Pedro \emph{et al.} propose a SVM-based NPC controller in a shooter game~\cite{melendez2009controlling}. The input is a three-dimensional vector, including left bullets, stamina, and near enemies. The output is the suggested behaviour, e.g., explore, attack, or run away. Obviously, the primary drawback of such an algorithm is limited state and behaviour classes and the flexibility in decision-making.
	
	Reinforcement learning is a classic machine learning algorithm on decision-making problems, which enables agents to automatically learn from the interaction experience with their surrounding environment. The agent behaviours will be given corresponding rewards. The desired behaviours are with a higher reward. Due to its excellent performance, reinforcement learning has been widely adopted in many games, e.g., shooter games~\cite{ponce2014hierarchical} and driving games~\cite{kovalsky2020neuroevolution}. It is worth noting that the objective of NPC designing is to increase the entertainment of the game, instead of maximising the ability of NPCs to beat human players~\cite{wang2009rl}. Hence, the reward function could be customised according to the game objective~\cite{glavin2015learning}. For example, Glavin \emph{et al.} develop a skill-balancing mechanism to dynamically adjust the skill level of NPCs according to players performance based on reinforcement learning~\cite{glavin2018skilled}.
	
	When the games are getting more and more complex, from 2D to 3D, the agent state becomes countless. Deep reinforcement learning, the combination of neural network and reinforcement learning is proposed to solve such problems. The most famous game based on deep reinforcement learning is chess with AlphaGo developed by DeepMind in 2015~\cite{wang2016does}. The state of chess is denoted as a matrix. Through the process of neural networks, the AlphaGo outputs the action with the highest possibility to win.
	
	\subsection{Autonomy of Avatar}
	Avatar refers to the digital representation of players in the metaverse, where players interact with the other players or the computer agents through the avatar~\cite{davis2009avatars}. A player may create different avatars in different applications or games. For example, the created avatar may be like a human shape, imaginary creatures, or animals~\cite{nijholt2017humans}. In social communication, relevant applications that require remote presence, facial and motion characteristics reflecting the physical human are essential~\cite{koutsabasis2012value}. Existing works in this area mainly focus on two problems: avatar creation and avatar modelling.
	
	To create more realistic virtual environments, a wide variety of avatar representations are necessary. However, in most video games, creators only rely on several specific models or allow players to create complete avatars with only several optional sub-models, e.g., nose, eyes, mouth, etc. Consequently, players' avatars are highly similar.
	
	Generative adversarial network (GAN) is a state-of-the-art deep learning model in learning the distribution of training samples and generate data following the same distribution~\cite{yi2019generative}. The core idea of GAN is the contest between a generator network and a discriminator network. Specifically, the generator network is used to output fake images with the learnt data distribution, while the discriminator network inputs the fake images and judge whether they are real. The generator network will be trained until these fake images are not recognised by the discriminator network. Then discriminator network will be trained to improve its recognition accuracy. During this procedure, these two networks learn from each other. Finally, we got a well-performing generator network. Several works~\cite{jin2017towards,li2019towards,hamada2018full} have applied GAN to automatically generate 2D avatars in games. Some works~\cite{chai2016autohair,niki2019semi,shi2019face} further introduce real-time processing 3D mesh and textures to generate 3D avatars. Chalas \emph{et al.} develop an autonomous 3D avatar generation application based on face scanning, instead of 2D images~\cite{chalas2017generating}
	
	Some video games allow players to leave behind their models of themselves when players are not in the game. For example, Forza Motorsport develops Drivatars, which learns players' driving style with artificial intelligence~\cite{drivatars}. When these players are not playing the game, other users can have race with their avatars. Specifically, the system collects players' driving data, including road position, race line, speed, brake, and accelerator. Drivatars learns from collected data and creates virtual players with the same driving style. It is worth noting that the virtual player is non-deterministic, which means the racing results for a given virtual player may be not the same in the same game. A similar framework is also realised with neural network in~\cite{munoz2010human}. 
	
	Gesler \emph{et al.} apply multiple machine learning algorithms in the first person shooter (FPS) game to learn players' shooting style, including moving direction, leap moment, and accelerator~\cite{geisler2002empirical}. Through extensive experiments, they find neural network outperforms other algorithms, including decision tree and Naive Bayes.
	
	For decision-making relevant games, reinforcement learning usually outperforms other AI algorithms. Mendoncca \emph{et al.} apply reinforcement learning in fighting games~\cite{mendoncca2015simulating}. They use the same fighting data to train reinforcement learning model and a neural network and find the reinforcement learning model performs much better.

	\section{Blockchain}\label{sec:blockchain}
	It is expected to connect everything in the world in the metaverse. Everything is digitised, including digital twins for physical entities and systems, avatars for users, large-scale, fine-grained map on various areas, etc. Consequently, unfathomably vast amounts of data are generated. Uploading such giant data to centralised cloud servers is impossible due to the limited network resources~\cite{xu2018survey}. Meanwhile, blockchain techniques are developing rapidly. It is possible to apply blockchains to the data storage system to guarantee the decentralisation and security in the metaverse~\cite{berg2019blockchain, cai2018decentralized}. 
	
\begin{figure}[t]
		\begin{center}
			\includegraphics[width=\columnwidth]{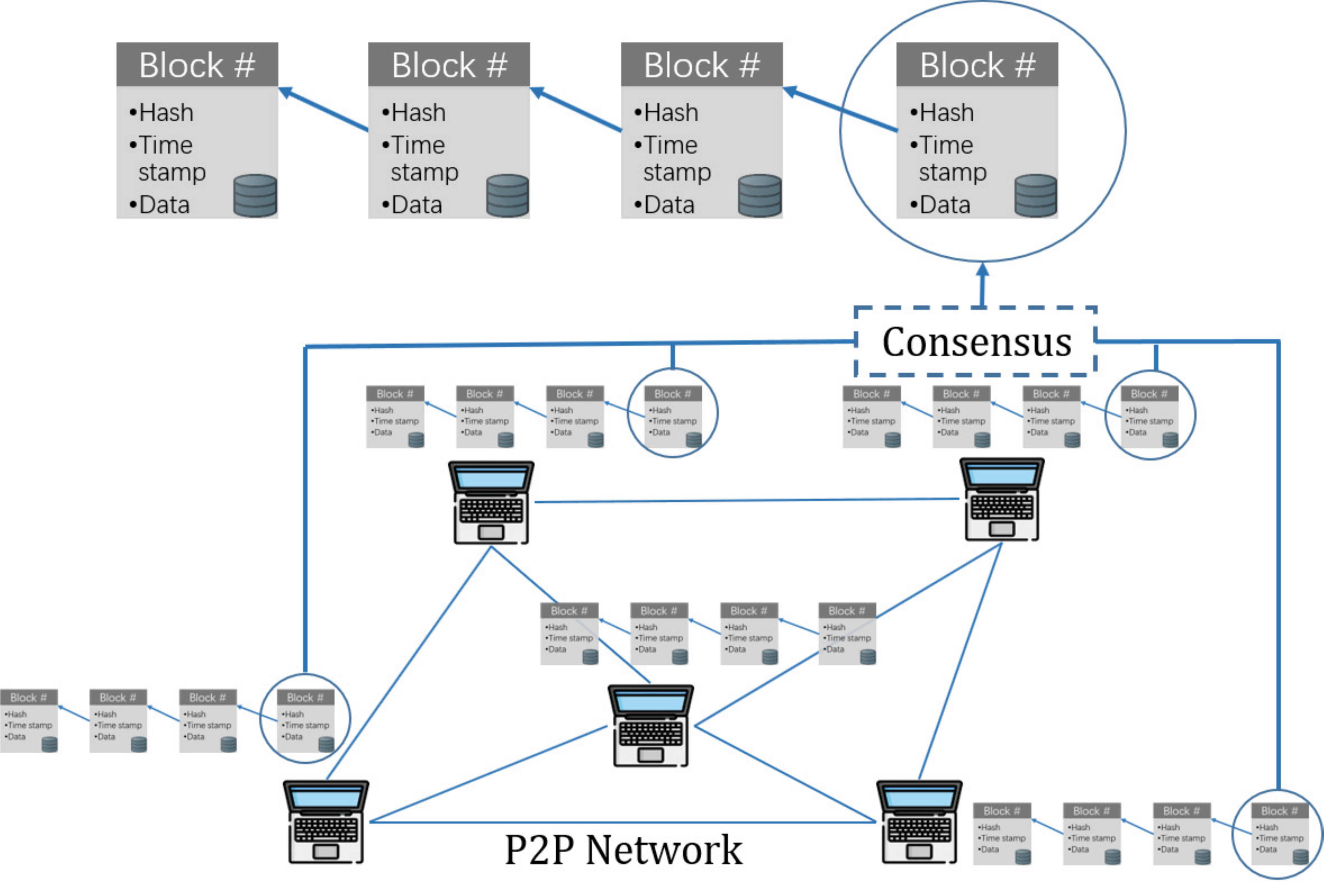}
		\end{center}
		\caption{Illustration of blockchain.}
		\label{blockchain}
	\end{figure}

	Blockchain is a distributed database, in which data is stored in blocks, instead of structured tables~\cite{nofer2017blockchain}. The architecture of blockchain is shown in Figure~\ref{blockchain}. The generated data by users are filled into a new block, which will be further linked onto previous blocks. All blocks are chained in chronological order. Users store blockchain data locally and synchronise them with other blockchain data stored on peer devices with a consensus model. Users are called nodes in the blockchain. Each node maintains the complete record of the data stored on the blockchain after it is chained. If there is an error on one node, millions of other nodes could reference to correct the error. Therefore, decentralisation and security are two of the obvious characteristics of blockchain~\cite{cai2018decentralized}. The most famous application of blochchain is Bitcoin, which is a digital currency proposed in 2009~\cite{urquhart2016inefficiency}. In this section, we discuss how blockchain is applied in the metaverse.
	%brief intro on blockchain, and then three problems: data interoperability, data sharing, data storage
	
	\subsection{Data storage}
	In the metaverse, various kinds of user data are digitised, collected, and stored. How to store such a massive amount of data is a crucial problem. Traditional data storage systems usually adopt the centralised architecture, which requires transmitting all data to a data centre. Considering such amount of data, extremely high storage capacity is essential, which is usually very expensive. Moreover, sensitive information may be included in such data, which may lead to potential privacy leakage issues. Blockchain, as a distributed database is just enough to handle these issues. %In blockchain, users 
	Users with blockchains can create data blocks and validate and record transactions cooperatively. Considering computation demands for mining, edge computing could also be applied, which will be discussed in Section~\ref{sec:edge}.
	
	Zyskind \emph{et al.} propose a distributive personal data management system based on blockchain~\cite{zyskind2015decentralizing}. There is a secure channel for data accessing. The data owner shares a key with all other users requesting the data. The authentication of requiters is done through blockchain, which guarantees the security of the data. However, the key is exposed to miners. Subsequently, Li \emph{et al.} apply certificate-less signature to solve the problem~\cite{li2018blockchain}. If the data are tampered with by malicious hackers, they could be recovered by local regenerative code technology~\cite{liang2020secure,ren2019secure}. The regenerative code is based on the redundancy of data. Once the data are found to be tampered with or damaged, data on other active nodes could be used to repair it in a multi-threaded manner. 
	
	Most smart devices, e.g., smartphones, have limited storage capability. If the amount of generated data is overwhelming, users may borrow storage space from other users, which may fail due to users' selfishness. Ren \emph{et al.} propose a blockchain-based incentive mechanism for data storage~\cite{ren2018incentive}. Specifically, there are two blockchains in this storage system. The first one is for data storage, while the second one is for access control. They propose to use a reasonable amount of stored data to replace the proof of work in mining, which could significantly reduce computation operations. 
	
	Recently, electronic voting is getting more and more popular. In an electronic voting system, people, no matter where they are, are able to participate in voting online. As a result, the voting records and results will be stored. Blockchain has great potential in preventing intentional tampering and accident on voting. However, there are some challenges in the application of blockchain as voting systems. The first challenge is the authentication. In blockchain-based voting systems, people use virtual identity to vote, while voting requires real identity. It is not easy to authenticate the validity of the voting results without knowing the voters' real identity. Bistarelli \emph{et al.} propose an end-to-end voting framework, which adopts the anonymous Kerberos to authenticate voters to solve the problem~\cite{bistarelli2017end}. The second challenge is the auditability of voting results. Blockchain is able to store all transaction records forever. However, private information of voters may be leaked during the auditing process. Meter \emph{et al.} apply asymmetric encryption and threshold encryption on voting content and private key respectively to solve such problem~\cite{meter2017design}.

	\subsection{Data sharing}
	% Blockchain-based data storage systems are of high scalability and flexibility. Users contribute their storage resources in blockchains. Each user could be both data requester and data provider. Moreover, the data is encrypted and relocated to anonymous node for storage, which further enhance the security of data. The data location is recorded by all nodes in blockchain. Data owner can access his/her data very conveniently. However, such data storage architecture is unfriendly for data sharing. Because the conventional sharing models are not supported in blockchain. Moreover, additional key management mechanism is needed to share encrypted data. 
	
	Blockchain-based data storage systems are of high scalability and flexibility. Users contribute their storage resources in blockchains. Each user could be both a data requester and a data provider. Moreover, the data is encrypted and relocated to an anonymous node for storage, further enhancing data security. All nodes in blockchains  record the data location. Thus, data owners can access their data very conveniently. However, such data storage architecture is unfriendly for data sharing, as blockchains do not support the conventional sharing models. Moreover, additional key management mechanisms are needed to share encrypted data. 
	
	Li \emph{et al.} design a key management mechanism for sharing data in blockchains, which is compatible with blockchain-based data storage systems~\cite{li2019meta}. The key is integrated with metadata and stored in blockchain. They also apply proxy re-encryption to protect the key in untrusted situations. Xia \emph{et al.} utilise the tamper-proof mechanism of blockchains to guarantee the security of shared data and introduce smart contract and access control to track the data accessing behaviour of all users~\cite{xia2017medshare}. Another similar approach is adopted in~\cite{wang2018blockchain}.
	
	\subsection{Data interoperability}
	Privacy and security are of utmost importance for managing the data in the metaverse. However, it is inevitable to access and operate on such data by multiple parties. %parities. 
	Consequently, conflicts occur. Blockchain provides a data platform with extremely high security, enabling different companies to share data. For example, banks and insurance companies can share the same customer data for their separate business through blockchain for interoperability~\cite{narayanan2017bitcoin}. 
	
	A typical application scenario of blockchain on data interoperability is smart healthcare. As we mentioned previously, digital twins would be created for patients based on their profile data for precise healthcare. Such digital twins could be accessed by multiple doctors. Some literature~\cite{shrier2016blockchain,liu2016medical} have proved the feasibility of applying distributed ledgers to storing patients' information from a theoretical aspect. Azaria \emph{et al.} design and implement a blockchain-based medical data management system~\cite{azaria2016medrec}. The system is able to provide authentication, interoperability, and confidentiality services. The operation of this system is similar to Bitcoin, which opens opportunities for aggregation and anonymisation through mining. 
	
	Remarkably, blockchain is also widely used in the financial field. Financial institutions all over the world are eager to reduce the clearing and settlement cycles and finally improve the efficiency of transactions and reduce the risk of mitigation. Singh~\emph{et al.} design a E-wallet architecture for secure payment across banks~\cite{singh2018interoperable}. In this architecture, banks are nodes in blockchain and deploy high-performance servers as miners. They adopt Proof of Stake (PoS) as the consensus model.

	\section{Computer Vision}\label{sec:cv}
	In this section, we examine the technical state of computer vision in interactive systems and its potential for the metaverse. Computer vision plays an important role in XR applications and lays the foundation for achieving the metaverse. 
	% A wide range of modalities, technologies, and devices are prepared for XR-driven interaction. %included, most 
	Most XR systems capture visual information through an optical see-through or video see-through display. This information is processed, and results are delivered via a head-mounted device or a smartphone, respectively. 
	By leveraging such visual information, computer vision plays a vital role in processing, analysing, and understanding visuals as digital images or videos to derive meaningful decisions and take actions.
	%By leveraging the visual information, computer vision plays a vital role in process, analyse, and understand the digital images or videos to derive meaningful decisions and take actions accordingly with the visual information. 
	In other words, computer vision allows XR devices to recognise and understand visual information of users activities and their physical surroundings, helping build more reliable and accurate virtual and augmented environments.  %Recently, with deep learning, computer vision has radically advanced the development of %changed the direction of 
	%VR/AR/MR/XR. 
	% In computer vision, it is meaningful to utilise visual information to acquire geometric or location information. This allows the interaction system to locate the position of the user and device to deliver spatial information. 
	
	Computer vision is extensively used in XR applications to build a 3D reconstitution of the user's environment and locate the position and orientation of the user and device.
	In Section~\ref{locmap}, we review the recent research works on 3D scene localisation and mapping in indoor and outdoor environments. Besides location and orientation, XR interactive system also needs to track the body and pose of users. We expect that in the metaverse, the human users will be tracked with computer vision algorithms and represented as avatars. %To interact with each other, it is required to estimate the users' body pose information. 
	With such intuition, in Section~\ref{body_gaze}, we analyse the technical status of human tracking and body pose estimation in computer vision.
	Moreover, the metaverse will also require to understand and perceive the user's surrounding environment based on scene understanding techniques. We discuss this topic in Section~\ref{scene_understand}. Finally, augmented and virtual worlds need to tackle the problems
	related to object occlusion, motion blur, noise, and the low-resolution of image/video inputs. Therefore, image processing is an important domain in computer vision, which aims to restore and enhance image/video quality for achieving better metaverse. We will discuss the state-of-the-art technologies in Section~\ref{image_processing}.

	\subsection{Visual Localisation and Mapping}\label{locmap}
	In the metaverse, human users and their digital representatives (i.e., avatars) will connect together and co-exist at the intersection between the physical and digital worlds.
	Considering the concept of digital twins and its prominent feature of interoperability, building such connections across physical and digital environments requires a deep understanding of human activities that may potentially drive the behaviours of one's avatar. 
	% Therefore, 
	% how can we map the 3D physical world, build a 3D visual map in the metaverse, and acquire the location of objects simultaneously?
	In the physical world, we acquire spatial information with our eyes and build a 3D reconstitution of the world in our brain, where we know the exact location of each object. 
	Similarly, the metaverse needs to acquire the 3D structure of an unknown environment and sense its motion. To achieve this goal, simultaneous Localisation and Mapping (SLAM) is a common computer vision technique that estimates device motion and reconstructs an unknown environment's~\cite{sturm2012benchmark,cadena2016past}. A visual SLAM algorithm has to solve several challenges simultaneously: (1) unknown space, (2) free-moving or uncontrollable camera, (3) real-time, and (4) robust feature tracking (drifting problem)~\cite{ouerghi2020comparative}. Among the diverse SLAM algorithms, the ORB-SLAM series, \eg, ORB-SLAM-v2~\cite{mur2017orb} have been shown to work well, \eg, in the AR systems~\cite{zeng2018orb,ouerghi2020comparative}. %Here we mainly discuss three crucial steps for visual SLAM, which are necessary for achieving the metaverse.
	
	\addtocounter{footnote}{-1}
	\begin{figure}[t!]
		\centering
		\includegraphics[width=\columnwidth]{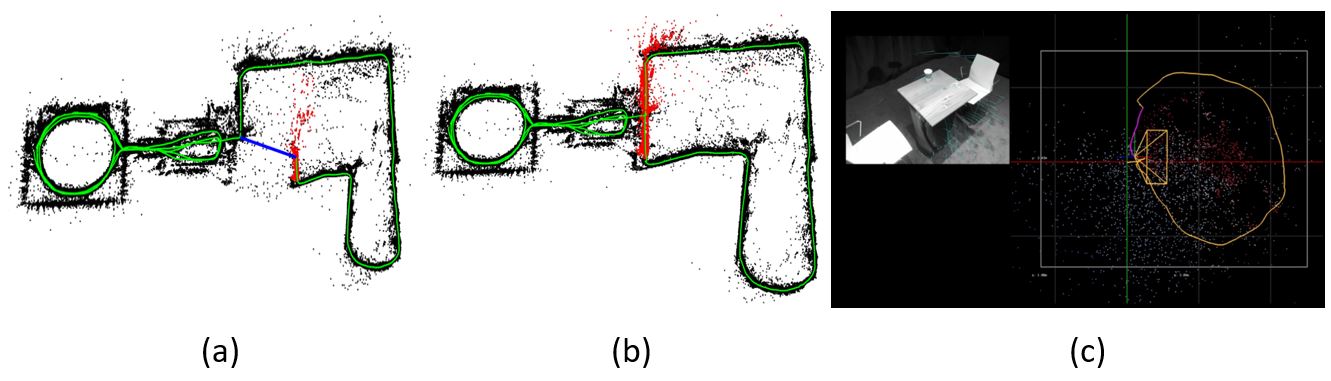}
		\caption{Mapping before (a) and after (b) close-loop detection in ORB-SLAM~\cite{mur2017orb}. The loop trajectory is drawn in green, and the local feature points for tracking are in red. (c) The visual SLAM demonstrate by ARCorev2 from Apple. The trajectory of loop detection is in yellow (Image source \protect\footnotemark).}
		\label{fig:slam}
	\end{figure}
	\footnotetext{\url{https://developer.apple.com/videos/play/wwdc2018/602}}
	
	Visual SLAM algorithms often rely on three primary steps: (1) feature extraction, (2) mapping the 2D frame to the 3D point cloud, and (3) close loop detection.
	
	The first step for many SLAM algorithms is to find feature points and generate descriptors~\cite{cadena2016past}. Traditional feature tracking methods, such as Scale-invariant feature transform (SIFT)~\cite{lowe2004distinctive}, detect and describe the local features in images; however, they often too slow to run in real-time. Therefore, most AR systems rely on computationally efficient feature tracking methods, such as feature-based detection~\cite{rublee2011orb} to match features in real-time without using GPU acceleration. Although recently, convolutional neural networks (CNNs) have been applied to visual SLAM and achieved promising performance for autonomous driving with GPUs~\cite{milz2018visual}, it is still challenging to apply to resource-constrained mobile systems.
	
	With the tracked key points (features), the second step for visual SLAM is how to map the 2D camera frames to get 3D coordinates or landmarks, which is closely related to camera pose estimation~\cite{reitmayr2010simultaneous}. When the camera outputs a new frame, the SLAM algorithm first estimates the key points. These points are then mapped with the previous frame to estimate the optical flow of the scene. Therefore, camera motion estimation paves the way for finding the same key points in the new frame. However, in some cases, the estimated camera pose is not precise enough. Some SLAM algorithms, \eg, ORB-SLAM~\cite{mur2017orb,nerurkar2020system} also add additional data to refine the camera pose by finding more key point correspondences. New map points are generated via triangulation of the matching key points from the connected frames. This process bundles the 2D position of key points in the frames and the translation and rotations between frames.   
	
	The last key step of SLAM aims to recover the camera pose and obtain a geometrically consistent map, also called close-loop detection~\cite{biswas2012depth}. As shown in Figure~\ref{fig:slam}(c) for AR, if a loop is detected, it indicates that the camera captures previously observed views. Accordingly, the accumulated errors in the camera motion can be estimated. In particular, ORB-SLAM~\cite{mur2017orb} checks whether the key points in a frame are matched with the previously detected key points from a different location. If the similarity exceeds a threshold, it means the user has returned to a known place. Recently, some SLAM algorithms also combined the camera with other sensors, \eg, the IMU sensor, to improve the loop detection precision~\cite{paul2017comparative}, and some works, \eg,~\cite{schonberger2018semantic}, have attempted to fuse the semantic information to SLAM algorithms to ensure the loop detection performance. 
	
	Although current state-of-the-art (SoTA) visual SLAM algorithms already laid a solid foundation for spatial understanding, the metaverse needs to understand more complex environments, especially the integration of virtual objects and real environments. Hololens has already started getting deeper in spatial understanding, and Apple has introduced ARKitv2\footnote{\url{https://developer.apple.com/videos/play/wwdc2018/602}} for 3D keypoint tracking, as shown in Figure~\ref{fig:slam}(c). In the metaverse, the perceived virtual universe is built in the shared 3D virtual space. Therefore, it is crucial yet challenging to acquire the 3D structure of an unknown environment and sense its motion. This could help to collect data for \eg, digital twin construction, which can be connected with AI to achieve auto conversion with the physical world. Moreover, in the metaverse, it is important to ensure the accuracy of object registration, and the interaction with the physical world. With these harsh requirements, we expect the SLAM algorithms in the metaverse to become more precise and computationally effective to use.  
	
	\subsection{Human Pose \& Eye Tracking} \label{body_gaze}
	In the metaverse, users are represented by avatars (see Section~\ref{sec:avatar}). Therefore, we have to consider the control of avatars in 
	3D virtual environments. 
	Avatar control can be achieved through human body and eye location and orientation in the physical world.
	Human pose tracking refers to the computer vision task of obtaining spatial information concerning human bodies in an interactive environment~\cite{barioni2018human}. In VR and AR applications, the obtained visual information concerning human pose can usually be represented as joint positions or key points for each human body part. These key points reflect the characteristics of human posture, which depict the body parts, such as elbows, legs, shoulders, hands, feet, etc.~\cite{uchiyama2012object,bauer2017anatomical}. In the metaverse, this type of body representation is simple yet sufficient for perceiving the pose of a user's body.
	
	Tracking the position and orientation of the eye and gaze direction can further enrich the user micro-interactions in the metaverse.
	Eye-tracking enables gaze prediction, and intent inference can enable intuitive and immersive user experiences, which can be adaptive to the user requirement for real-time interaction in XR environments~\cite{pfeiffer2014eyesee3d,kapp2021arett,gardony2020eye}. In the metaverse, it is imperative for eye tracking to operate reliably under diverse users, locations, and visual conditions. Eye tracking requires real-time operations within the power and computational limitations imposed by the devices.  
	
	Achieving significant milestones of the above two techniques relies on releasing
	several high-quality body and eye-tracking datasets~\cite{zhang2020atari,krafka2016eye,andriluka2018posetrack,bazarevsky2020blazepose} combined with the recent advancement
	in deep learning. In the following subsections, we review and analyse body pose and eye-tracking methods developed for XR, and derive their potential benefits for the metaverse.

	\subsubsection{Human Pose Tracking}
	When developing methods to track human poses in the metaverse, we need to consider several challenges. First, a pose tracking algorithm needs to handle the self-occlusions of body parts. Second, the robustness of tracking algorithms can impact the sense of presence, especially in multi-user scenarios.
	Finally, a pose tracking algorithm needs to track the human body even in vastly diverse illumination conditions,
	\eg, in the too bright or dark scenes. Considering these challenges, most body pose tracking methods combine the RGB sensor with infrared or depth sensors~\cite{barioni2018human,shao2013computer,nunez2017real,wang2019coaug} to improve the detection accuracy. Such sensor data are relatively robust to abrupt illumination changes and convey depth information for the tracked pixel. For XR applications, Microsoft Kinect\footnote{\url{https://developer.microsoft.com/en-us/windows/kinect/}} and Open Natural Interaction (OpenNI)\footnote{\url{https://structure.io/openni}} are two popular frameworks for body pose estimation. 
	
	In recent years, deep learning methods have been continuously developed in the research community to extract 2D human pose information from the RGB camera data~\cite{dang2019deep,cao2019openpose,fang2017rmpe} or 3D human pose information from RGB-D sensor data~\cite{mehta2017vnect,wang2021deep,hu2020fingertrak}. Among the SoTA methods for 2D pose tracking, OpenPose~\cite{cao2019openpose} has been broadly used by researchers to track  users' bodies in various virtual environments such as %user's body in the interaction platform, such as 
	VR~\cite{huang20193d,d2020markerless}, AR~\cite{bajireanu2019mobile,nuzzi2020hands,wang2020avatarmeeting}, and metaverse~\cite{shin2021non}. For 3D pose tracking,  FingerTrack~\cite{hu2020fingertrak} recently presented a 3D finger tracking and hand pose estimation method, which displays high potential for XR applications and the metaverse.  
	
	Compared to single body pose tracking, multi-person tracking is more challenging. The tracking algorithm needs to count the number of users and their positions and group them by classes~\cite{zheng2020deep}. In the literature, many methods have been proposed for VR~\cite{silva2019tensorpose,czesak2016fusion} and AR~\cite{marchand2015pose,su2019deep,nagpal2021pose}. In the metaverse, both single-person and multi-person body pose tracking algorithms are needed in different circumstances. Reliable and efficient body pose tracking algorithms are needed to ensure the close ties between the metaverse and the physical world and people.
	
	\begin{figure}[t!]
		\centering
		\includegraphics[width=\columnwidth]{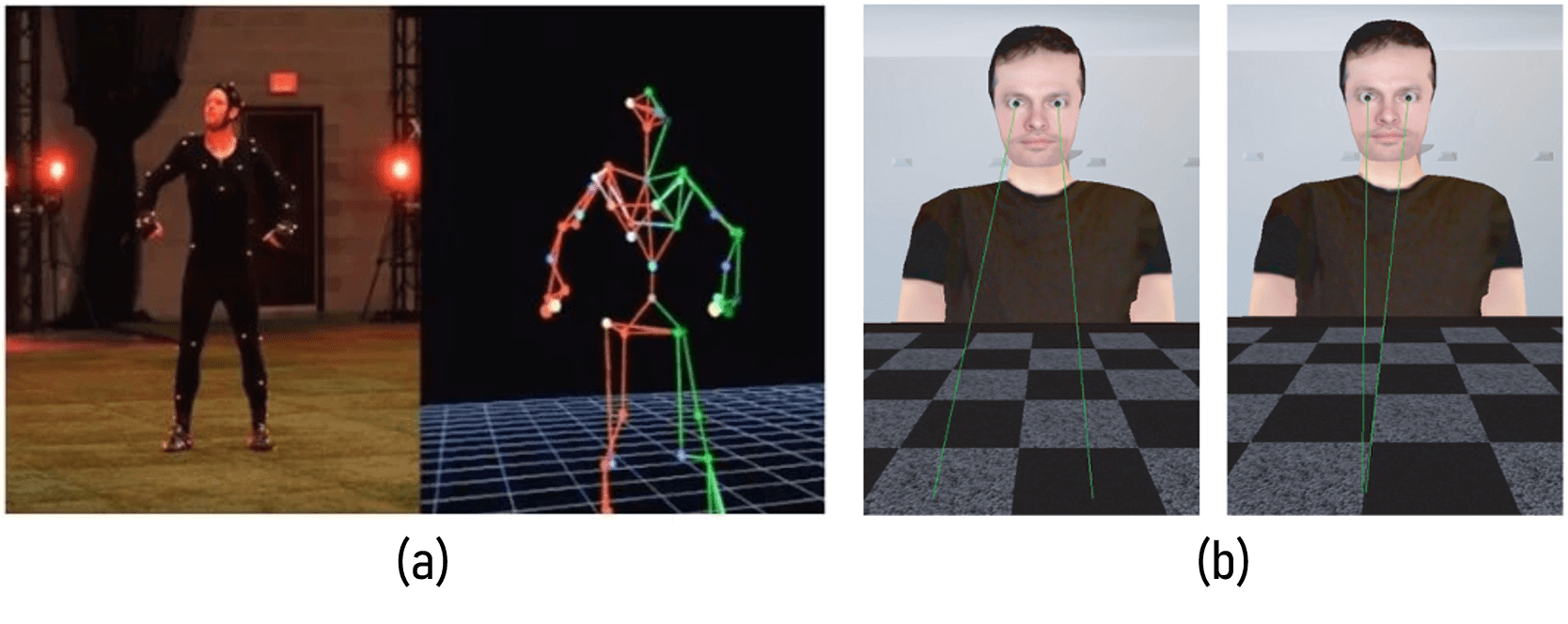}
		\vspace{-15pt}
		\caption{Visual examples of pose and eye tracking. (a) body pose tracking results from Openpose~\cite{cao2019openpose} and (b) eye tracking with no eye convergence (left) and eye convergence (right)~\cite{murray2007assessment}.}
		\label{fig:pose_eye_track}
	\end{figure}

	\subsubsection{Eye Tracking}
	Eye-tracking is another challenging topic in achieving the metaverse as the human avatars need to `see' the immersive 3D environment. %Therefore, the task is to track a user's eye movement information and calculate the gaze information in the immersive environment. 
	Eye tracking is based on continuously measuring the distance between the pupil centre and the refection of the cornea~\cite{haffegee2009eye}. The angle of the eyes converges at a certain point where the gaze intersects. The region displayed within the angle of the eyes is called `vergence'~\cite{tanriverdi2000interacting} -- the distance changes with regard to the angle of the eye. Intuitively, the computer vision algorithms in eye-tracking should be able to measure the distance by deducing from the angle of the eyes where the gaze is fixed~\cite{murray2007assessment}. To measure the distance, one representative way is to leverage infrared cameras, which can record and track the eye movement information, as in the HMDs. In VR, the HMD device is placed close to the eyes, making it easy to display the vergence. However, the device cannot track the distance owning to the 3D depth information. Therefore, depth estimation for the virtual objects in the immersive environment is one of the key problems. 
	
	Eye-tracking can bring lots of benefits for immersive environments in the metaverse. One of them is reducing the computation cost in rendering the virtual environment. Eye tracking makes it possible to only render the contents in the view of users. As such, it can also  facilitate the integration of the virtual and real world. 
	However, there are still challenges in eye tracking. First of all, the lack of focus blur can lead to an incorrect perception of the object size and distance in the virtual environment~\cite{clay2019eye}. Another challenge for eye tracking is to ensure precise distance estimation with incomplete gaze due to the occlusion~\cite{clay2019eye}.  Finally, eye tracking may lead to motion sickness and eye fatigue~\cite{peissl2018eye}. In the metaverse, the requirements for eye tracking can be much higher than traditional virtual environments. This opens up some new research directions, such as understanding human behaviour accurately and creating more realistic eye contact for the avatars, similar to the physical eye contact, in the 3D immersive environment.
	
	\subsection{Holistic Scene Understanding} \label{scene_understand}
	%What does it take to understand both the physical environment and the metaverse?
	In the physical world, we understand the world by answering four fundamental questions: what is my role? What are the contents around me? How far am I from the referred object? What might the object be doing? In computer vision, holistic scene understanding aims to answers these questions~\cite{wang2021psat}. A person's role is already clear in the metaverse as they are projected through an avatar. However, the second question in computer vision is formulated based on semantic segmentation and object detection.
	Regarding the third question, we estimate the distance to the reference objects based on our eyes in the physical world. This way of scene perception in computer vision is called stereo matching and depth estimation. The last question requires us to interpret the physical world based on our understanding. For instance, `a rabbit is eating a carrot'. We need first to recognise the rabbit and the carrot and then predict the action accordingly to interpret the scene. 
	The metaverse requires us to interact with other objects and users in both the physical and virtual world.
	Therefore, holistic scene understanding plays a pivotal role in ensuring the operation of the metaverse.

	\begin{figure*}[t!]
		\centering
		\includegraphics[width=\textwidth]{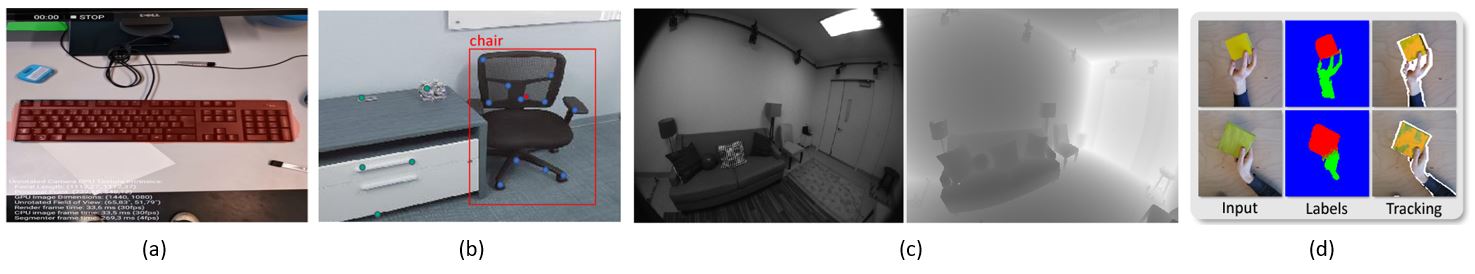}
		\vspace{-15pt}
		\caption{Visual examples for holistic scene understanding. (a) Semantic segmentation in AR environment~\cite{turkmen2019scene}; (b) scale estimation in object detection (the blue dots are generated by the detector)~\cite{li2020object}; (c) Stereo depth estimation result (right) for VR~\cite{chaurasia2020passthrough+}; (d) Deep learning-based hand action recognition based on labels~\cite{schroder2017deep}. }
		\label{fig:seg_det}
	\end{figure*}
	
	\subsubsection{Semantic Segmentation and Object Detection}
	Semantic segmentation is a computer vision task to categorise an image into different classes based on the per-pixel information~\cite{wang2021evdistill,wang2021dual}, as shown in Figure~\ref{fig:seg_det}(a). It is regarded as one of the core techniques to understand the environment fully~\cite{tanzi2021real}. In computer vision, a semantic segmentation algorithm should efficiently and quickly segment each pixel based on the class information. Recent deep learning-based approaches~\cite{chen2018encoder,wang2021dual,wang2021evdistill} have shown a significant performance enhancement in urban driving datasets designed for autonomous driving. However, performing accurate semantic segmentation in real-time  remains challenging. For instance, AR applications require semantic segmentation algorithms to run with a speed of around 60 frames per second (fps)~\cite{ko2020novel}. Therefore, semantic segmentation is a crucial yet challenging task for achieving the metaverse.
	
	Object detection is another fundamental scene understanding task aiming to localise the objects in an image or scene and identify the class information for each object~\cite{liu2019edge}, as shown in Figure~\ref{fig:seg_det}(b). Object detection is widely used in XR and is an indispensable task for achieving the metaverse. For instance, in VR, face detection is a typical object detection task, while text recognition is a common object detection task in AR. In a more sophisticated application, AR object recognition aims to attach a 3D model to the physical world~\cite{li2020object}. This requires the object detection algorithms to precisely locate the position of objects and correctly recognise the class. By placing a 3D virtual object and connecting it with the physical object, users can manipulate and relocate it. AR object detection can help build a richer and more immersive 3D environment in the metaverse. 
	In the following, we analyse and discuss the SoTA semantic segmentation and object detection algorithms for achieving the metaverse.
	
	The early attempts of semantic segmentation mostly unitise the feature tracking algorithms, \eg, SIFT~\cite{schonberger2018semantic} that aim to segment the pixels based on the classification of the handcrafted features, such as the support vector machine (SVM)~\cite{noble2006support}. These algorithms have been applied to VR~\cite{lin2016virtual} and AR~\cite{schutt2019semantic}. However, these conventional methods suffer from limited segmentation performance. Recent research works have explored the potential of CNNs for semantic segmentation. These methods have been successfully applied to AR~\cite{ko2020novel,tanzi2021real,zhang2020slimmer,turkmen2019scene}. Some works have shown the capability of semantic segmentation for tackling the occlusion problems in MR~\cite{kido2021assessing,roxas2018occlusion}. However, as image segmentation deals with each pixel, it leads to considerable computation and memory load.
	
	To tackle this problem, recent endeavours focus on real-time semantic segmentation. Theses methods explore the image crop/resizing~\cite{zhao2018icnet} or efficient network design ~\cite{siam2018rtseg,mehta2018espnet} or transfer learning~\cite{liu2019structured,wang2021knowledge}. Through these techniques, some research works managed to achieve real-time semantic segmentation in MR~\cite{kido2020mobile,gajic2020egocentric,chen2020context}.  
	
	In the metaverse, we need more robust and real-time semantic segmentation methods to understand the pixel-wise information in a 3D immersive world. More adaptive semantic segmentation methods are needed because due to the diversity and complexity of virtual and real objects, contents, and human avatars. In particular, in the interlaced metaverse world, the semantic segmentation algorithms also need to distinguish the pixels of the virtual objects from the real ones. The class information can be more complex in this condition, and the semantic segmentation models may need to tackle  unseen classes.
	
	Object detection in the metaverse can be classified into two categories: detection of  specific instances (\eg, face, marker, text) and detection of  generic categories (\eg, cars, humans). Text detection methods have been broadly studied in XR,~\cite{hbali2016face,mangiarua2020scalable}. These methods have already matured and can be directly applied to achieving the metaverse. Face detection has also been studied extensively in recent years, and the methods have shown to be robust in various recognition scenarios in XR applications, \eg,~\cite{lu2020exploration,kojic2020user,golnari2020deepfacear,marques2019adaptive,svensson2018object}. 
	
	In the metaverse, users are represented as avatars, and multiple avatars can interact with each other. The face detection algorithms need to detect the real faces (from the physical world) and the synthetic faces (from the virtual world). Moreover, the occlusion problems, sudden face pose changes, and illumination variations in the metaverse can make it more challenging to detect faces in the metaverse. Another problem for face detection is the privacy problem. Several research works have studied this problem in AR application~\cite{acquisti2014face,acquisti2011faces,dick2020address}. In the metaverse, many users can stay in the 3D immersive environment; hence, privacy in face detection can be more stringent. Future research should consider the robustness of face detection, and better rules or criteria need to be studied for face detection in the metaverse.   
	
	The detection of the generic categories has been studied massively in recent years by the research community. Much effort using deep learning has been focused on the detection of multiple classes. The two-stage detector, FasterRCNN~\cite{ren2015faster}, was one of the SoTA methods in the early development stage using deep learning. Later on, the Yolo series and SSD detectors~\cite{redmon2018yolov3,bochkovskiy2020yolov4,liu2016ssd} have shown wonderful detection performance on various scenes with multiple classes. These detectors have been successfully applied to AR~\cite{mahurkar2018integrating,simony2018complex,bahri2019accurate,li2020object}.
	
	From the above review, we can see that the SoTA object detection methods have already been shown to work well for XR. However, there are still some challenges for achieving the metaverse. The first challenge is the smaller or tiny object detection. This is an inevitable problem in the 3D immersive environment as many contents co-exist in the shared space. With variations of Field of View (FoV) of the camera, some contents and objects will become smaller, making it hard to detect. Therefore, the object detector in the metaverse should be reinforced to detect these objects regardless of the capture hardware. The second one is the data and class distribution issues. In general, it is easy to collect large-scale datasets with more than 100 classes; however, it is not easy to collect datasets with a diverse scene and class distribution in the metaverse. 
	The last one is the computation burden for object detection in the metaverse. The 3D immersive world in the metaverse comprises many contents and needs to be shared even in remote places. With the increment of class, the computation burden is increased accordingly. To this end, more efficient and lightweight object detection methods are expected in the research community.  
	
	\subsubsection{Stereo Depth Estimation}
	Depth estimation using stereo matching is a critical task in achieving the metaverse. %Estimating the distance to the physical space, virtual objects, and other users in the immersive environment is indispensable. 
	The estimated distance directly determines the position of contents in the immersive environment. The common way to estimate depth is using a stereo camera~\cite{el2019distance}, as shown in Figure~\ref{fig:seg_det}(c). 
	In VR, stereo depth estimation is conducted in the  virtual space. Therefore, depth estimation estimates the absolute distance between a virtual object to the virtual camera (first-person view) or the referred object (third-person view). The traditional methods first extract feature points and then us them to compute the cost volumes, which is used to estimate the disparity~\cite{scharstein2002taxonomy}. In recent years, extensive research has been focused on exploring the potential of deep learning to estimate depth in VR, \eg, ~\cite{lai2019real,li2020unsupervised}.
	
	In XR, one of the critical issues is to ensure that depth estimation is done based on both virtual and real objects. In this way, the XR users can place the virtual objects in the correct positions. 
	Early methods in the literature for depth estimation in AR/MR rely on the absolute egocentric depth~\cite{Wildenbeest2013}, indicating how far it is from a virtual object to the viewer. The key techniques include ``blind walking''~\cite{lampton1995distance}, imagined blind walking~\cite{loomis2003visual}, and triangulation by walking~\cite{willemsen2004effects}. Recently, deep learning-based methods have been applied to XR~\cite{prokopetc2019towards,badias2020real,ping2020effects}, showing much precise depth estimation performance. Stereo cameras have been applied to some HMDs, \eg, the Oculus Rift,~\cite{kanbara2000stereoscopic}. Infrared camera sensor are also embedded in some devices, such as HoloLens, enabling easier depth information collection. 
	
	\begin{figure*}[t!]
		\centering
		\includegraphics[width=\textwidth]{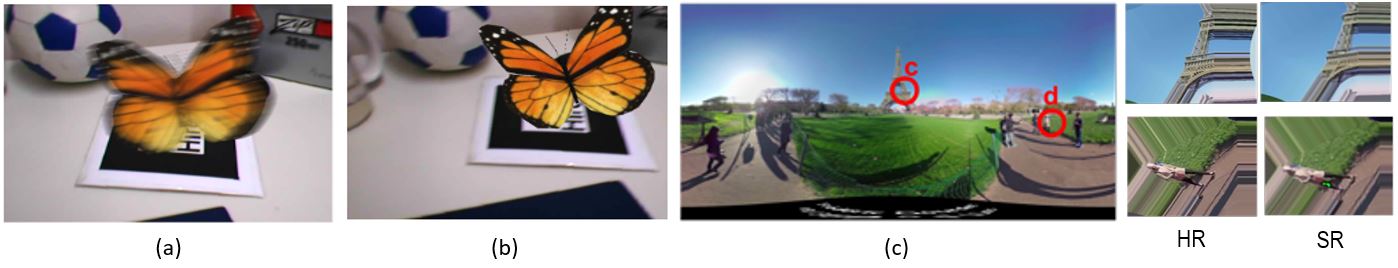}
		\vspace{-12pt}
		\caption{Visual examples for image restoration and enhancement. (a) Motion blur image and (b) no motion blurred image~\cite{fischer2006handling}; (c) Super-resolution image with the comparison of HR and SR image patches~\cite{li2021applying}. }
		\label{fig:image_enh}
	\end{figure*}
	
	In the metaverse, depth estimation is a key task in ensuring the precise positioning of objects and contents. In particular, all users own their respective avatars, and both the digital and real contents are connected. Therefore, depth estimation in such a computer-generated universe is relatively challenging. Moreover, the avatars representing human users in the physical world are expected to experience heterogeneous activities in real-time in the virtual world, thus requiring more sophisticated sensors and algorithms to estimate depth information. 
	
	\subsubsection{Action Recognition}
	In the metaverse, a human avatar needs to recognise the action of the other avatars and contents. In computer vision, understanding a person's action is called action recognition, which involves localising and predicting human behaviours~\cite{zhu2020comprehensive}, as illustrated in Figure~\ref{fig:seg_det}(d). In XR, HMDs such as Hololens, usually needs to observe and recognise the user's actions and generate action-specific feedback in the 3D immersive environment. For instance, it is often necessary to capture and analyse the user's motion with a camera for interaction purposes. With the advent of the Microsoft Kinect, there have been many endeavours to capture human body information and understand the action~\cite{sieluzycki2016microsoft,wang2019coaug}. The captured body information is used to recognise the view-invariant action~\cite{huang2010virtual,rao2001view}. For instance, one aspect of action recognition is finger action recognition~\cite{lee2011vision}.
	
	Recently, deep learning has been applied to action recognition in AR based on pure RGB image data~\cite{schroder2017deep,dong2019gesture} or multi-modal data via sensor fusion~\cite{chung2019sensor}. It has also shown potential for emotion recognition in VR~\cite{marin2020emotion}. When we dive deeper into the technical details of the success of action recognition in XR, we find that it is important to generate context-wise feedback based on the local and global information of the captured pose information.
	
	In the metaverse, action recognition can be very meaningful. A human avatar needs to recognise the action of other avatars or objects so that the avatar can take the correct action accordingly in the 3D virtual spaces. Moreover, human avatars need to emotionally and psychologically understand others and the 3D virtual world in the physical world. More adaptive and robust action recognition algorithms need to be explored. The most challenging step of action recognition in the metaverse is recognising the virtual contents across different virtual worlds. Users may create and distribute virtual content from a virtual world to the other. The problem of catastrophic forgetting for AI models on multi-modal data for activity recognition should also be tackled~\cite{kwon2021exploring}.

	\subsection{Image Restoration and Enhancement} \label{image_processing}
	The metaverse is connected seamlessly with the physical environments in real-time. In such a condition, an avatar needs to work with a physical person; therefore, it is important to display the 3D virtual world with less noise, blur, and high-resolution (HR) in the metaverse. In adverse visual conditions, such as haze, low or high luminosity, or even rainy weather conditions, the interactive systems in the metaverse still needs to show the virtual universe. 
	
	In computer vision, these problems are studied under two aspects: image restoration and image enhancement~\cite{wang2021joint,wang2020eventsr,wang2022semi,wang2019event}. Image restoration aims to reconstruct a clean image from the degraded one (\eg, noisy, blur image). In contrast, image enhancement focuses on improving image quality. 
	In the metaverse, image restoration and enhancement are much in need. For instance, the captured body information and the generated avatars may suffer from blur and noise when the user moves quickly. The  system thus needs to denoise and deblur the users' input signals and output clean visual information. Moreover, when the users are far from the camera, the generated avatar may be in a low-resolution (LR). It is necessary to enhance the spatial resolution and display the avatar in the 3D virtual environment with HR. 
	
	\subsubsection{Image Restoration}
	
	Image restoration has been shown to be effective for VR display. For instance,~\cite{xu2021artistic} focuses on colour VR based on image similarity restoration. In~\cite{zhao2017optimization,qiao2020situ,fischer2006handling}, optimisation-based methods are proposed to recover the textural details and remove the artefacts of the virtual images in VR, as shown in Figure~\ref{fig:image_enh}(b). These techniques can be employed as \textit{Diminished Reality} (DR)~\cite{mori2017survey}, which allows human users to view the blurred scenes of the metaverse with `screened contents'. Moreover,~\cite{vzuvzi2018impact} examines how image dehazing can be used to restore clean underwater images, which can be used for marker-based tracking in AR. Another issue is blur, which leads to registration failure in XR. The image quality difference between the real blurred images and the virtual contents could be apparent in the see-through device, \eg, Microsoft Hololens. Considering this problem,~\cite{okumura2006augmented,okumura2006image} proposes first to blur the real images captured by the camera and then render the virtual objects with blur effects. 
	
	Image restoration has been broadly applied in VR and AR. In the metaverse, colour correction, texture restoration, and blur estimation also play important roles in ensuring a realistic 3D environment and correct interaction among human avatars. However, it is worth exploring more adaptive yet effective restoration methods to deal with the gap between real and virtual contents and the correlation with the avatars in the metaverse. In particular, the physical world, the users, and the virtual entities are connected more closely in the metaverse than those of AR/VR. Therefore, image restoration should be subtly merged with the interaction system in the metaverse to ensure effectiveness and efficiency. 
	
	\subsubsection{Image Enhancement}
	Image enhancement, especially image super-resolution, has been extensively studied for XR displays. Image resolution has a considerable impact on user's view quality, which is related to the motion sickness caused by HMDs. Therefore, extensive research has been focused on optics SR~\eg,~\cite{clini2014augmented,grabovivckic2017super} and image SR~\cite{narasimhan2018ultra,feng2020applying,li2021applying} for the display in VR/AR. An example of image SR for 360 images for VR is shown in Figure~\ref{fig:image_enh}(c). Recently,~\cite{narasimhan2018ultra,feng2020applying,korinevskaya2018fast,Sevom2018360PS} applied deep learning and have achieved promising performance on VR displays. These methods overcome the resolution limitations that cause visible pixel artefacts in the display.      
	
	In the metaverse, super-resolution display affects the perception of the 3D virtual world. In particular, to enable a fully immersive environment, it is important to consider the display's image quality, for the sake of realism~\cite{realism-MR-TVCG}. This requires image super-resolution not only in optical imaging but also in the image formation process. Therefore, future research could consider the display resolution for the metaverse. Recently, some image super-resolution methods, \eg,~\cite{song2021addersr} have been directly applied to HR display, and we believe these techniques could help facilitate the technological development of the optical and display in the metaverse. Moreover, the super-resolution techniques in the metaverse can also be unitised to facilitate the visual localisation and mapping, body and pose tracking, and scene understanding tasks. Therefore, future research could jointly learn the image restoration/enhancement methods and the end-tasks to achieve the metaverse.

	\section{Edge and Cloud}\label{sec:edge} 
	\begin{figure*}[!t]
		\centering
		\includegraphics[width=.8\linewidth]{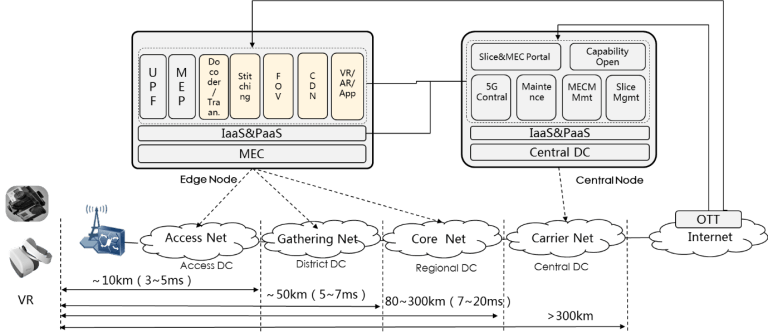}
		\caption{AR/VR network latency from the edge to the cloud~\cite{gsmacloudedge}.}
		\label{fig:edgecloudarvr}
	\end{figure*}
	With continuous, omnipresent, and universal interfaces to information in the physical and virtual world~\cite{7435333}, the metaverse encompasses the reality-virtuality continuum and allows user's seamless experience in between. To date, the most attractive and widely adopted metaverse interfaces are mobile and wearable devices, such as AR glasses, headsets, and smartphones, because they allow convenient user mobility. However, the intensive computation required by the metaverse is usually too heavy for mobile devices. Thus offloading is necessary to guarantee the timely processing and user experience. The traditional cloud offloading faces several challenges: user experienced latency, real-time user interaction, network congestion, and user privacy. In this section, we review the rising edge computing solution and its potential to tackle these challenges.
	
	\subsection{User Experienced Latency}
	In the metaverse, it is essential to guarantee an immersive feeling for the user to provide the same level of experience as reality. One of the most critical factors that impact the immersive feeling is the latency, e.g., motion to photon (MTP) latency\footnote{MTP latency is the amount of time between the user's action and its corresponding effect to be reflected on the display screen.}. Researchers have found that MTP latency needs to be below the human perceptible limit to allow users to interact with holographic augmentations seamlessly and directly~\cite{10.1145/1012551.1012559}. For instance, in the registration process of AR, large latency often results in virtual objects lagging behind the intended position~\cite{holloway1997registration}, which may cause sickness and dizziness. As such, reducing latency is critical for the metaverse, especially in scenarios where real-time data processing is demanded, e.g., real-time AR interaction with the physical world such as AR surgeries~\cite{fuchs1998augmented,soler2004virtual,rhienmora2010augmented}, or real-time user interactions in the metaverse such as multiplayer interactive exhibit in VR~\cite{li2016virtual} or multiple players' battling in Fortnite.
	
	As mentioned earlier, the metaverse often requires too intensive computation for mobile devices and thus further increases the latency. To compensate for the limited capacity of graphics and chipsets in the mobile interfaces (AR glasses and VR headsets etc.), offloading is often used to relieve the computation and memory burden at the cost of additional networking latency~\cite{braud2020multipath}. Therefore a balanced tradeoff is crucial to make the offloading process transparent to the user experience in the virtual worlds. But it is not easy. For example, rendering a locally navigable viewport larger than the headset's field of view is necessary to balance out the networking latency during offloading~\cite{yaqoob2021combined}. However, there is a tension between the required viewport size and the networking latency: longer latency requires a larger viewport and streaming more content, resulting in even longer latency~\cite{mehrabi2021multi}. Therefore, a solution with physical deployment improvement may be more realistic than pure resource orchestration.
	
	Due to the variable and unpredictable high latency~\cite{10.1145/1879141.1879143,cloudlet,9163287,u10}, cloud offloading cannot always reach the optimal balance and causes long-tail latency performance, which impacts user experience~\cite{40801}. Recent cloud reachability measurements have found that the current cloud distribution is able to deliver network latency of less than 100 ms. However, only a small minority (24 out of 184) of countries reliably meet the MTP threshold~\cite{corneo2021surrounded} via wired networks and only China (out of 184) meets the MTP threshold via wireless networks~\cite{nitinderimc}. Thus a complementary solution is demanded to guarantee a seamless and immersive user experience in the metaverse.
	
	Edge computing, which computes, stores, and transmits the data physically closer to end-users and their devices, can reduce the user-experienced latency compared with cloud offloading~\cite{lee2020augmented,shi2019mobile}. As early as 2009, Satyanarayanan~\emph{et al.}~\cite{cloudlet} recognized that deploying powerful cloud-like infrastructure just one wireless hop away from mobile devices, i.e., so-called cloudlet, could change the game, which is proved by many later works. For instance, Chen~\emph{et al.}~\cite{10.1145/3132211.3134458} evaluated the latency performance of edge computing via empirical studies on a suite of applications. They showed LTE cloudlets could provide significant benefits (60\% less latency) over the default of cloud offloading. Similarly, Ha~\emph{et al.}~\cite{ha2014towards} also found that edge computing can reduce the service latency by at least 80 ms on average compared to the cloud via measurements. Figure~\ref{fig:edgecloudarvr} depicts a general end-to-end latency comparison when moving from the edge to the cloud for an easier understanding.
	\begin{figure*}[!t]
		\centering
		\includegraphics[width=.8\linewidth]{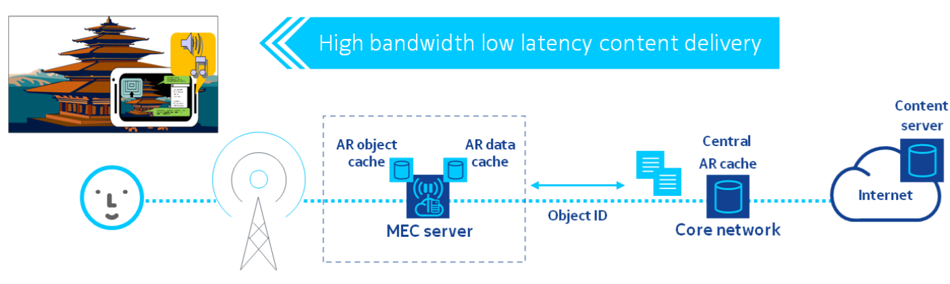}
		\caption{An example MEC solution for AR applications~\cite{hu2015mobile}.}
		\label{fig:edgear}
	\end{figure*}
	
	Utilising the latency advantage of edge computing, researchers have proposed some solutions to improve the performance of metaverse applications. For instance, EdgeXAR, Jaguar, and EAVVE target mobile AR services. EdgeXAR offers a mobile AR framework taking the benefits of edge offloading to provide lightweight tracking with 6 Degree of Freedom and hides the offloading latency from the user's perception~\cite{edgexar}. Jaguar pushes the limit of mobile AR's end-to-end latency by leveraging hardware acceleration on edge cloud equipped with GPUs~\cite{zhang2018jaguar}. EAVVE proposes a novel cooperative AR vehicular perception system facilitated by edge servers to reduce the overall offloading latency and makes up the insufficient in-vehicle computational power~\cite{zhou2018arve,9163287}. Similar approaches have also been proposed for VR services. Lin~\emph{et al.}~\cite{lin2021resource} transformed the problem of energy-aware VR experience to a Markov decision process and realised immersive wireless VR experience using pervasive edge computing. Gupta~\emph{et al.}~\cite{gupta2019millimeter} integrated scalable 360-degree content, expected VR user viewport modelling, mmWave communication, and edge computing to realise an 8K 360-degree video mobile VR arcade streaming system with low interactive latency. Elbamby~\emph{et al.}~\cite{elbamby2018edge} proposed a novel proactive edge computing and mmWave communication system to improve the performance of an interactive VR network game arcade which requires dynamic and real-time rendering of HD video frames. As the resolution increases, edge computing will play a more critical role to reduce the latency of 16K, 24K, or even higher resolution of the metaverse streaming.
	\subsection{Multi-access edge computing}
	The superior performance on reducing latency in virtual worlds has made edge computing an essential pillar in the metaverse’s creation in the eyes of many industry insiders. For example, Apple uses Mac with an attached VR headset to support 360-degree VR rendering~\cite{imac}. Facebook Oculus Quest 2 can provide VR experiences on its own without a connected PC thanks to its powerful Qualcomm Snapdragon XR2 chipset~\cite{oculuschipset}. However, its capacity is still limited compared with a powerful PC, and thus the standalone VR experience comes at the cost of lower framerates and hence less detailed VR scenes. By offloading to an edge server (e.g., PC), users can enjoy a more interactive and immersive experience at higher framerates without sacrificing detail. The Oculus Air Link~\cite{oculusair} announced by Facebook in April 2021 allows Quest 2 to offload to the edge at up to 1200 Mbps over the home Wi-Fi network, enabling a lag-free VR experience with better mobility. These products, however, are constrained to indoor environments with limited user mobility.
	
	To allow users to experience truly and fully omnipresent metaverse, seamless outdoor mobility experience supported by cellular networks is critical. Currently, last mile access is still the latency bottleneck in LTE networks~\cite{hadvzic2017edge}. With the development of 5G (promising down to 1 ms last mile latency) and future 6G, Multi-access edge computing (MEC) is expected to boost metaverse user experience by providing standard and universal edge offloading services one-hop away from the cellular-connected user devices, e.g., AR glasses. MEC, proposed by the European Telecommunications Standards Institute~(ETSI), is a telecommunication-vendor centric edge cloud model wherein the deployment, operation, and maintenance of edge servers is handled by an ISP operating in the area and commonly co-located with or one hop away from the base stations~\cite{mohan2018anveshak}. Not only can it reduce the round-trip-time (RTT) of packet delivery~\cite{zhou20215g}, but also it opens a door for near real-time orchestration for multi-user interactions~\cite{ren2020edge,jia2018delay}. MEC is crucial for outdoor metaverse services to comprehend the detailed local context and orchestrate intimate collaborations among nearby users or devices. For instance, 5G MEC servers can manage nearby users' AR content with only one-hop packet transmission and enable real-time user interaction for social AR applications such as 'Pokémon GO'~\cite{chen2021ar}. An example MEC solution proposed by ETSI~\cite{hu2015mobile} is depicted in Figure~\ref{fig:edgear}. 
	
	Employing MEC to improve metaverse experience has acquired academic attention. Dai~\emph{et al.}~\cite{dai2019view} designed a view synthesis-based 360-degree VR caching system over MEC-Cache servers in Cloud Radio Access Network (C-RAN) to improve the QoE of wireless VR applications. Gu~\emph{et al.}~\cite{gu2021reliability} and Liu~\emph{et al.}~\cite{liu2018mec} both utilised the sub-6 GHz links and mmWave links in conjunction with MEC resources to tackle the limited resources on VR HMDs and the transmission rate bottleneck for normal VR and panoramic VR video (PVRV) delivery, respectively.
	
	In reality, metaverse companies have also started to employ MEC to improve user experience. For instance, DoubleMe, a leading volumetric capture company, announced a proof of concept project, Holoverse, in partnership with Telefónica, Deutsche Telekom, TIM, and MobiledgeX, to test the optimal 5G Telco Edge Cloud network infrastructure for the seamless deployment of various services using the metaverse in August 2021~\cite{twinworld}. The famous Niantic, the company which has developed ‘Ingress’, ‘Pokémon GO’ and ‘Harry Potter: Wizards Unite’, envisions building a ``Planet-Scale AR''. It has allied with worldwide telecommunications carriers, including Deutsche Telekom, EE, Globe Telecom, Orange, SK Telecom, SoftBank Corp., TELUS, Verizon, and Telstra, to boost their AR service performance utilising MEC~\cite{nianticlabs}. With the advancing 5G and 6G technologies, the last mile latency will get further reduced. Hence MEC is promising to improve its benefit on the universal metaverse experience.
	
	\subsection{Privacy at the edge}\label{subsec:edgeprivacy}
	%zero-trust security architectures
	The metaverse is transforming how we socialise, learn, shop, play, travel, etc. Besides the exciting changes it's bringing, we should be prepared for how it might go wrong. And because the metaverse will collect more than ever user data, the consequence if things go south will also be worse than ever. One of the major concerns is the privacy risk~\cite{leenes2007privacy,falchuk2018social}. For instance, the tech giants, namely Amazon, Apple, Google (Alphabet), Facebook, and Microsoft, have advocated password-less authentication~\cite{alqubaisi2020should,haber2020passwordless} for a long time, which verifies identity with a fingerprint, face recognition, or a PIN. The metaverse is likely to continue this fashion, probably with even more biometrics such as audio and iris recognition~\cite{lodge2013nameless,boddington2021internet}. Before, if a user lost the password, the worst case is the user lost some data and made a new one to guarantee other data's safety. However, since biometrics are permanently associated with a user, once they are compromised (stolen by an imposter), they would be forever compromised and cannot be revoked, and the user would be in real trouble~\cite{ratha2006cancelable,ouda2010bioencoding}. 
	
	Currently, the cloud collects and mines the data of end-users and at the service provider side and thus has a grave risk of serious privacy leakage~\cite{ryan2011cloud,cuzzocrea2014privacy,liu2015survey}. In contrast, edge computing would be a better solution for both security and privacy by allowing data processing and storage at the edge~\cite{mollah2017security}. Edge service can also remove the highly private data from the application during the authorization process to protect user privacy. For instance, federated learning, a distributed learning methodology gaining wide attention, trains and keeps user data at local devices and updates the global model via aggregating local models~\cite{bonawitz2019towards}. It can run on the edge servers owned by the end users and conduct large-scale data mining over distributed clients without demanding user private data uploaded other than local gradients updates. This solution (train at the edge and aggregate at the cloud) can boost the security and privacy of the metaverse. For example, the eye-tracking or motion tracking data collected by the wearables of millions of users can be trained in local edge servers (ideally owned by the users) and aggregated via a federated learning parameter server. Hence, users can enjoy services such as visual content recommendations in the metaverse without leaking their privacy.
	
	Due to the distinct distribution and heterogeneity characteristics, edge computing involves multiple trust domains that demand mutual authentication for all functional entities~\cite{zhang2018data}. Therefore, edge computing requires innovative data security and privacy-preserving mechanisms to guarantee its benefit. Please refer to Section~\ref{sec:grand} for more details.
	
	\subsection{Versus Cloud}
	As stated above, the edge wins in several aspects: lower latency thanks to its proximity to the end-users, faster local orchestration for nearby users' interactions, privacy-preservation via local data processing. However, when it comes to long-term, large-scale metaverse data storage and economic operations, the cloud is still leading the contest by far. The primary reason is that the thousands of servers in the cloud datacenter can store much more data with better reliability than the edge. This is critical for the metaverse due to its unimaginably massive amount of data. As reasoned by High Fidelity~\cite{highfidelity}, the metaverse will be 1,000 times the size of earth 20 years from now, assuming each PC on the planet only needs to store and serve and simulate a much smaller area than a typical video game. For this reason, robust cloud service is essential for maintaining a shared space for thousands or even millions of concurrent users in such a big metaverse. 
	
	Besides, as the Internet bandwidth and user-device capacity increase, the metaverse will continue expansion and thus demand expanding computation and storage capacity. It is much easier and more economical to install additional servers at the centralised cloud warehouses than the distributed and space-limited edge sites. Therefore, the cloud will still play a vital role in the metaverse era. On the other hand, edge computing can be a complementary solution to enhance real-time data processing and local user interaction while the cloud maintains the big picture. 
	
	To optimise the interaction between the cloud and the edge, an efficient orchestrator is a necessity to meet diversified and stringent requirements for different processes in the metaverse~\cite{ravindra2017mathbb,carnevale2018cloud,wu2020cloud}. For example, the cloud runs extensive data management for latency-tolerant operations while the edge takes care of real-time data processing and exchange among nearby metaverse users. The orchestrator in this context can help schedule the workload assignment and necessary data flows between the cloud and the edge for better-integrated service to guarantee user's seamless experience. For example, edge services process real-time student discussions in a virtual classroom at a virtual campus held by the cloud. Or, like mentioned in Section~\ref{subsec:edgeprivacy}, the edge stores private data such as eye-tracking traces, which can leak user's interests to various types of visual content, while the cloud stores the public visual content.
	
	Several related works have been proposed lately to explore the potential of edge cloud collaborations for the metaverse. Suryavansh~\emph{et al.}~\cite{suryavansh2019tango} compared hybrid edge and cloud with baselines such as only edge and only cloud. They analyzed the impact of variation of WAN bandwidth, cost of the cloud, edge heterogeneity, and found that the hybrid edge cloud model performs the best in realistic setups. On the other hand, Younis~\emph{et al.} and Zhang~\emph{et al.} proposed solutions for AR and VR, respectively. More specifically, Younis~\emph{et al.}~\cite{younis2020latency} proposed a hybrid edge cloud framework, MEC-AR, for MAR with a similar design to Figure~\ref{fig:edgear}. In MEC-AR, MEC processes incoming edge service requests and manages the AR application objects. At the same time, the cloud provides an extensive database for data storage that cannot be cached in MEC due to memory limits. Zhang~\emph{et al.}~\cite{zhang2017towards} focused on the three main requirements of VR-MMOGs, namely stringent latency, high bandwidth, and supporting a large number of simultaneous players. They correspondingly proposed a hybrid gaming architecture that places local view change updates and frame rendering on the edge and global game state updates on the cloud. As such, the system cleverly distributes the workload while guaranteeing immediate responses, high bandwidth, and user scalability.
	
	In summary, edge computing is a promising solution to complement current cloud solutions in the metaverse. It can 1) reduce user experienced latency for metaverse task offloading, 2) provide real-time local multi-user interaction with better mobility support, and 3) improve privacy and security for the metaverse users. Indeed, the distribution and heterogeneity characteristics of edge computing also bring additional challenges to fully reach its potential. We briefly outline several challenges in Section~\ref{sec:grand}.
	
	% \subsection{Network Slicing}\label{sec:slicing}

	\section{Network}\label{sec:network}
	
	By design, a metaverse will rely on pervasive network access, whether to execute computation-heavy tasks remotely, access large databases, communicate between automated systems, or offer shared experiences between users. To address the diverse needs of such applications, the metaverse will rely heavily on future mobile networking technologies, such as 5G and beyond.

	\subsection{High Throughput and Low-latency}
	
	Continuing on the already established trends of real-time multimedia applications, the metaverse will require massive amounts of bandwidth to transmit very high resolution content in real-time. Many interactive applications consider the motion-to-photon latency, that is the delay between an action by the user and its impact on-screen~\cite{zhao2017estimating}, as one of the primary drivers of user experience. 
	
	The throughput needs of future multimedia applications are increasing exponentially. The increased capabilities of 5G (up to 10Gb/s~\cite{alliance20155g}) have opened the door to a multitude of applications relying on the real-time transmission of large amounts of data (AR/VR, cloud gaming, connected vehicles).
	By interconnecting such a wide range of technologies, the metaverse's bandwidth requirements will be massive, with high-resolution video flows accounting for the largest part of the traffic, followed by large amounts of data and metadata generated by pervasive sensor deployments~\cite{lu2019collaborative}. In a shared medium such as mobile networks, the metaverse will not only require a significant share of the available bandwidth, but also likely compete with other applications. As such, we expect the metaverse's requirements to exceed 5G's available bandwidth~\cite{braud2020multipath}.
	Latency requirements highly depend on the application. In the case of highly interactive applications such as online and cloud gaming, 130\,ms is usually considered as the higher threshold~\cite{lampe2013assessing}, while some studies exhibit drops in user performance for latencies as low as 23\,ms~\cite{10.1145/2702123.2702432}. Head-mounted displays such as see-through AR or VR, as well as haptic feedback devices exhibit motion-to-photon latency requirements down to the millisecond to preserve the user's immersion~\cite{lincoln2016motion,challacombe2003trans}. 
	
	Many factors contribute to the motion-to-photon latency, among which the hardware sensor capture time (e.g., frame capture time, touchscreen presses~\cite{10.1145/3083187.3083191}), and the computation time. For applications requiring latency in the order of the millisecond, the OS context switching frequency (often set between 100Hz and 1500Hz~\cite{time}), and memory allocation and copy times between different components (e.g. copy between CPU and GPU memory spaces) also significantly affect the overall motion-to-photon latency~\cite{hestness2015gpu}. In such constrained pipeline, network operations introduce further latency. Although 5G promised significant latency improvements, recent measurement studies show that the radio access network (RAN) itself displays very similar latency to 4G, while most of the improvements come from the communication between the gNB and the operator core network~\cite{10.1145/3387514.3405882}. However, it is important to note that most 5G networks are implemented in Non Standalone (NSA) mode, where only the RAN to the gNB use 5G radio, while the operator core network remains primarily 4G. Besides, despite standardising RAN latency to 4\,ms for enhanced Mobile Broadband (eMBB) and 0.5\,ms for Ultra-Reliable Low-Latency Communication (uRRLC -- still not implemented)~\cite{38.913}, the communication between the gNB and the core network account for most of the round trip latency (between 10 and 20\,ms), with often little control from the ISP~\cite{10.1145/3387514.3405882}. As such, unless servers are directly connected to the 5G gNB, the advantages of edge computing over cloud computing may be significantly limited~\cite{10.1145/3442381.3449854}, especially in countries with widespread cloud deployments~\cite{9472847}. Another consideration for reduced latency could be for content providers to control the entire end-to-end path~\cite{ammar2017vision}, by reaching inside the ISP using to network virtualization~\cite{8737635}. Such a vision requires commercial agreements between ISPs and content providers that would be more far-reaching than peering agreements between AS. One of the core condition for the metaverse to succeed will be the complete coordination of all actors (application developers, ISPs, content providers) towards ensuring a stable, low-latency and high throughput connection.
	
	\begin{figure}[t]
		\centering
		\includegraphics[width=\columnwidth]{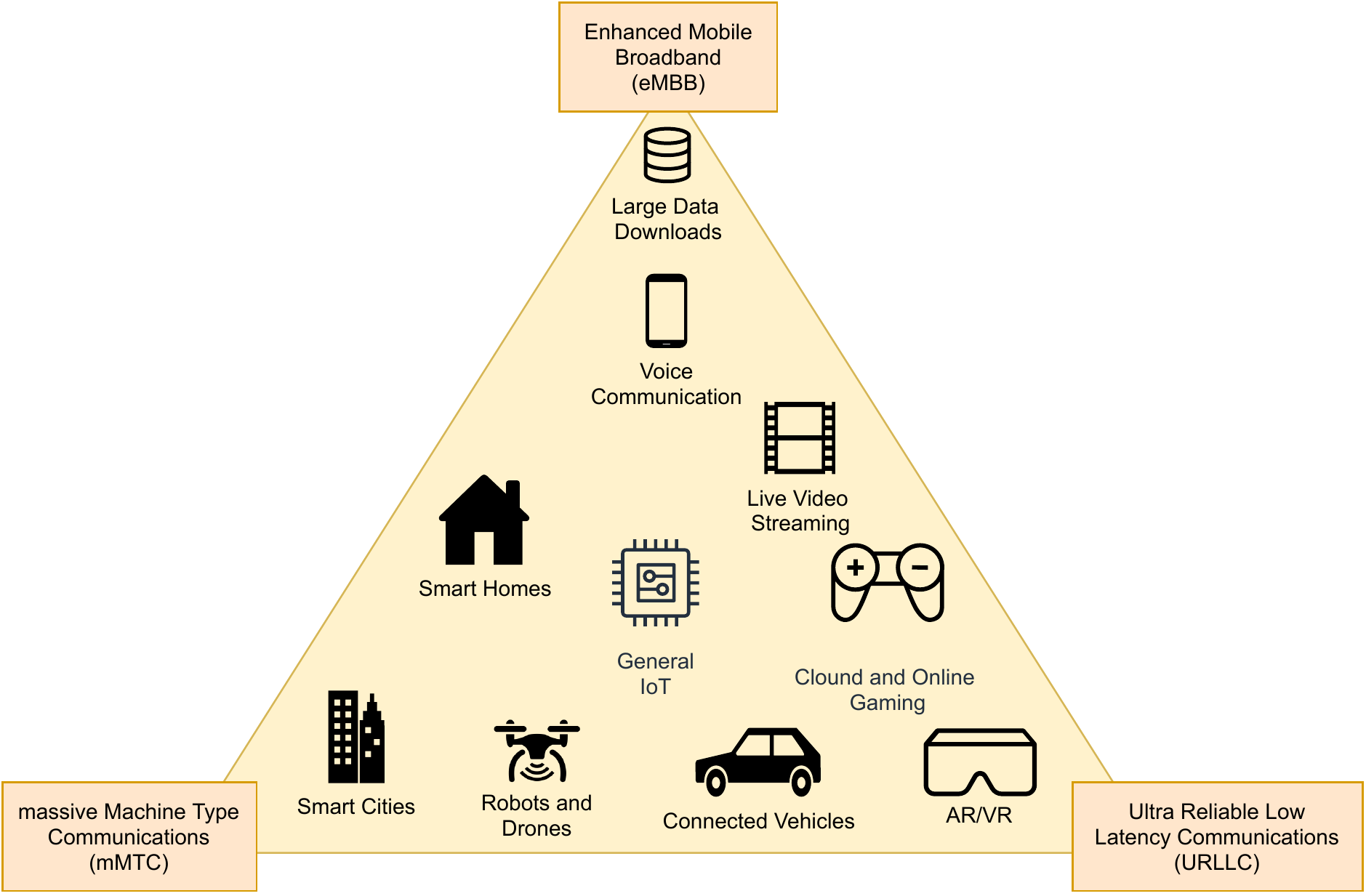}
		\caption{Metaverse applications and 5G service classes.}
		\label{fig:5gapplications}
	\end{figure}
	
	At the moment, 5G can therefore barely address the latency requirements of modern multimedia applications, and displays latency far too high for future applications such as see-through AR or VR. The URLLC service class promises low latency and high reliability, two often conflicting goals, with a standardised 0.5\,ms RAN latency.  However, URLLC is still currently lacking frameworks encompassing the entire network architecture to provide latency guarantees from client to server~\cite{8472907}. As such, no URLLC has so far %not 
	been commercially deployed. Besides, we expect uRRLC to prioritize applications for which low-latency is a matter of safety, such as healthcare, smart grids, or connected vehicles, over entertainment applications such as public-access AR and VR. The third service class provided by the 5G specification is massive Machine Type Communication (mMTC). This class targets specifically autonomous machine-to-machine communication to address the growing number of devices connected to the Internet~\cite{osseiran20165g}. Numerous applications of the metaverse will require mMTC to handle communication between devices outside of the users' reach, including smart buildings and smart cities, robots and drones, and connected vehicles. Future mobile networks will face significant challenges to efficiently share the spectrum between billions of autonomous devices and human-type applications~\cite{mahmood2020white,braud2021machine}. We summarize the application of these service classes in Figure~\ref{fig:5gapplications}
	Network slicing will also be a core enabler of the metaverse, by providing throughput, jitter, and latency guarantees to all applications within the metaverse~\cite{alliance2016description}. However, similar to URLLC, deploying network slicing in current networks will most likely target mission-critical applications, where network conditions can significantly affect the safety of the equipment or the users~\cite{8550640,8246845}. Besides, network slicing still needs to address the issue of efficiently orchestrating network resources to map the network slices with often conflicting requirements to the finite physical resources~\cite{8758980}. Finally, another feature of 5G that may significantly improve both throughput and latency is the usage of new frequency bands. The Millimeter wave band (24GHz-39GHz) allows for wide channels (up to 800MHz) providing large throughput while minimizing latency below 1\,ms. mmWave frequencies suffer from low range and obstacle penetration. As such, mmWave has been primarily used through dense base station deployments in crowded environments such as the PyeongChang olympics in 2018 (Korea) or Narita airport (Japan)~\cite{sakaguchi2017and}. Such dense deployments allowed to serve a significantly higher number of users simultaneously, while preserving high throughput and low latency at the RAN.
	
	\subsection{Human- and user-centric networking}
	
	The metaverse is a user-centric application by design. As such, every component of the multiverse should place the human user at its core. In terms of network design, such consideration can take several forms, from placing the user experience at the core of traffic management, to enabling user-centric sensing and communication.
	
	To address these issues, the network community has been increasingly integrating metrics of user experience in network performance measures, under the term Quality of Experience (QoE). QoE aims to provide a measurable way to estimate the user's perception of an application or a service~\cite{brunnstrom2013qualinet}. Most studies tend to use the term QoE as a synonym for basic Quality of Service (QoS) measures that may affect the user experience (e.g., latency, throughput). However, several works attempt to formalise the QoE through various models combining network- and application-level metrics. Although these models represent a step in the right direction, they are application-specific, and can be affected by a multitude of factors, whether human, system, or context~\cite{liotou2015quality}. Measuring QoE for a cloud gaming application run on a home video game console such as Sony PS Now\footnote{https://www.playstation.com/en-us/ps-now/} is significantly different from a mobile XR application running on a see-through headset. Besides, many studies focus on how to estimate the video quality as close as possible to the user's perception~\cite{5687947,chen2014qos}, and most do not consider other criteria such as usability or the subjective user perception~\cite{barakovic2010qoe}. The metaverse will need to integrate such metrics to handle user expectations and proactively manage traffic to maximise the user experience.
	
	Providing accurate QoE metrics to assess the user experience is critical for user-centric networked applications. The next step is to integrate QoE in how the network handles traffic. QoE can be integrated at various levels on the network. First, the client often carries significant capabilities in sensing the users, their application usage, and the application's context of execution. Besides, many applications such as AR or live video streaming may generate significant upload traffic. As such, it makes sense to make the client responsible for managing network traffic from an end-to-end perspective~\cite{braud2017future, 10.1145/3386290.3396935}. The server-side often carries more computing power, and certain applications are download-heavy, such as 360 video or VR content streaming. In this case, the server may use the QoE measurements communicated by the client to adapt the network transmission accordingly. Such approach has been used for adapting the quality of video streaming based on users' preferences~\cite{10.1145/3304109.3306231}, using client's feedback~\cite{8456342}. Finally, it is possible to use QoE measures to handle traffic management in core network, whether through queuing policies~\cite{8705295,6879998}, software defined network~\cite{liotou2018qoe}, or network slicing~\cite{yousaf2017network}. To address the stringent requirements leading to a satisfying user experiences, the metaverse will likely require to skirt the traditional layered approach to networks. The lower network layers may communicate information on network available resources for the application layer to adapt the amount of data to transmit, while measurement of QoE at application-level may be considered by the lower layers to adapt the content transmission~\cite{braud2017future}.
	
	Making networks more human-centric also means considering human activities that may affect nework communication. Mobility and handover are one of the primary factor affecting the core network parameters' stability. Handover have always been accompanied with a transient increase in latency~\cite{wylie2010throughput}. Although many works attempt to minimise handover latency in 5G~\cite{heinonen2016mobility, erel2019road}, such latency needs to be accounted for when designing ultra-low-latency services in mobile scenarios. The network conditions experienced by a mobile user are also directly related to the heterogeneity of mobile operator infrastructure deployment. A geographical measurement study of 4G latency in Hong Kong and Helsinki over multiple operators showed that mobile latency was significantly impacted by both the ISP choice and the physical location of the user~\cite{9066080}. Overall, user mobility significantly affects the network parameters that drive the user experience, and should be accounted for in the design of user-centric applications.
	
	Another aspect of human-centric networking lies within the rise of embodied sensors. In recent years, sensor networks have evolved from fixed environment sensors to self-arranging sensor networks~\cite{pottie1998wireless}. Many of such sensors were designed to remain at the same location for extended durations, or in controlled mobility~\cite{natalizio2013controlled}. In parallel, embodied sensors have long been thought to sense only the user. However, we are now witnessing a rise in embodied sensors sensing the entire environment of the user, raising the question of how such sensors may communicate in the already-crowded communication landscape. Detecting and aggregating redundant information between independent sensors may be critical to release important resources on the network~\cite{randhawa2017data}.
	
	% https://ieeexplore.ieee.org/abstract/document/9522162
	% https://ieeexplore.ieee.org/abstract/document/7528428
	% https://ieeexplore.ieee.org/abstract/document/6785339
	
	\subsection{Network-aware applications}
	
	In the previous section, we saw how the transmission of content should be driven by QoE measurements at the application layer. While this operation enables a high accuracy in estimating user experience by combining network metrics with application usage measures, the lower network layers only have limited control on the content to be transmitted. In many applications of the metaverse, it would make more sense for the application layer to drive the amount of data to transmit, as well as the priority of the content to the lower network layers~\cite{braud2020multipath}. Network-aware applications were proposed in the late 1990s to address such issues~\cite{miller2000collecting,bolliger1998framework}. Many framework were proposed, for both fixed and mobile networks~\cite{1265689}. More recently, network-aware applications have been proposed for resource provisioning~\cite{santos2019towards}, distributed learning optimization~\cite{wang2021network}, and content distribution~\cite{jiang2018cachalot,zhang2021sextant}. 
	
	With the rapid deployment of 5G, there is a renewed interest in network-aware applications~\cite{huang-alto-mowie-for-network-aware-app-03}. 5G enabled many user-centric applications to be moved to the cloud, such as cloud gaming, real-time video streaming, or cloud VR. These applications rely extensively on the real-time transmission of video flows, which quality can be adapted to the network conditions. The 5G specification includes network capability exposure, where the gNB can communicate the RAN conditions to the user equipment~\cite{38.913}. In edge computing scenarios where the edge server is located right after the gNB, the user equipment is thus made aware of the conditions of the entire end-to-end path. When the server is located further down the network, network capability exposure stills addresses one of the most variable components of the end-to-end path, providing valuable informations to drive the transmission. Such information from the physical and access layer can then be propagated to the network layer, where path decisions may be taken according to the various networks capabilities, the transport layer to proactively address potential congestion~\cite{zhu2017radio}, and the application layer to reduce or increase the amount of data to transmit and thus maximise the user experience~\cite{ramadan2021case}. 
	
	\begin{figure}[t]
		\centering
		\includegraphics[width=.98\columnwidth]{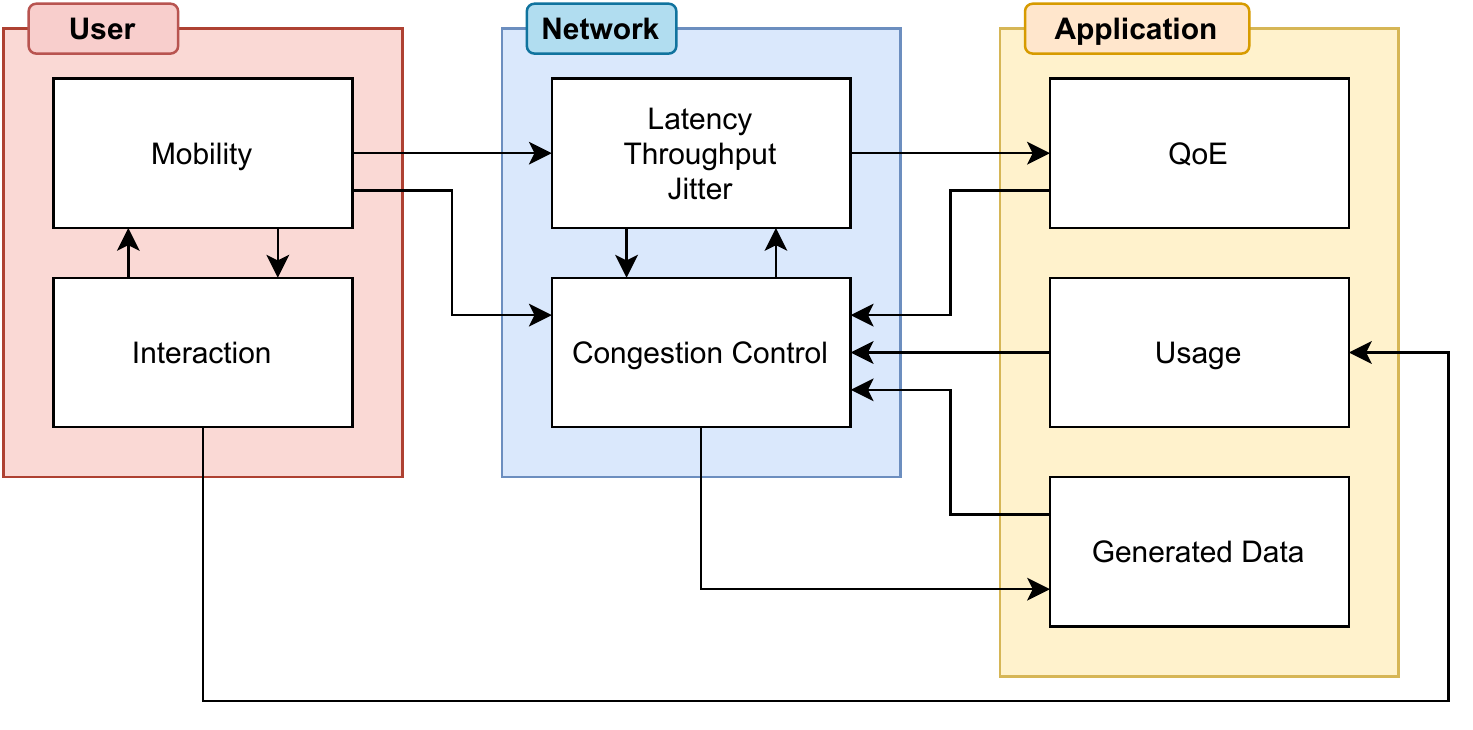}
		\caption{Network- and User-aware applications in the metaverse. A synergy between the traditional network layers and the application-level measures of user experience allow for maximising the user experience given the actual network conditions.}
		\label{fig:networkuser}
	\end{figure}
	
	Figure~\ref{fig:networkuser} summarises how a synergy between user-centric and network-aware applications can be established to maximize the user experience. The application communicates QoE and application usage metrics to the lower layers in order to adapt the transmission and improve the user experience. In parallel, the network layers communicate the network conditions to the application, which in turns regulates the amount of content to transmit on the network, for instance, by reducing the resolution of a video stream.

	\begin{figure*}[!t]
		\centering
		\includegraphics[width=\linewidth]{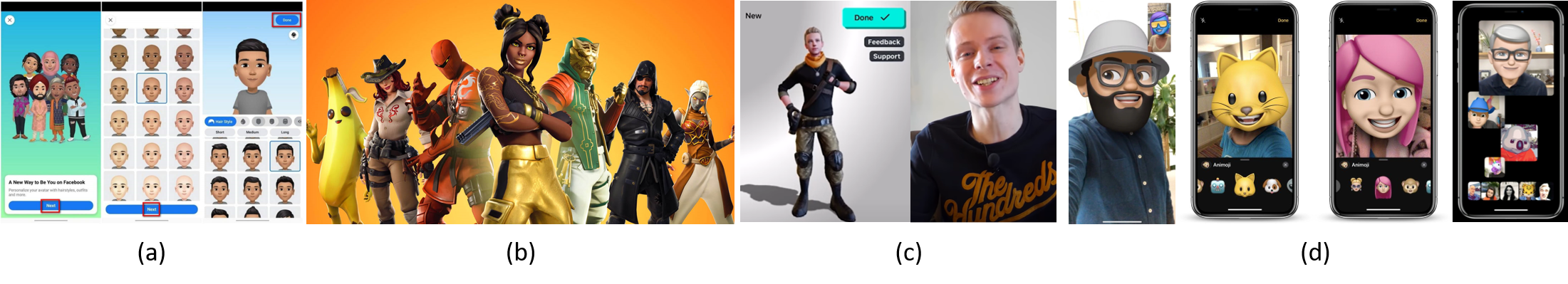}
		\caption{Several real-life examples of avatars, as a `second-identity' on a wide spectrum of virtual worlds: 
			(a) \textit{Facebook Avatar -- users can edit their own avatars in social media;}
			(b) \textit{Fortnite -- a multiplayer game that allows game players to create and edit their own worlds;}
			(c) \textit{VR Chat -- a VR game, and;}
			(d) \textit{Memoji -- virtual meetings with cartoonised faces during FaceTime on Apple iOS devices, regarded as an example of AR.}
		}
		\label{fig:avatar-real-life}
	\end{figure*}

	\section{Avatar}\label{sec:avatar}%no responsible person yet!

	%Appearance, Social influence, Continual identity, Human-avatar interaction, Collective experience.
	
	The term \textit{Avatar} is originated from the Hindu concept that describes the incarnation of a Hindu god, appearing as humans or animals in the ordinary world\footnote{\url{https://www.merriam-webster.com/dictionary/avatar}}. Avatars appear in a broad spectrum of digital worlds. First, it has been commonly used as profile pictures in various chatrooms (e.g., \textit{ICQ}), forums (e.g., \textit{Delphi}), blogs (e.g., \textit{Xanga}), as well as social networks (e.g., \textit{Facebook}, Figure~\ref{fig:avatar-real-life}(a)). 
	Moreover, game players, with very primitive metaverse examples such as \textit{AberMUD} and \textit{Second Life}, leverage the term avatar to represent themselves. Recently, game players or participants in virtual social networks can modify and edit the appearance of their avatars, with nearly unlimited options~\cite{DIS-practice-avatar}, for instance, \textit{Fortnite}, as shown in Figure~\ref{fig:avatar-real-life}(b). Also, VR games, such as VR Chat (Figure~\ref{fig:avatar-real-life}(c)), allow users to scan their physical appearance, and subsequently choose their virtual outfits, to mimic the users' real-life appearances. Figure~\ref{fig:avatar-real-life}(d) shows that online meetings, featured with AR, enable users to convert their faces into various cartoon styles. Research studies have also attempted to leverage avatars as one's close friends, coaches, or an imaginary self to govern oneself and goal setting such as learning and nutrition~\cite{nutrition-avatar, avatar-education-self}.
	
	Under the domain of computer science and technology, avatars denote the digital representation of users in virtual spaces, as above mentioned, and other physical embodied agents, e.g., social robots, regardless of form sizes and shapes~\cite{dark-robot}. This section focuses the discussion on the digital representationsn. However, it is worthy of pinpointing that the social robots could be a potential communication channel between human users and virtual entities across the real world and the metaverse, for instance, robots can become aware of the user's emotions and interact with the users appropriately in a conversation~\cite{10.1145/2912150}, or robots can serve as service providers as telework (telepresence workplace) in physical worlds~\cite{robot-disable}. 
	
	The digital representation of a human user aims to serve as a mirrored self to represent their behaviours and interaction with other users in the metaverse. 
	The design and appearance of avatars could impact the user perceptions, such as senses of realism~\cite{avatar-realism} and presence~\cite{avatar-missed-finger-presence}, trust~\cite{avatar-behavior}, body ownership~\cite{avatar-body-ownership}, and group satisfaction~\cite{avatar-group-sat}, during various social activities inside the metaverse, which are subject to a bundle of factors, such as the details of the avatar's face~\cite{avatar-face-tracking} and the related micro-expression~\cite{avatar-micro-expression}, the completeness of the avatar's body~\cite{avatar-missed-finger-presence}, the avatar styles~\cite{avatar-style}, representation~\cite{realism-another-avatar}, colour~\cite{avatar-color} and positions~\cite{avatar-size}, fidelity~\cite{avatar-high-fed}, the levels of detail in avatars' gestures~\cite{avatar-detail}, shadow~\cite{avatar-shadow}, the design of avatar behaviours~\cite{avatar-behavior}, synchronisation of the avatar's body movements~\cite{avatar-syn-motion}, Walk-in-Place movements~\cite{leg-movements}, ability of recognising the users' self motions reflected on their avatars~\cite{avatar-fedelity-sex}, cooperation and potential glitches among multiple avatars~\cite{avatar-coordination}, and to name but a few.
	As such, avatars has the key role of shaping how the virtual social interaction performs in the multi-user scenarios inside the metaverse~\cite{DIS-practice-avatar}. 
	However, the current computer vision techniques are not ready to capture and reflect the users' emotions, behaviours and their interaction in real-time, as mentioned in Section~\ref{sec:cv}, 
	Therefore, additional input modality can be integrated to improve the granularity of avatars. For instance, the current body sensing technology is able to enrich the details of the avatar and reflect the user's reactions in real-time. In~\cite{avatar-pupillary}, an avatar's pupillary responses can reflect its user's heartbeat rate. 
	In the virtual environments of \textit{VR Chat}, users in the wild significantly rely on body sensing technology (i.e., sensors attached on their body) to express their body movements and gestural communication, which facilitate non-verbal user interaction (i.e., voice, gestures, gaze, and facial expression) emulating the indispensable part of real-life communication~\cite{Knapp1972NonverbalCI}.

	When avatars become more commonplace in vastly diversified virtual environments, the studies of avatars should go beyond the sole design aspects as above. We briefly discuss six under-explored issues related to the user interaction through avatars with virtual environments -- 1) in-the-wild user behaviours, 2) the avatar and their contexts of virtual environments, 3) avatar-induced user behaviours, 4) user privacy, 5) fairness, and 6) connections with physical worlds.
	First, as discussed in prior sections, metaverse could become independent virtual venues for social gatherings and other activities. The user behaviours in the wild (i.e., outside laboratories), on behalf of the users' avatars, need further investigation, and the recently emerging virtual worlds could serve as a testing bed for further studies. 
	For instance, it is interesting to understand the user behaviours, in-group dynamics, between-group competitions, inside the virtual environments encouraging users to earn NFTs through various activities. Second, We foresee that 
	users with avatars will experience various virtual environments, representing diversified contexts. The appearance of avatars should fit into such contexts. For instance, avatars should behave professionally to gain trust from other stakeholders in virtual work environments~\cite{avatar-context}. 
	Third, it is necessary to understand the changes and dynamics of user behaviours induced by the avatars in virtual environments. A well-known example is the Proteus Effect~\cite{Yee2007ThePE} that describes the user behaviours within virtual worlds are influenced by the characteristics of our avatar.
	Similarly, supported by the \textit{Self-perception theory}, user's behaviours in virtual environments are subjects to avatar-induced behavioural and attitudinal changes through a shift in self-perception~\cite{Avatar-Proteus-Effect}. 
	
	Furthermore, when the granularity of the avatars can be truly reflected by advancing technologies, avatar designers should consider privacy-preserving mechanisms to protect the identity of the users~\cite{privacy-photo-obfuscation-shared-photos}. 
	Next, the choices of avatars should represent a variety of populations. The current models of avatars may lead to biased choices of appearances~\cite{avatar-biased}, for instance, a tall and white male~\cite{cannnot-ignore-avatar-real-identity}. Avatar designers should offer a wide range of choices that enables the population to equally choose and edit their appearance in virtual environments.
	Finally, revealing metaverse avatars in real-world environments are rarely explored. Revealing avatars in the real world is able to enhance the presence (i.e., co-presence of virtual humans in the real world~\cite{co-presence-virtual-humans}), especially when certain situations prefer the physical presence of an avatar that represents a specific person, e.g., lectures~\cite{avatar-virtual-physical}. Interaction designers should explore various ways of displaying the avatar on tangible devices (three examples as illustrated in Figure~\ref{fig:alter-display}) as well as social robots.

	\section{Content Creation}\label{sec:content-create}
	
	\begin{figure*}[!t]
		\centering
		\includegraphics[width=\linewidth]{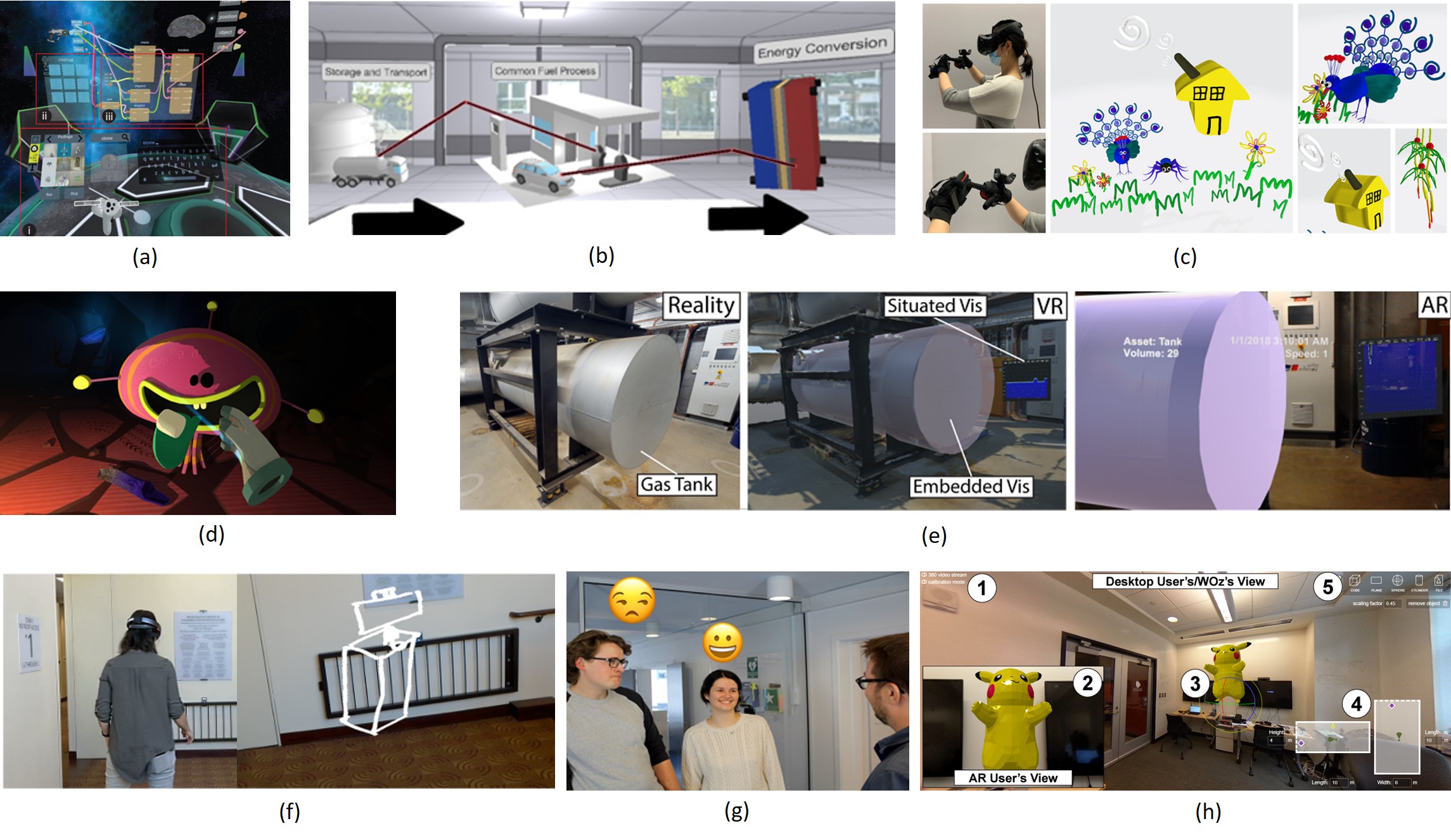}
		\caption{Authoring systems with various virtual environments across extended reality (e) \& (h), VR (a) -- (d), and AR (f) -- (g): 
			(a) FlowMatic~\cite{Zhang2020FlowMaticAI},
			(b) VR nuggets with patterns~\cite{Horst2019VirtualRF},
			(c) HandPainter~\cite{Lau-chi21} for VR artistic painting,
			(d) Authoring Interactive VR narrative~\cite{Cutler2020AuthoringIV},
			(e) Corsican Twin~\cite{Prouzeau2020CorsicanTA} as an example of digital twins,
			(f) PintAR~\cite{Gasques2019PintARSS} for low-fidelity AR sketching,
			(g) Body LayARs~\cite{Pohl2020BodyLA} creates AR emojis according to the detected faces, 
			(h) Creating medium-fidelity AR/VR experiences with 360 degree theatre~\cite{Speicher2021DesignersTS}. }
		\label{fig:authorinh-create}
	\end{figure*}
	
	This section aims to describe the existing authoring systems that support content creation in XR, and then discuss censorship in the metaverse and a potential picture of creator culture. 
	
	\subsection{Authoring and User Collaboration}\label{ssec:authoring}
	
	In virtual environments, authoring tools enable users to create new digital objects in intuitive and creative manners. Figure~\ref{fig:authorinh-create} illustrates several examples of XR/AR/VR authoring systems in the literature. In VR~\cite{Zhang2020FlowMaticAI, Lau-chi21, Horst2019VirtualRF, Cutler2020AuthoringIV}, the immersive environments provides virtual keyboards and controllers that assist users in accomplishing complicated tasks, e.g., constructing Functional Reactive Programming (FRP) diagram as shown in Figure~\ref{fig:authorinh-create}(a). In addition, re-using existing patterns can speed up the authoring process in virtual environments, such as a presentation (Figure~\ref{fig:authorinh-create}(b)). Also, users can leverage smart wearables to create artistic objects, e.g., smart gloves in Figure~\ref{fig:authorinh-create}(c). Combined with the above tools, users can design interactive AI characters and their narratives in virtual environments (Figure~\ref{fig:authorinh-create}(d)). 
	In AR or MR, users can draw sketches and paste overlays on physical objects and persons in their physical surroundings~\cite{Leiva2020ProntoRA, Nebeling2018ProtoARRP, Chidambaram2021ProcessARAA, Gasques2019PintARSS, Pohl2020BodyLA}. Augmenting the physical environments can be achieved by drawing a new sketch in mid-air~\cite{Leiva2020ProntoRA, Gasques2019PintARSS}, e.g., Figure~\ref{fig:authorinh-create}(f), detecting the contexts with pre-defined AR overlays ((Figure~\ref{fig:authorinh-create}(g)), recording the motions of real-world objects to simulate their physical properties in AR~\cite{Mller2021SpatialProtoER}, inserting physical objects in AR (Figure~\ref{fig:authorinh-create}(h)), or even using low-cost objects such papers~\cite{Nebeling2019360protoMI} and polymer clay~\cite{Nebeling2018ProtoARRP}.

	Although the research community is increasingly interested in XR/AR/VR authoring systems~\cite{Freitas2020ASR}, such authoring tools and platforms mainly assist users in creating and inserting content without high technological barriers. Additionally, it is important to note that AI can play the role of automatic conversion of entities from the physical world to virtual environments (Section~\ref{sec:art-int}).
	As such, UI/UX designers and other non-coders feel more accessible to content creation in virtual environments, on top of virtual world driven by the AI-assisted conversion. Nevertheless, to build the metaverse at scale, three major bottlenecks exist: 1) organising the new contents in interactive and storytelling manners~\cite{Ashtari2020CreatingAA}, 2) allowing collaborative works among multiple avatars (i.e., human users)~\cite{Krau2021CurrentPC}, and 3) user interaction supported by multiple heterogeneous devices~\cite{XD-AR-Cross}. 
	To the best of our knowledge, only limited work attempts to resolve the aforementioned bottleneck, and indicate the possibility of role-based collaborative content creation~\cite{Speicher2021DesignersTS, XR-studio, CHI-Studio-ARplsuVR-XR}. As depicted by Speichers~\emph{et al.}~\cite{Speicher2021DesignersTS}, the peer users can act in different roles and work collaboratively in virtual environments, such as \textit{wizards}, \textit{observers}, \textit{facilitators}, \textit{AR and VR users as content creators}, and so on. Similarly, Nebeling~\emph{et al.} consider three key roles of \textit{directors}, \textit{actors}, and \textit{cinematographers} to create complex immersive scenes for storytelling scenarios in virtual environments.

	Although we cannot speculate all the application scenarios of the authoring techniques and solutions, human users can generate content in various ways, i.e., user-generated content, in the metaverse. 
	It is important to note that 
	such authoring systems and their digital creation are applicable to two apparent use cases. First, remote collaboration on physical tasks~\cite{WANG2021102071} and virtual tasks~\cite{He2020CollaboVRAR} enable users to give enriched instructions to their peers, and accordingly create content for task accomplishment remotely. 
	Second, the content creation can facilitate the video conference or equivalent virtual venues for social gathering, which are the fundamental function of the metaverse.
	Since 2020, the unexpected disruption by the global pandemic has sped up the digital transformation, and hence virtual environments are regarded as an alternative for virtual travelling, social gathering and professional conferencing~\cite{10.1145/3411763.3451740, Sarkady2020VirtualRA}. 
	Online Lectures and remote learning are some of the most remarkable yet impactful examples, as schools and universities suspend physical lessons globally. Students primarily rely on remote learning and obtaining learning materials from proprietary online platforms. Teachers choose video conferencing as the key reaching point with their students under this unexpected circumstance. 
	However, such online conferences would require augmentations to improve their effectiveness~\cite{sara-chi21-onlineLecture}. 
	\textit{XRStudio} demonstrates the benefits from the additions of virtual overlays (AR/VR) in video conferencing between instructors and students. 
	Similarly, digital commerce relies heavily on online influencers to stimulate sales volumes. Such online influencers share user-generated content via live streaming, for instance, tasting and commenting on foods online~\cite{food-KOL}, gain attention and interactions with viewers online. 
	According to the above works, we foresee that the future of XR authoring systems can serve to augment the participants (e.g., speakers) during their live streaming events. The enriched content, supported by virtual overlays in XR, can facilitate such remote interaction. The speakers can also invite collaborative content creations with the viewers. The metaverse could serve as a medium to knit the speakers (the primary actor of user-generated content) and the viewers virtually onto a unified landscape.

	\subsection{Censorship}\label{ssec:censorship} 
	
	Censorship is a common way of suppressing ideas and information when certain stakeholders, regardless of individuals or groups, as well as authorities may find such ideas and information are objectionable, dangerous, or detrimental~\cite{IMC-10.1145/2663716.2663720, 10.1109/ICDCS.2010.46, 179184}.
	In the real world, censorship brings limited access to specific websites, controlling the dissemination of information electronically, restricting the information disclosed to the public, facilitating religious beliefs and creeds, and reviewing the contents to be released, so as to guarantee the user-generated contents would not violate rules and norms in a particular society, with the potential side effects of sacrificing freedom of speech or certain digital freedom (e.g., discussions on certain topics)~\cite{9381296}. 
	Several censorship techniques (e.g., DNS manipulation and HTTP(S)-layer interference) are employed digitally~\cite{IMC-10.1145/2663716.2663720, 10.1109/ICDCS.2010.46, 179183, 179184, 9381296, 10.1145/2504730.2504763, censored-planet}: 1) entire subnets are blocked by using IP-filtering techniques; 2) certain sensitive domain is limited to block the access of specific websites; 3) certain keywords become the markers of targeting certain sensitive traffic, 4) Specific contents and pages are specified as the sensitive or restricted categories, perhaps with manual categorisations. 
	
	Other prior works of censorship in the Internet and social networks have reflected the censorship employed in %China~\cite{10.1109/ICDCS.2010.46}, 
	Iran~\cite{179184}, Egypt, Sri Lanka, Norway~\cite{censored-planet}, Pakistan~\cite{179183}, Syria~\cite{IMC-10.1145/2663716.2663720} and other countries in the Arab world~\cite{10.1145/2504730.2504763}. The majority of these existing works leverages the probing approaches -- the information being censored is identified by the events of requests of generating new content and subsequently the actual blocking of such requests. Although the probing approaches allow us to become more aware of censorship in particular regions, it poses two key limitations: 1) limited observation size (i.e., limited scalability) and 2) difficult identification of the contents being censored (i.e., primarily by inference or deduction). 
	
	Once the metaverse becomes a popular place for content creations, numerous user interaction traces and new content will be created. For instance, Minecraft has been regarded as a remarkable virtual world in which avatars have a high degree of freedom to create new user-generated content. Minecraft also supports highly diversified users who intend to meet and disseminate information in such virtual worlds. 
	%In addition, such virtual worlds feature with highly diversified users that intend to meet and disseminate information with each other. 
	In 2020, Minecraft acted as a platform to hold the first library for censored information, named \textit{The Uncensored Library}\footnote{\url{https://www.uncensoredlibrary.com/en}}, with the emphasis of `\textit{A safe haven for press freedom, but the content you find in these virtual rooms is illegal}'. 
	Analogue to the censorship employed on the Internet, we conjecture that similar censorship approaches will be exerted in the metaverse, especially when the virtual worlds in the metaverse grow exponentially, for instance, blocking the access of certain virtual objects and virtual environments in the metaverse. 
	It is projected that censorship may potentially hurt the interoperability between virtual worlds, e.g., will the users' logs and their interaction traces be eradicated in one censored virtual environment? As such, do we have any way of preserving the ruined records? Alternatively, can we have any instruments temporarily served as a haven for sensitive and restricted information?
	Also, other new scenarios will appear in the virtual 3D spaces. For example, censorship can be applied to restrict certain avatar behaviours, e.g., removal of some keywords in their avatars' speeches, forbidding avatars' body gestures, and other non-verbal communication means~\cite{talk-wo-voice}. 
	
	Although we have no definitive answer to the actual implementation of the censorship in the metaverse and the effective solutions to alleviate such impacts, we advocate a comprehensive set of metrics to reflect the degree of censorship in multitudinous virtual worlds inside the metaverse, which could serve as an important lens for the metaverse researchers to understand the root cause(s) and its severity and popularity of the metaverse censorship. The existing metrics for the Internet, namely \textit{Censored Planet}, perform a global-scale censorship observatory that helps to bring transparency to censorship practices, and supports the human rights of Internet users through discovering key censorship events.

	\subsection{Creator Culture}\label{ssec:creator-culture}
	
	The section on content creation ends with a conjecture of creator culture, as we can only construct our argument with the existing work related to creators and digital culture to outline a user-centric culture on a massive scale inside the metaverse. 
	First, as every participant in the metaverse would engage in creating virtual entities and co-contribute to the new assets in the metaverse, we expect that the aforementioned authoring systems should remove barriers for such co-creation and co-contribution. In other words, the digital content creation will probably let all avatars collaboratively participate in the processes, instead of a small number of professional designers~\cite{EUE-behavioral-patterns}.
	Investigating the design space of authoring journeys and incentive schemes designated for amateur and novice creators to actively participate in the co-creation process could facilitate the co-creation processes~\cite{CHI17-create-consumer}. The design space should further extend to the domain of human-AI collaboration, in which human users and AI can co-create instances in the metaverse~\cite{co-create-AI}.
	Also, one obvious incentive could be token-based rewards. For instance, in the virtual environment \textit{Alien Worlds}, coined as a token-based pioneer of the metaverse, allows players' efforts, through accomplishing missions with their peers, to be converted into NFTs and hence tangible rewards in the real world.
	
	% On the other hand, 
	It is projected that the number of digital contents in the metaverse will proliferate, as we see the long-established digital music and arts~\cite{MM-ART-2015, music-big-data-infra}. For instance, Jiang~\emph{et al.}~\cite{Lau-chi21} offer a virtual painting environment that encourages users to create 3D paintings in VR. Although we can assume that computer architectures and databases should own the capacity to host such growing numbers of digital contents, we cannot accurately predict the possible outcomes when the accumulation of massive digital contents exceed the capacity of the metaverse -- the outdated contents will be phased out or be preserved. This word capacity indicates the computational capacity of the metaverse, and the iteration of the virtual space. An analogy is that real-world environments cannot afford an unlimited number of new creations due to resource and space constraints. For example, an old street painting will be replaced by another new painting. 
	
	Similarly, the virtual living space containing numerous avatars (and content creators) may add new and unique contents into their virtual environments in iterative manners. 
	In virtual environments, the creator culture can be further enhanced by establishing potential measurements for the preservation of outdated contents, for instance, a virtual museum to record the footprint of digital contents\cite{knowledge-management, digital-culture}. 
	The next issue is how the preserved or contemporaneous digital contents should appear in real-world environments. Ideally, everyone in physical environments can equally access the fusing metaverse technology, sense the physical affordances of the virtual entities~\cite{10.1145/3313831.3376614}, and their contents in public urban spaces~\cite{creator-centric}. Also, the new virtual culture can influence the existing culture in the real world, for instance, digital cultures can influence working relationships in workspaces~\cite{vultures, VR-organisational-drivers}.

	\section{Virtual Economy}\label{sec:ntf-econ}
	
\begin{figure}[!t]
		\centering
		\includegraphics[width=\linewidth]{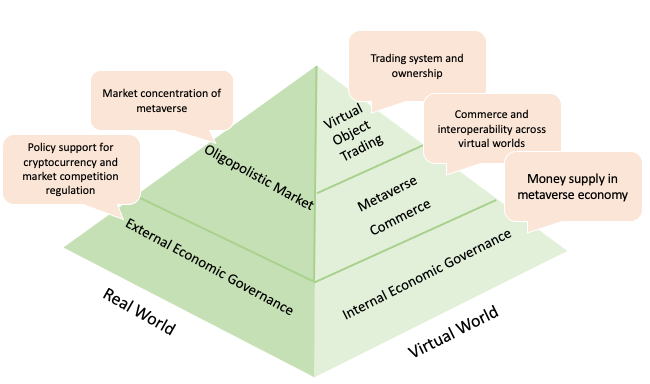}
		\caption{A breakdown of sub-topics discussed in the section of Virtual Economy, where they can be separated into two strands depending on whether they are related to real or the virtual world. Amongst them, internal/ external economic governance forms the bedrock of the virtual economy. Building upon, the section discusses the metaverse industry's market concentration in the real world and commerce, specifically trading in the virtual world.}
		\label{fig:econ_sum}
	\end{figure}

	Evident in Figure~\ref{fig:econ_sum}, this section first introduces readers to the economic governance required for the virtual worlds. Then, we discuss the metaverse industry’s market structure and details of economic support for user activities and content creation discussed in the previous section. 
	
	\subsection{Economic Governance}
	
	Throughout the past two decades, we have observed several instances where players have created and sustained in-game economic systems. The space theme game \textit{EVE} quintessentially distinguishes itself from others with a player-generated sophisticated cobweb of an economic system, where players also take up some roles in economic governance, as demonstrated by their monthly economic reports\footnote{\url{https://bit.ly/3o49mgM}}. This is not to say, however, metaverse developers can simply mimic \textit{EVE}'s success and delegate all economic governance to their users. For one, one of the main underlying difficulties of realising cryptocurrency as a formal means of transaction is its association with potential deflationary pressure. Specifically, whereas players control currency creation in \textit{EVE}\footnote{\url{https://bit.ly/3u6PiLP}}, cryptocurrency is characterised by a steady and relatively slow money supply growth due to how the `mining' process is set up. Unlike the current world we reside in, where central banks can adjust money supply through monetary instruments and other financial institutions can influence money supply by creating broad money, cryptocurrency in its nascent form simply lacks such a mechanism. Consequently, the quantity theory of money entails that if money velocity is relatively stable in the long term, one is justified to be concerned about deflationary pressure as the money supply fails to accommodate the growing amount of transactions in a thriving metaverse~\cite{harwick2016cryptocurrency}. Though some may posit that issuing new cryptocurrency is a viable remedy to address the relatively static money supply, such a method will only be viable if the new currency receives sufficient trust to be recognised as a formal currency. To achieve such an end, users of the metaverse community will have to express some level of acceptance towards the new currency, either endogenously motivated or through developers' intervention. However, suppose an official conversion rate between the newly launched cryptocurrency and the existing one was to be enforced by developers. In that case, they could find themselves replaying the failure of bimetallism as speculators in the real world are incentivised to exploit any arbitrage, leading to `bad' crypto drives out `good' crypto under Gresham's Law~\cite{macleod1858elements}. Therefore, to break this curse, some kind of banking system is needed to enable money creation through fractional reserve banking~\cite{harwick2016cryptocurrency} instead of increasing the monetary base. Meaning that lending activities in the metaverse world can increase the money supply. There are already several existings platforms such as BlockFi that allow users to deposit their cryptocurrency and offer an interest as reward. Nevertheless, the solution does not come with no hitch, as depositing cryptocurrency with some establishments can go against the founding ideas of decentralisation~\cite{harwick2016cryptocurrency}. Alternative to introducing a banking system, others have proposed different means to stabilise cryptocurrency. An example can be stabilisation through an automatic rebasing process to national currency or commodity prices~\cite{ametrano2016hayek}. A pegged cryptocurrency is not an imaginary concept in nowadays world. A class of cryptocurrency known as stablecoin that pegs to sovereign currencies already exists, and one study have shown how arbitrage in one of the leading stablecoins, Tether, has produced a stabilising effect on the peg~\cite{lyons2020keeps}. Even more, unlike the potential vulnerability of stablecoins to changes in market sentiment on the sufficiency of collateral to maintain the peg~\cite{lyons2020keeps}, a commonly recognised rebasing currency may circumvent such hitch as it does not support a peg through the use of collateral. Nonetheless, it is worth mentioning that there has not yet been a consensus on whether cryptocurrency's deflationary feature should be considered as its shortcoming nor the extent of deflationary pressure will manifest in cryptocurrency in future. Additionally, another major doubt on cryptocurrency becoming a standard form of means of transaction arises from its highly speculative attribute. Thus, developers should consider the economic governance required to tune cryptocurrency into a reliable and robust currency to be adopted by millions of metaverse users. Similarly, we have also noticed the need of internal governance in areas such as algorithmic fairness~\cite{Procedural-justice, qualitative-exploration-fairness}, which we will discuss in detail in Section~\ref{sec:fair}.
	
	Furthermore, another potential scope for economic governance emerges at a higher level: governments in our real world. As we will show in the next section, degrees of competition between metaverse companies can affect consumer welfare. Therefore, national governments or even international bodies should be entrusted to perform their roles in surveilling for possible collusion between these firms as they do in other business sectors. In extreme cases, the governments should also terminate mergers and acquisitions or even break apart metaverse companies to safeguard the welfare of consumers, as the social ramification being at stake (i.e., control over a parallel world) is too great to omit. That being said, economic governance at (inter) national level is not purely regressive towards the growth of metaverse business. Instead, state intervention will play a pivotal role in buttressing cryptocurrency's status as a trusted medium of exchange in the parallel world. This is because governments' decisions can markedly shape market sentiment. This is seen in the two opposing instances of Turkey's restriction\footnote{\url{https://reut.rs/3AEuttF}} on cryptocurrency payment and El Salvador's recognition of Bitcoin as legal tender\footnote{\url{https://cnb.cx/39COl4m}}, which both manifest as shocks to the currency market. Therefore, even in lack of centralised control, governments' assurances and involvements in cryptocurrency that promise political stability towards the currency can in return brings about stability in the market as trust builds in. Indeed, government involvement is a positive factor for trust in currency valued by interviewees in a study~\cite{zarifis2014consumer}. Though it may not wholly stabilise the market, it removes the uncertainty arising from political factors. Furthermore, national and international bodies' consents will also be essential for financial engineering, such as fractional reserve banking for cryptocurrency. Building such external governance is not a task starting from scratch; One can learn from past regulations on cryptocurrency and related literature discussions~\cite{motsi2018financial, schaupp2018cryptocurrency}. Nonetheless, the establishment of the cryptocurrency banking system has another fallibility in robustness as authorities can face tremendous hardship in acting as lender of last resort to forestall the systematic collapse of this new banking system~\cite{wall_2019}, which only increases their burden on top of tackling illegal activities associated with decentralised currency~\cite{foley2019sex}.
	
	\subsection{Oligopolistic Market}
	
	\addtocounter{footnote}{-1}
	\begin{figure}[!ht]
		\centering
		\includegraphics[width=\linewidth]{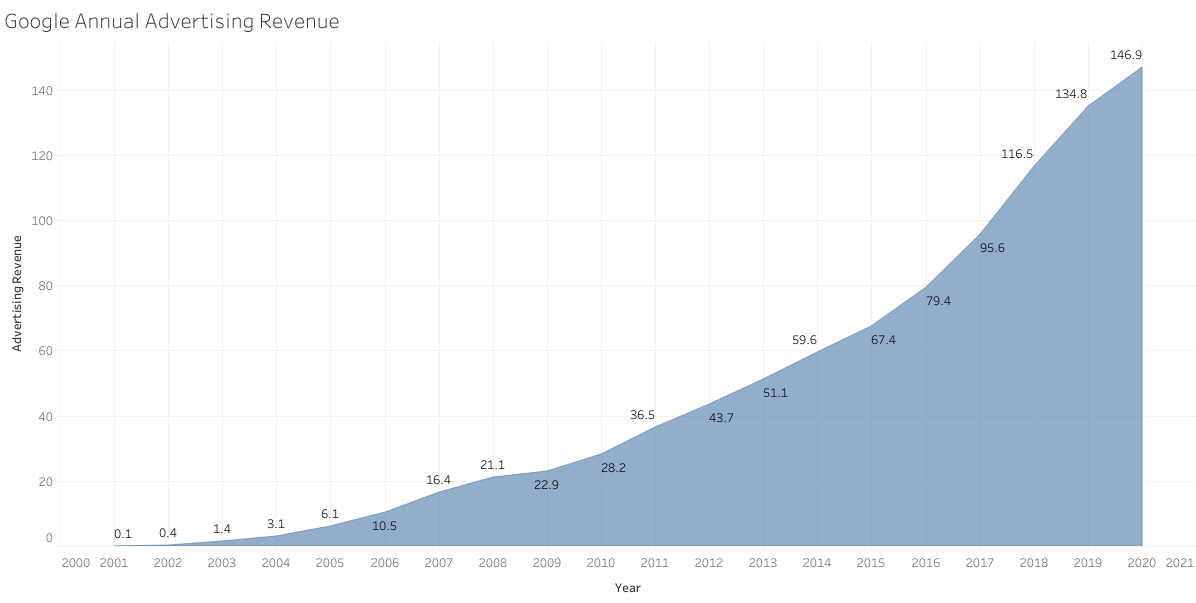}
		\caption{Historical trend of Google's annual advertising revenue\protect\footnotemark.}
		\label{fig:google_ads}
	\end{figure}
	\footnotetext{{\url{https://bit.ly/3o2wGeM}}}
	
\begin{figure}[!ht]
		\centering
		\includegraphics[width=\linewidth]{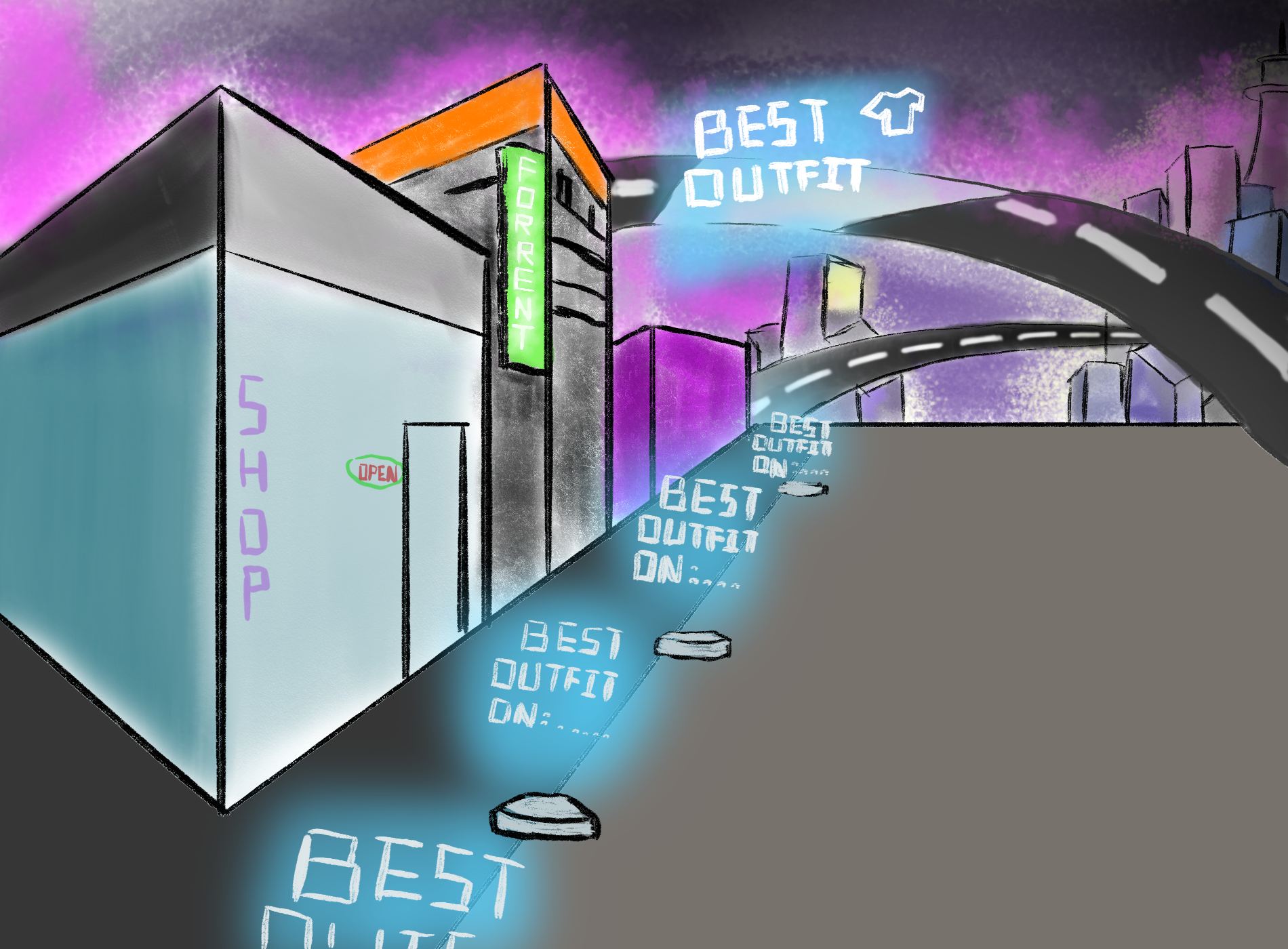}
		\caption{A scenario of a virtual world filled where advertisements are ubiquitous. Hence demonstrating how companies in the metaverse industry, especially when the market is highly concentrated, could possibly flood individuals' metaverse experiences with advertisements. The dominant player in the metaverse could easily manipulate the user understanding of `good' commerce.}
		\label{fig:monopoly_power}
	\end{figure}

	Observing the dominance of big tech companies in our real world, it is no surprise for individuals like Tim Sweeney, founder of Epic Games, to call for an `\textit{open metaverse}'\footnote{\url{https://bit.ly/3Cwaj5w}}. With the substantial cost involved in developing a metaverse, however, whether a shift in the current paradigm to a less concentrated market for metaverse will take place is questionable. Specifically, empirical findings have shown that sunk cost is positively correlated to an industry's barriers to entry~\cite{kessides1990market}. In the case of the metaverse, sunk cost may refer to companies' irretrievable costs invested in developing a metaverse system. In fact, big corporate companies like Facebook and Microsoft have already put their skins in the game\footnote{\url{https://www.washingtonpost.com/technology/2021/08/30/what-is-the-metaverse/}}$^,$\footnote{\url{https://bit.ly/3kCFOVi}}. Hence, unless the cost of developing and maintaining a metaverse world capable of holding millions of users drastically decreases in the future either due to institutional factors or simply plain-vanilla technological progress, late coming startups with a lack of financing will face significant hardship in entering the market. With market share concentrated at the hands of a few leading tech companies, the metaverse industry can become an oligopolistic market. Though it is de jure less extreme when compared to having our parallel world dominated by a gargantuan monopoly, the incumbent oligopolies can still wield great power, especially at the third stage of metaverse development (i.e., the surreality). With tech giants like Alphabet generating a revenue of 147 billion dollars from Google's advertisements alone\footnote{\url{https://cnb.cx/3kztchN}} in real-life (Figure~\ref{fig:google_ads}) shows Google's historical growth of advertising revenue), the potential scope for profit in a metaverse world at the last stage of development cannot be neglected. The concern about ``From the moment that we wake up in the morning, until we go to bed, we're on those handheld tablets''\footnote{\url{https://wapo.st/3EKDns8}} does expose not only privacy concerns but also the magnitude of the business potential of owning and overseeing such a parallel world (as demonstrated in Figure~\ref{fig:monopoly_power}). However, an oligopolistic market is not entirely malevolent. Letting alone its theoretical capability of achieving a Pareto efficient outcome, we indeed see more desirable outcomes specifically for rivalling tech giants' consumers in recent years\footnote{\url{https://econ.st/3i03Sjq}}. Such a trend is accompanied by the rise of players who once were outsiders to a particular tech area but with considerable financial strength decidedly challenge established technology firms. Therefore, despite leading tech companies like the FANG group (Facebook, Amazon, Netflix, and Alphabet) may prima facie be the most prominent players in making smooth transitions to a metaverse business, it does not guarantee they will be left uncontested by other industrial giants which root outside of tech industry. In addition, economic models on oligopolistic markets also provide theoretical bedrocks for suggesting a less detrimental effect of the market structure on consumers' welfare provided that products are highly differentiated and firms do not collude~\cite{dixit1982oligopoly}. The prior is already evident at the current stage of metaverse development. Incumbent tech players, though recognising metaverse's diversity in scope, have approached metaverse in differentiated manners. Whereas Fortnite inspired Sweeney's vision of metaverse\footnote{\url{https://bit.ly/3EJGsIS}}, Mark Zuckerberg's recent aim was to test out VR headsets for work\footnote{\url{https://cnet.co/2XG0ovg}}. It is understandable that given metaverse's uncertainties and challenges, companies choose to approach it in areas where they hold expertise first and eventually converges to similar directions. Having different starting points may still result in differentiation in how each company's metaverse manifests. In addition, the use of different hardware such as AR glasses and VR headsets by different companies can also contribute to product differentiation. The latter, however, will largely depend on economic governance, albeit benevolent intentions held by some firms\footnote{\url{https://econ.st/2ZpMwpL}}.
	
	\subsection{Metaverse commerce}
	
	\addtocounter{footnote}{-6}
	\begin{figure*}[!t]
		\centering
		\includegraphics[width=\linewidth]{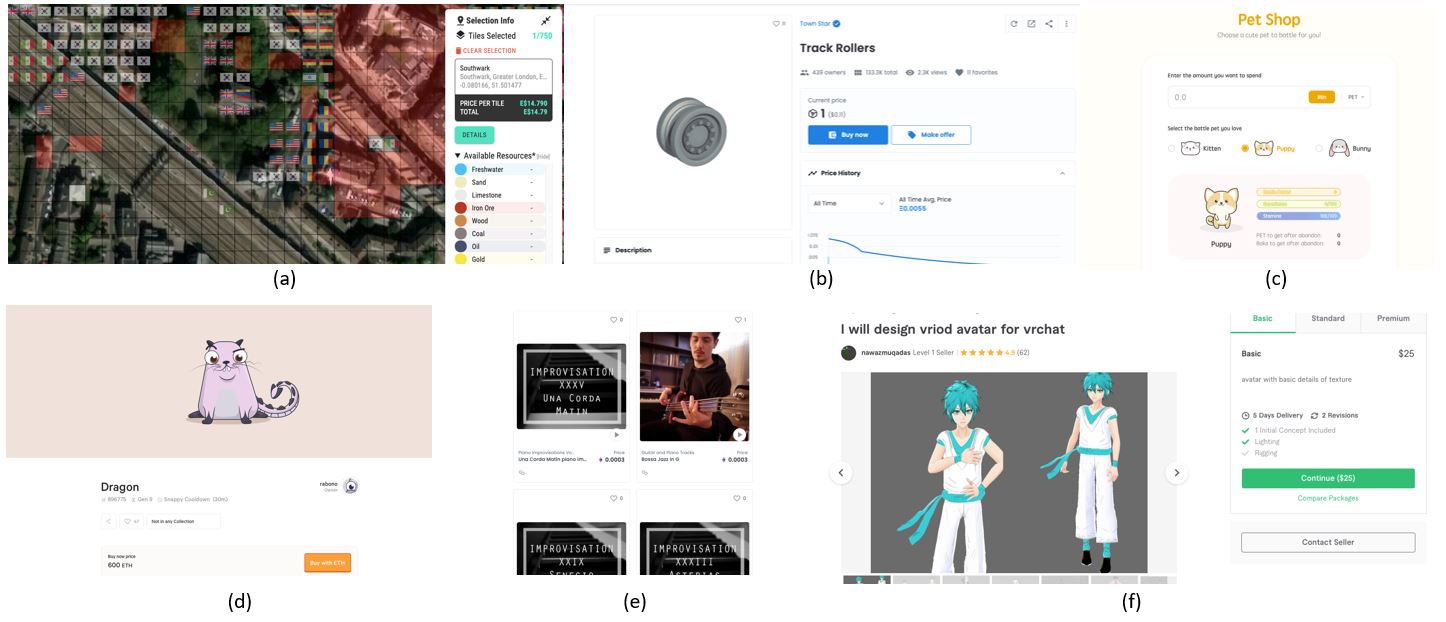}
		\caption{A collection of various virtual objects currently traded online: 
			(a) Plots of land in London offered on Earth 2, an virtual replica of our planet Earth\protect\footnotemark,
			(b) A virtual track roller listed on OpenSea\protect\footnotemark,
			(c) Virtual pet on Battle Pets\protect\footnotemark,
			(d) CryptoKitties\protect\footnotemark,
			(e) Sound tracks listed on OpenSea\protect\footnotemark,
			(f) Custom made virtual avatar on Fiverr\protect\footnotemark.}
		\label{fig:virtual_objects}
	\end{figure*}
	\addtocounter{footnote}{-5}
	\footnotetext{\url{https://earth2.io/}}
	\stepcounter{footnote}\footnotetext{\url{https://bit.ly/3i0ElXw}}
	\stepcounter{footnote}\footnotetext{\url{https://www.battlepets.finance/\#/pet-shop}}
	\stepcounter{footnote}\footnotetext{\url{https://bit.ly/39vMfDp}}
	\stepcounter{footnote}\footnotetext{\url{https://opensea.io/collection/music}}
	\stepcounter{footnote}\footnotetext{\url{https://bit.ly/3EHJPA5}}

	As an emerging concept, metaverse commerce refers to trading taking place in the virtual world, including but not limited to user-to-user and business-to-user trade. As commerce takes place digitally, the trading system can largely borrow from the established e-commerce system we enjoy now. For instance, with a net worth of 48.56 Billion USD\footnote{\url{https://www.macrotrends.net/stocks/charts/EBAY/ebay/net-worth}}, eBay is a quintessential example of C2C e-commerce for the metaverse community to transplant from. Nonetheless, metaverse commerce is not tantamount to the existing e-commerce. Not only do the items traded differs, which will be elaborated in the next section, but the main emphasis of metaverse commerce is also interoperability: users' feasibility to carry their possessions across different virtual worlds\footnote{\url{https://bit.ly/3CAbwbZ}}.
	The system of the metaverse is not about creating one virtual world, but many. Namely, users can travel around numerous virtual worlds to gain different immersive experiences as they desire. Therefore, as individuals can bring their possessions when they visit another country for vacation, developers should also recreate such experiences in the digital twin. At the current stage, most video games, even those offered by the same providers, do not proffer players with full interoperability from one game to another. Real-life, however, does offer existing games with some elements of interoperability, albeit in lesser forms. To illustrate, games like Monster Hunter and Pokémon allow players to transfer their data from Nintendo 3DS to Nintendo Switch\footnote{\url{https://bit.ly/3hSzRll}}$^,$\footnote{\url{https://bit.ly/3AzUZEp}}. 
	Nevertheless, such transfers tend to be unilateral (e.g., from older to the newer game) and lacks an immersive experience as they typically take place outside the actual gameplay. Another class of games arguably reminiscent of interoperability can be games with downloadable contents (DLC) deriving from purchases of other games from the same developer. A case in point could be Capcom's `Monster Hunter Stories 2''s bonus contents, where players of the previous Capcom's game `Monster Hunter Rise' can receive an in-game outfit that originated in `Monster Hunter Rise'\footnote{\url{https://bit.ly/3Cvjymo}}. However, having some virtual item bonus that resembles users' virtual properties in another game is not the same as complete interoperability. 
	An additional notable case for interoperability for prevailing games is demonstrated in Minecraft: gamers can keep their avatars' `skin'\footnote{\url{https://minecraft.fandom.com/wiki/Skin}} and `cape'\footnote{\url{https://minecraft.fandom.com/wiki/Cape\#Obtaining}} when logging onto different servers, which can be perceived as a real-world twin of metaverse players travelling between different virtual worlds. After inspecting all three types of existing game functions that more or less link to the notion of interoperability, one may become aware of the lack of user freedom as a recurring theme. 
	Notably, inter-game user-to-user trade is de facto missing, and the type of content, as well as the direction of flow of contents between games, are strictly set by developers. More importantly, apart from the Minecraft case, there is a lack of smoothness in data transfer as it is not integrated as part of a natural gaming experience. That is, the actions of transferring or linking game data is not as natural as real life behaviour of carrying or selling goods from one place to another. Therefore, metaverse developers should factor in the shortcomings of existing games in addressing interoperability and promote novel solutions. 
	While potentially easier for metaverse organised by a sole developer, such solutions may be more challenging to arrive at for smaller and individual developers in a scenario of `open metaverse'. As separate worlds can be built in the absence of a common framework, technical difficulties can impede users' connections between different virtual spaces, let alone the exchange of in-game contents. With that being said, organisations like the Open Metaverse Interoperability Group have sought %seek 
	to connect individual virtual spaces with a common protocol\footnote{\url{https://omigroup.org/home/}}. Hence, perhaps like the emergence of TCP/IP protocol (i.e., a universal protocol), we need common grounds of some sort to work on for individual metaverse developers.

	\subsection{Virtual Objects Trading}
	
\begin{figure}[!t]
		\centering
		\includegraphics[width=\linewidth]{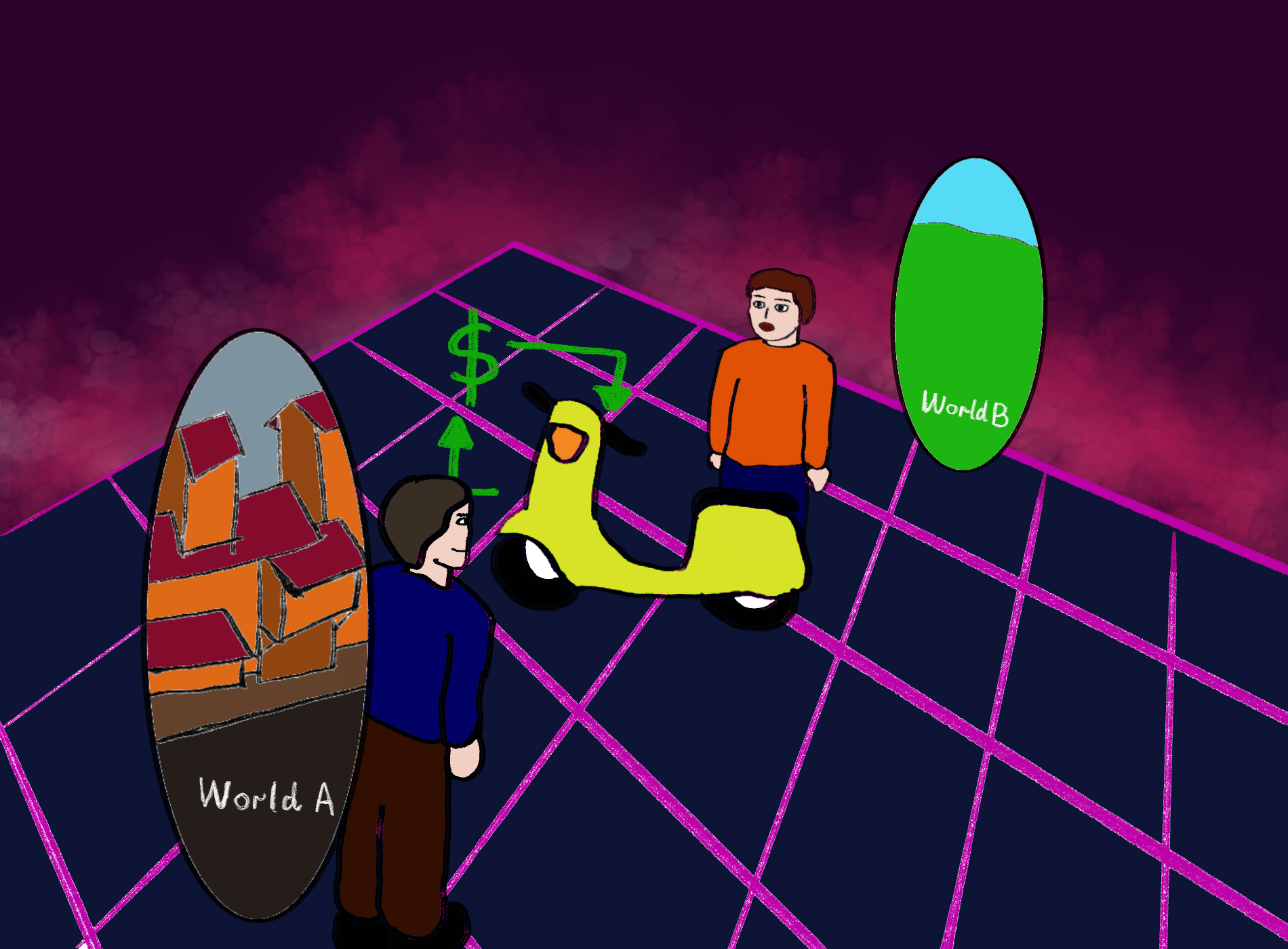}
		\caption{Our conjecture of how virtual object trading may look like. This figure shows two users from different virtual worlds entering a trading space through portals (the two ellipse-shaped objects), where they trade a virtual moped.}
		\label{fig:virtual-trading}
	\end{figure}

	As briefly hinted in the preceding section, virtual objects trading is about establishing a trading system for virtual objects between different stakeholders in the metaverse. Since human kinds first began barter trading centuries ago, trading has been an integral part of our mundane lives. Hence, the real-world's digital twins should also reflect such eminent physical counterparts. Furthermore, the need for a well-developed trading system only deepens as we move from the stage of digital twins to digital natives, where user-created virtual contents begin to boom. Fortunately, the existence of several real-life exemplars sheds light on the development of the metaverse trading system. Trading platforms for Non-Fungible Tokens (NFTs), such as OpenSea and Rarible, allow NFT holders to trade with one another at ease, similar to trading other conventional objects with financial values. As demonstrated in Figure~\ref{fig:virtual_objects}, a wide range of virtual objects are being traded at the moment. Some have gone further by embedding NFT trading into games: Battle Pets\footnote{\url{https://www.battlepets.finance/\#/}} and My DeFi Pet\footnote{\url{https://yhoo.it/3kxSNrD}} allow players to nurture, battle and trade their virtual pets with others. Given the abundance of real-life NFT trading examples, metaverse developers can impose these structures in the virtual world to create a marketplace for users to exchange their virtual contents. In addition, well-known real-life auctioning methods for goods with some degrees of common values such as Vickrey-Clarke-Groves mechanism ~\cite{sessa2017exploring} and Simultaneous Multiple Round Auction\cite{Milgrom2004PuttingAT} can also be introduced in the virtual twin for virtual properties like franchises for operating essential services in virtual communities such as providing lighting for one's virtual home. However, similar to difficulties encountered with metaverse commerce, existing trading systems also need to be fine-tuned to accommodate the virtual world better. One potential issue can be trading across different virtual worlds. Particularly, an object created in world A may not be compatible in world B, especially when different engines power the two worlds. Once again, as virtual object trading across different worlds is intertwined with interoperability, the call for a common framework becomes more salient. At the current stage, some have highlighted inspirations for constructing an integrated metaverse system can be obtained from retrospecting existing technologies such as the microverse architecture\footnote{\url{https://spectrum.ieee.org/open-metaverse?utm_campaign=post-teaser&utm_content=1kp270f8}}$^,$\footnote{\url{https://microservices.io/}}. In Figure~\ref{fig:virtual-trading}, we conjecture how tradings between two different virtual worlds may look like.

	With more virtual objects trades at the digital natives stage and more individuals embracing a lifestyle of \textit{digital nomad}, the virtual trading market space should also be competent in safeguarding ownership of virtual objects. In spite of the fact that a NFT cannot be appropriated by other users from the metaverse communities, counterfeits can always be produced. Specifically, after observing a user-generated masterpiece listed on the virtual trading platform, individuals with mischievous deeds may attempt to produce counterfeits of it and claim for its originality. NFT-related defraud is not eccentric, as reports have shown several cases where buyers were deluded into thinking they were paying for legitimate pieces from famous artists, where trading platforms lack sufficient verification\footnote{\url{https://bit.ly/3CwcZ3c}}$^,$\footnote{\url{https://www.cnn.com/style/article/banksy-nft-fake-hack/index.html}}. This can be particularly destructive to a metaverse community given the type of goods being traded. Unlike necessities traded in real life: such as staples, water and heating, where a significant proportion of values of these goods derive from their utilitarian functions to support our basic needs, virtual objects' values can depend more on their associated social status. In other words, the act of possessing some rare NFTs in the virtual world may be similar to individuals consumption of Veblen goods~\cite{veblen2017theory} like luxurious clothing and accessories. Therefore, the objects' originality and rareness become a significant factor for their pricing. Hence, a trading market flooded with feigned items will deter potential buyers. With more buyers' concerns about counterfeit items and consequently becoming more reserved towards offering a high price, genuine content creators are disincentivised. This coincides with George Akerlof's `market of lemon', leading to undesirable market distortion~\cite{akerlof1978market}. 
	
	Given the negative consequences, the question to be asked is: which stakeholder should be responsible for resolving such a conundrum? Given that consumers tend not to possess the best information and capacity to validate listed items, they should not be forced to terminate their metaverse experience to conduct an extensive search of the content creator's credibility in real life. Similarly, content creators are not most capable of protecting themselves from copyright infringement as they may be unable to price in their loss through price discrimination and price control~\cite{klein2002economics}. Therefore, metaverse developers should address the ownership issue to upkeep the market order. So far, some studies have attempted to address art forgery %frogery 
	with the use of neural networks by examining particular features of an artwork~\cite{wang2016fake, elgammal2018picasso}. Metaverse developers may combine conventional approaches by implementing a more stringent review process before a virtual object is cleared for listing as well as utilising neural networks to flag items that are highly similar to items previously listed on the platform, which may be achieved by building upon current achievements in applications of neural networks in related fields~\cite{wang2014learning, bell2015learning}.

	\section{Social Acceptability}\label{sec:social}%tristan or abhishek (pls discuss before anyone begins this section)
	
	This section discusses a variety of design factors influencing the social acceptability of the metaverse. The factors include privacy threats, user diversity, fairness, user addiction, cyberbullying, device acceptability, cross-generational design, acceptability of users' digital copies (i.e., avatars), and green computing (i.e., design for sustainability).
	
	\subsection{Privacy threats}
	Despite the novel potentials which could be enabled by the metaverse ecosystem, it will need to address the issue of potential privacy leakage in the earlier stage when the ecosystem is still taking its shape, rather than waiting for future when the problem is so entrenched in the ecosystem that any solution to address privacy concerns would require redesign from scratch. An example of this issue is the third-party cookies based advertisement ecosystem, where the initial focus was to design for providing utilities. The entire revenue model was based on cookies which keep track of users in order to provide personalised advertisements, and it was too late to consider privacy aspects. Eventually, they were enforced by privacy regulations like GDPR, and the final nail to the coffin came from Google's decision to eliminate third-party cookies from Chrome by 2022, which have virtually killed the third-party cookies based advertisement ecosystem. Also, we have some early signs of how society might react to the ubiquitous presence of technologies that would enable the metaverse from the public outcry against the Google Glass, when their concerns (or perceptions) are not taken into account. Afterwards, many solutions were presented to respect of the privacy of bystanders and non-users~\cite{aditya2016pic, shu2018cardea}. However, all of them rely on the good intentions of the device owners because there is no mechanism, either legal or technical, in place to verify whether the bystanders' privacy was actually respected. Coming up with a verifiable privacy mechanism would be one of the foremost problems to be solved in order to receive social acceptability. 
	
\begin{figure}[!t]
		\centering
		\includegraphics[width=\linewidth]{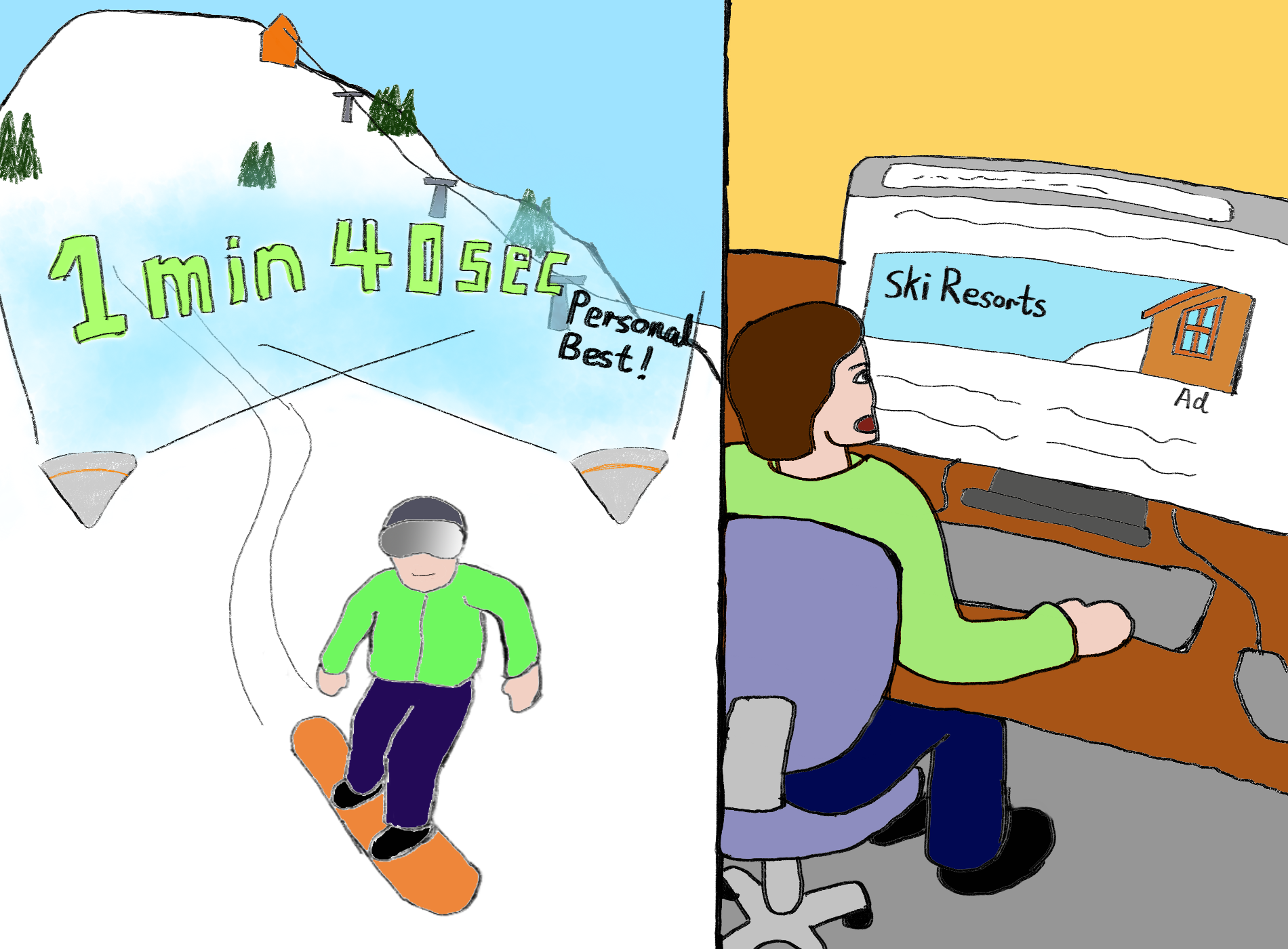}
		\caption{The figure pictorially depicts uncontrolled data flow to every activity under the metaverse. The digitised world with MR and the user data are collected in various activities (Left); Subsequently, the user data is being sold to online advertising agents without the user's prior consents (Right).}
		\label{fig:ad_data}
	\end{figure}

	Another dimension of privacy threat in the context of social acceptability comes from the privacy paradox, where users willingly share their own information, as demonstrated in Figure \ref{fig:ad_data}. For the most part, users do not pay attention to how their public data are being used by other parties, but show very strong negative reactions when the difference between the actual use of their data and the perceived use of data become explicit and too contrast. For example, many people shared their data on Facebook willingly. Still, the Facebook and Cambridge Analytica Data Scandal triggered a public outcry to the extent that Facebook was summoned by the U.S. Congress and the U.K. Parliament to hearings, and Cambridge Analytica went bankrupt soon after. One solution would be not to collect any user's data at all. However it will greatly diminish the potential innovations which the ecosystem could enable. Another solution which has also been advocated by world leaders like the German chancellor Angela Merkel, is to enable user-consented privacy trading, where users can sell their personal data in return for benefits, either monetary or otherwise. Researchers have already provided their insights on the economics of privacy~\cite{acquisti2016economics}, and the design for an efficient market for privacy trading~\cite{UCAM-CL-TR-925, 9162015}. This approach will enable the flow of data necessary for potential innovations, and at the same time, it will also compensate users fairly for their data, thereby paving the path for broader social acceptability~\cite{smart-data-sell}.

	%\cite{10.1145/3386290.3396935}
	\subsection{User Diversity}
	
	%User Acceptance  of Virtual Worlds: The Hedonic Framework ~\cite{10.1145/1314234.1314250}
	
	% People with disabilities
	
	% Children
	
	% Old
	
	% Bystander
	
	%write a shit here, please use
	As stated in a visionary design of human-city interaction~\cite{Lee2020TowardsAR}, the design of mobile AR/MR user interaction in city-wide urban should consider various stakeholders. Similarly, the metaverse should be inclusive to everyone in the community, regardless of race, gender, age and religion, such as children, elderly, disabled individuals, and so on. In the metaverse, various contents can appear and we have to ensure the contents are appropriate to vastly diversified users. In addition, it is important to consider personalised content display in front of users~\cite{A2W-LAM2021}, and promote the fairness of the recommendation systems, in order to minimise the biased contents and thus impact the user behaviours and decision making~\cite{diversified-fairness} (More details in Section~\ref{sec:fair}). The contents in virtual worlds can lead to higher acceptance by delivering factors of enjoyment, emotional involvement, and arousal~\cite{10.1145/1314234.1314250}. `\textit{How to design the contents to maximise the acceptance level under the consideration of user diversity}', i.e., design for user diversification, would be a challenging question.

	\subsection{Fairness}\label{sec:fair}
	Numerous virtual worlds will be built in the metaverse, and perhaps every virtual world has its separate rules to govern the user behaviours and their activities. As such, the efforts of managing and maintaining such virtual worlds would be enormous. We expect that autonomous agents, support by AI (Section~\ref{sec:art-int}), will engage in the role of governance in virtual worlds, to alleviate the demands of manual workload. 
	It is important to pinpoint that the autonomous agents in virtual worlds rely on machine learning algorithms to react to the dynamic yet constant changes of virtual objects and avatars. It is well-known that \textit{no model can perfectly describe the real-world instance}, and similarly, an unfair or biased model could systematically harm the user experiences in the metaverse. The biased services could put certain user groups in disadvantageous positions. 
	
	On social networks, summarising user-generated texts by algorithmic approaches can make some social groups being under-represented. In contrast, fairness-preserving summarisation algorithms can produce overall high-quality services across social groups~\cite{summarizing-ugcs-fairness}. This real-life example sheds light on the design of the metaverse.
	For this reason, the metaverse designers, considering the metaverse as a virtual society, should include the algorithmic fairness as the core value of the metaverse designs~\cite{qualitative-exploration-fairness}, and accordingly maintain the procedural justices when we employ algorithms and computer agents to take managerial and governance roles, which requires a high degree of transparency to the users and outcome control mechanisms. In particular, outcome controls refer to the users' adjustments to the algorithmic outcomes that they think is fair~\cite{Procedural-justice}. Unfavourable outcomes to individual users or groups could be devastating. This implies the importance of user perceptions to the fairness of such machine learning algorithms, i.e., perceived fairness. However, leaning to perceived fairness could fall into another trap of \textit{outcome favorability} bias~\cite{factors-algorithm-outcomes}. 
	Additionally, metaverse designers should open channels to collect the voices of diversified community groups and collaboratively design solutions that lead to fairness in the metaverse environments~\cite{qualitative-exploration-fairness}. 
	
	\subsection{User Addiction}\label{ssec:addiction}
	Excessive use with digital environments (i.e., user addictions) would be an important issue when the metaverse becomes the most prevalent venue for people to spend their time in the virtual worlds. In the worst scenario, users may leverage the metaverse to help them `escaping' from the real world, i.e., escapism~\cite{10.1145/1314234.1314250}. Prior works have found shreds of evidence of addictions to various virtual cyberspaces or digital platforms such as social networks~\cite{OSN-addictions}, mobile applications~\cite{mobileapplication-addiction}, smartphones~\cite{smartphone-addiction}, VR~\cite{VR-addiction}, AR~\cite{AR-addiction}, and so on.
	User addictions to cyberspaces could lead to psychological issues and mental disorders, such as depression, loneliness, as well as user aggression~\cite{game-addiction-psy}, albeit restrictions on screentime had been widely employed~\cite{screen-time}. 
	Knowing that the COVID-19 pandemic has prompted a paradigm shift from face-to-face meetings or social gatherings to various virtual ways, most recent work has indicated that the prolonged usage of such virtual meetings and gatherings could create another problem -- abusive use or addiction to the Internet~\cite{scale-addiction}. 
	
	Therefore, we have questioned whether `\textit{the metaverse will bring its users to the next level of user addiction}'. We discuss the potential behaviour changes through reviewing the existing AR/VR platforms, based \textit{not-at-all} on evidence.
	First, \textit{VR Chat}, known as a remarkable example of metaverse virtual worlds, can be considered as a pilot example of addiction to the metaverse\footnote{\url{https://www.worldsbest.rehab/vrchat-addiction/}}. Meanwhile, VR researchers studied the relationship among such behavioural addiction in VR, root causes, and corresponding treatments~\cite{Segawa2019VirtualR}. 
	Also, AR games, e.g., Pokemon Go, could lead to the behaviour changes of massive players, such as spending behaviours, group-oriented actions in urban areas, dangerous or risky actions in the real world, and such behaviour changes could lead to discernible impacts on the society~\cite{AR-pokemon, chasing-pokemon}. 
	A psychological view attempts to support the occurrence of user addiction, which explains the extended self of a user, including person's mind, body, physical possessions, family, friends, and affiliation groups, encourages user to explore the virtual environments and pursue rewards, perhaps in an endless reward-feedback loop, in virtual worlds~\cite{BELK201650}. We have to pinpoint that we raise the issues of addictions of immersive environments (AR/VR) here, aiming at provoking debates and drawing research attentions. 
	In the metaverse, the users could experience~\textit{super-realism} that allows users to experience various activities that highly resemble the real world. 
	Also, the highly realistic virtual environments enable people to try something impossible in their real life (e.g., replicating an event that are immoral in our real life~\cite{VR-slavery} or experiencing racist experience~\cite{anchorhold-afference} ), with a bold assumption that such environments can further exacerbate the addictions, e.g., longer usage time. Further studies and observation of in-the-wild user behaviours could help us to understand the new factors of user addiction caused by the super-realistic metaverse. 
	
	\subsection{Cyberbullying}\label{ssec:cyberbully}
	Cyberbullying refers to the misbehaviours such as sending, posting, or sharing negative, harmful, false, or malevolent content about victims in cyberspaces, which frequently occurs on social networks~\cite{cyber-aggression}. 
	We also view the metaverse as gigantic cyberspace. As such, another unignorable social threat to the ecosystem could be cyberbullying in the metaverse. 
	The metaverse would not be able to run in long terms, and authorities will request to shut down some virtual worlds in the metaverse, according to the usual practice -- shutdown the existing cyberbullying cyberspace\footnote{\url{https://www.change.org/p/shut-down-cyberbullying\break -website-ask-fm-in-memory-of-izzy-dix-12-other-teens-globally}}. 
	Moreover, considering the huge numbers of virtual worlds, the metaverse would utilise cyberbullying detection approaches are driven by algorithms~\cite{stochastic-cyberbullying}. The fairness of such algorithms~\cite{fairness-cyberbullying} will become the crucial factors to deliver perceived fairness to the users in the metaverse. After identifying any cyberbullying cases, mitigation solutions, such as care and support, virtual social supports, and self-disclosures, should be deployed effectively in virtual environments~\cite{participatory-design-cyberbullying, cyber-bullying-beyond-self-disclose}. 
	However, recognising cyberbullying in the game-alike environment is far more complicated than social networks. For instance, the users' misbehaviour can be vague and difficult to identify~\cite{cyber-toxic}. Similarly, 3D virtual worlds inside the metaverse could further complicate the scenarios and hence make difficult detection of cyberbullying at scale. 
	
	\subsection{Other Social Factors}\label{ssec:other-social}
	
	First, social acceptability to the devices connecting people with the metaverse needs further investigation, which refers to the acceptability of the public or bystanders' to such devices, e.g., mobile AR/VR headsets~\cite{mobile-CHI-social-accept}. Additionally, the user safety of mobile headsets could negatively impact the users and their adjacent bystanders, causing breakdowns of user experience in virtual worlds~\cite{dao-breakdown-chi21}. To the best of our knowledge, we only found limited studies on social acceptability to virtual worlds~\cite{user-accept-v-worlds}, but not digital twins as well as the metaverse. 
	
	Moreover, the gaps in cross-generation social networks also indicate that Gen Z adults prefer Instagram, Snapchat and Tiktok over Facebook. Rather, Facebook retains more users from Gen X and Y~\cite{7427661}. Until now, social networks have failed to serve all users from multiple demographics in one platform. From the failed case, we have to prepare for the user design of cross-generational virtual worlds, especially when we consider the metaverse with the dynamic user cohorts in a unified landscape. 
	
	Besides, we should consider the user acceptability of the avatars, the digital copy of the users, at various time points. For instance, once a user passes away, what is the acceptability of the user's family members, relatives, or friends to the avatars? This question is highly relevant to the virtual immorality that describes storing a person's personality and their behaviours as a digital copy~\cite{Why-Not-Immortality}. The question could also shape the future of \textit{Digital Humanity}~\cite{10.1145/3383583.3398551} in the metaverse, as we are going to iterate the virtual environments, composed of both virtual objects and avatars, as separate entities from the real world, e.g., should we allow the new users talking with a two-centuries-long avatar representing a user probably passed away? 
	
	Furthermore, the metaverse, regarded as a gigantic digital world, will be supported by countless computational devices. As such, the metaverse can generate huge energy consumption and pollution. Given that the metaverse should not deprive future generations, the metaverse designers should not neglect the design considerations from the perspective of green computing. Eco-friendliness and environmental responsibility could impact the user affection and their attitudes towards the metaverse, and perhaps the number of active users and even the opposers~\cite{green-computing-TAM}. Therefore, sourcing and building the metaverse with data analytics on the basis of sustainability indices would become necessary for the wide adoption of the metaverse~\cite{sustainability,sustain-consideration}.
	
	Finally, we briefly mention other factors that could impact the user acceptability to the metaverse, such as in-game injuries, unexpected horrors, user isolation, accountability and trust (More details in Section~\ref{sec:trust}), identity theft/leakage, virtual offence, manipulative contents inducing user behaviours (e.g., persuasive advertising), to name but a few~\cite{Slater2020TheEO, XR-advertising}.
	
	\begin{figure}[!t]
		\centering
		\includegraphics[width=\linewidth]{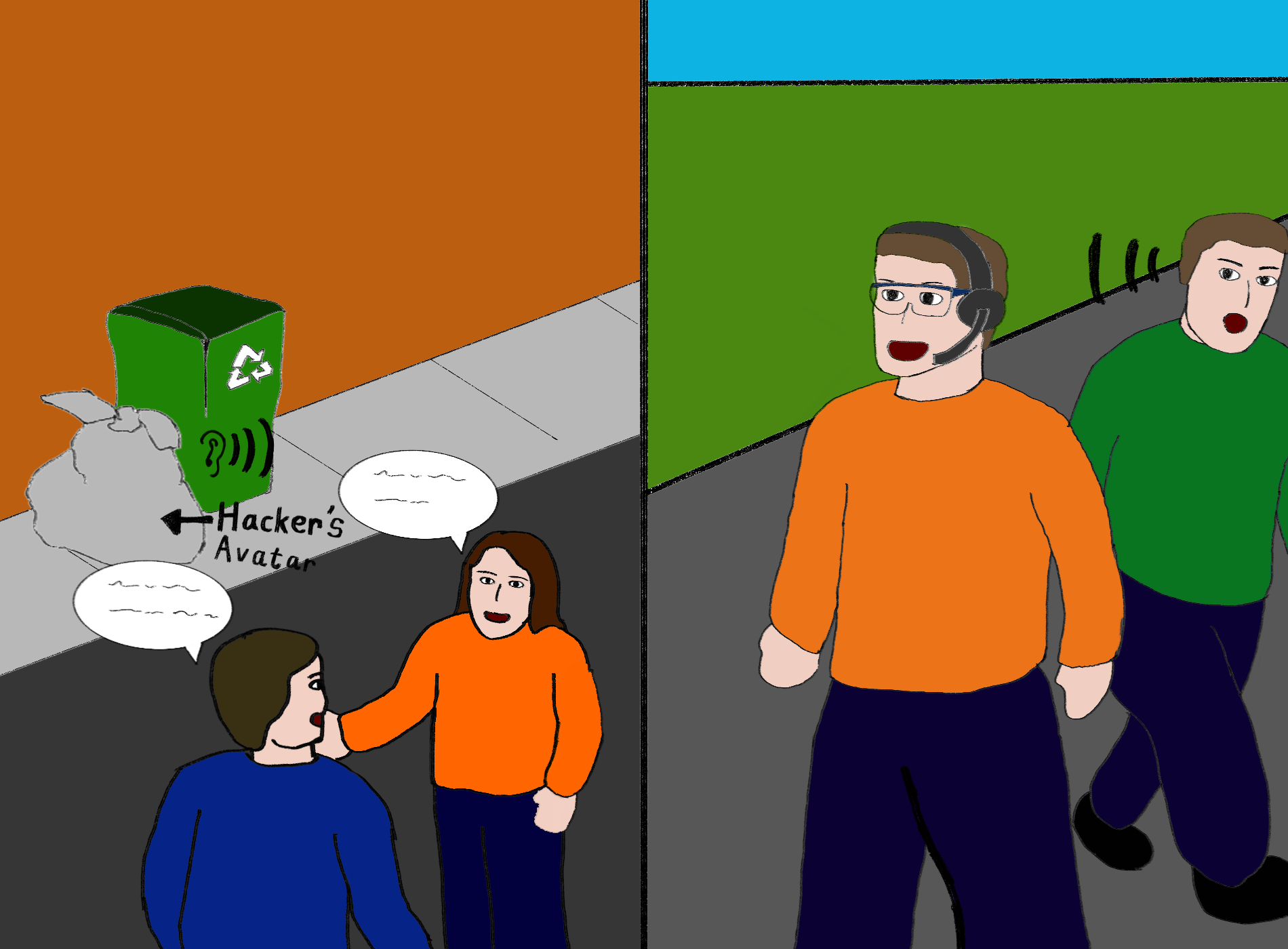}
		\caption{when the metaverse is enabled by numerous technologies and sensors, the highly digitalized worlds, regardless of completely virtual (left, a malicious avatar as a camouflage of garbage next to a garbage bin) or merged with the physical world (right, an adjacent attack observes a user's interaction with immersive environments, similar to shoulder surfing attack.), will be easily monitored (or eavesdrop) by attackers/hackers.}
		\label{fig:eavesdropping}
	\end{figure}
	
	\section{Privacy and Security}\label{sec:privacy-security}%carlos
	
	Internet-connected devices such as wearables allow monitoring and collect users' information. This information can be interpreted in multiple ways. In most situations, such as in smart homes, we are not even aware of such ubiquitous and continuous recordings, and hence, our privacy can be at risk in ways we cannot foresee. These devices can collect several types of data: personal information (e.g., physical, cultural, economic), users' behaviour (e.g., habits, choices), and communications (e.g., metadata related to personal communications). In many situations, users accept the benefits in comparison with the possible privacy and security risks of using these smart devices or services~\cite{leenes2007privacy}. For example, GPS positioning is used to search for nearby friends~\cite{al2009locating}. In the case of VR - the primary device used to display the metaverse - the new approaches to enable more immersive environments (e.g., haptic devices, wearables to track fine-grained users' movements) can threaten the users in new ways.

	The metaverse can be seen as a digital copy of what we see in our reality, for example, buildings, streets, individuals. However, the metaverse can also build things that do not exist in reality, such as macro concerts with millions of spectators (Figure~\ref{fig:eavesdropping}). The metaverse can be perceived as a social microcosmos where players (individuals using the metaverse) can exhibit realistic social behaviour. In this ecosystem, the privacy and security perceptions of individuals can follow the real behaviours. In this section, we will elaborate on the privacy and security risks that individuals can face when using the metaverse. We start with an in-depth analysis of the users’ behaviour in the metaverse and the risks they can experience, such as invasion of privacy or continuous monitoring, and privacy attacks that individuals can suffer in the metaverse such as deep-fakes and alternate representations. Second, we evaluate how designers and developers can develop ethical approaches in the metaverse and protect digital twins. Finally, we focus on the biometric information that devices such as VR headsets and wearables can collect about individuals when using the metaverse.

	\subsection{Privacy behaviours in the metaverse}
	
	In the metaverse, individuals can create avatars using similar personal information, such as gender, age, name, or entirely fictional characters that do not resemble the physical appearance or include any related information with the real person. For example, in the game named \textit{Second Life} -- an open-world social metaverse - the players can create their avatars with full control over the information they want to show to other players. 
	
	However, due to the nature of the game, any player can monitor the users’ activities when they are in the metaverse (e.g., which places they go, whom they talk to). Due to the current limitations of VR and its technologies, users cannot be fully aware of their surroundings in the metaverse and who is following them. The study by~\cite{leenes2007privacy} shows that players do behave similarly in the metaverse, such as \textit{Second Life}, and therefore, their privacy and security behaviours are similar to the real one. As mentioned above, players can still suffer from extortion, continuous monitoring, or eavesdropping when their avatars interact with other ones in the metaverse. 
	
	\textit{A solution} to such privacy and security threats can be the use of multiple avatars and privacy copies in the metaverse~\cite{leenes2007privacy}. The first technique focuses on creating different avatars with different behaviour and freedom according to users’ preferences. These avatars can be placed in the metaverse to confuse attackers as they will not know which avatar is the actual user. The avatars can have different configurable (by the user) behaviours. For example, when buying an item in the metaverse, the user can generate another avatar that buys a particular set of items, creating confusion and noise to the attacker who will not know what the actual avatar is. The second approach creates temporary and private copies of a portion of the metaverse (e.g., a park). In this created and private portion, attackers can not eavesdrop on the users. The created copy from the main fabric of the metaverse will or not create new items (for example, store items). Then in the case the private portion use resources from the main fabric, the metaverse API should address the merge accordingly from the private copy to the main fabric of the metaverse. For example, if the user creates a private copy of a department store, the bought items should be updated in the store of the main fabric when the merge is complete. This will inherently create several challenges when multiple private copies of the same portion of the metaverse are being used simultaneously. Techniques that address the parallel use of items in the metaverse should be implemented to avoid inconsistencies and degradation of the user experience (e.g., the disappearance of items in the main fabric because they are being used in a private copy). Finally, following the creation of privacy copies, the users can also be allowed to create invisible copies of their avatar so they can interact in the metaverse without being monitored. However, this approach will suffer from similar challenges as the private copies when the resources of the main fabric are limited or shared.
	
\begin{figure}[!t]
		\centering
		\includegraphics[width=\linewidth]{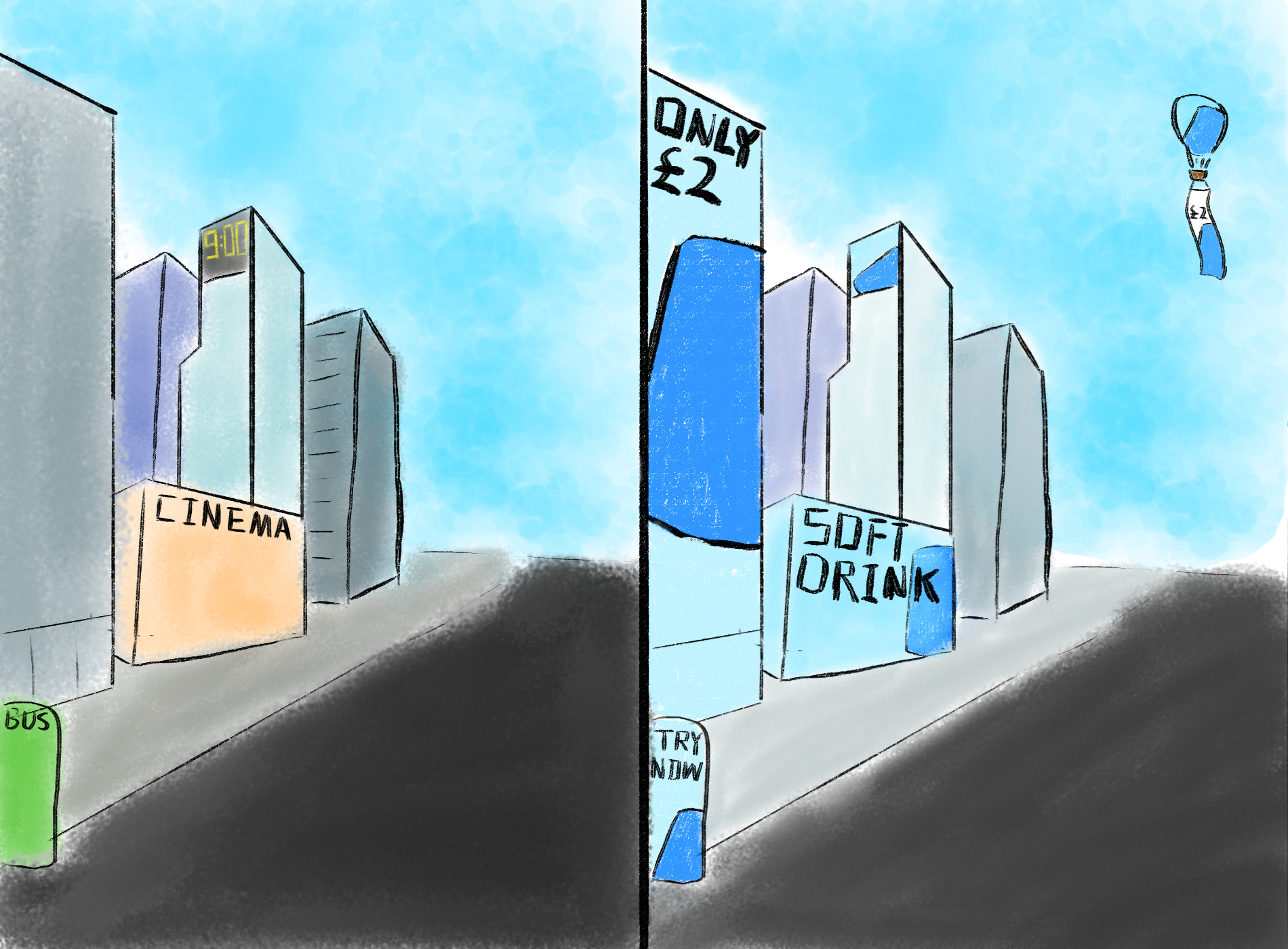}
		\caption{One possible undesirable outcome in the metaverse -- occupied by advertising content. The physical world (left) is occupied with full of advertising content in immersive views (right). This may apply to the users without a premium plan, i.e., free users. Users have to pay to remove such unwanted content. More importantly, if the digital content appears in the real world with the quality of \textit{super-realism}, the human users may not be able to distinguish the tangible content in the real world. User perceptions in the real world can be largely distorted by the dominant players in the metaverse.}
		\label{fig:AR_ads}
	\end{figure}

	In these virtual scenarios, the use of deep-fakes and alternate representations can have a direct %direr 
	effect on users' behaviours, e.g., Figure~\ref{fig:AR_ads}. In the metaverse, the generated virtual worlds can open potential threats to privacy more substantial than in the real world. For example, `deep-fakes' can have more influence in users' privacy behaviours. Users can have trouble differentiating authentic virtual subjects/objects from deep-fakes or alternate representations aiming to `trick' users. The attackers can use these techniques to create a sense of urgency, fear, or other emotions that lead the users to reveal personal information. For example, the attacker can create an avatar that looks like a friend of the victim to extract some personal information from the latter. In other cases, the victim's security can be at stake, such as physically (in the virtual world) assaulting the victim. Finally, other more advanced techniques can use techniques, such as dark patterns, to influence users into unwanted or unaware decisions by using prior logged observations in the metaverse. For example, the attacker can know what the users like to buy in the metaverse, and he/she will design a similar virtual product that the user will buy without noticing it is not the original product the user wanted. Moreover, machine learning techniques can enable a new way of chatbots and gamebots in the metaverse. These bots will use the prior inferred users' traits (e.g., personality) to create nudged~\cite{acquisti2015privacy} social interactions in the metaverse.

	\subsection{Ethical designs}
	
	As we mentioned above, the alternate representations and deep-fakes that attackers can deliver in the metaverse should be avoided. First, we discuss how the metaverse can be regulated and even the governance possibilities in the metaverse.
	
	For example, \textit{Second Life} is operated in the US, and therefore, it follows US regulations in terms of privacy and security. However, the metaverse can achieve worldwide proportions creating several challenges to protect the users in such a broad spectrum. 
	The current example of \textit{Second Life} shows an in-world (inside the metaverse) regulation and laws. In this environment, regulations are enforced using code and continuous monitoring of players (e.g., chat logs, conversations). The latter can help the metaverse developers to ban users after being reported by others. However, as we can observe, this resembles some governance. This short of governance can interfere with the experience in the metaverse, but without any global control, the metaverse can become anarchy and chaos. This governance will be in charge of decisions such as restrictions of a particular player that has been banned. 
	
	In the end, we can still face the worldwide challenges of regulations and governance in the metaverse to have some jurisdiction over the virtual world. We can foresee that the following metaverse will follow previous approaches in terms of regulations (according to the country the metaverse operates) and a central government ruled by metaverse developers (using code and logs). 
	
	Some authors~\cite{leenes2007privacy} have proposed the gradual implementation of tools to allow groups to control their members similarly as a federated model. Users in the metaverse can create neighbours with specific rules. For example, users can create specific areas where only other users with the same affinities can enter. 
	Technologies such as blockchain can also allow forcing users of the metaverse to not misbehave according to some guidelines, with the corresponding punishment (maybe by democratic approaches). However, the regulations of privacy and security and how to enforce them are out of this section's scope.

	\subsubsection{Digital twins protection}
	
	Digital twins are virtual objects created to reflect physical objects. These digital objects do not only resemble the physical appearance but can also the physical performance or behaviour of real-world assets. Digital twins will enable clones of real-world objects and systems. Digital twins can be the foundation of the metaverse, where digital objects will behave similarly to physical ones. The interactions in the metaverse can be used to improve the physical systems converging in a massive innovation path and enhanced user experience.

	In order to protect digital twins, the metaverse has to ensure that the digital twins created and implemented are original~\cite{rasheed2020digital}. In this regard, the metaverse requires a trust-based information system to protect the digital twins. Blockchain is a distributed single chain, where the information is stored inside cryptographic blocks~\cite{nofer2017blockchain}. The validity of each new block (e.g., creation of a new digital twin) is verified by a peer-to-peer network before adding the new record to the chain. Several works~\cite{reyna2018blockchain,sghaier2020capability,chen2019blockchain} propose the use of blockchain systems to protect the digital twins in the metaverse. In~\cite{chen2019blockchain}, the authors propose a blockchain-based system to store health data electronically (e.g., biometric data) health records that digital twins can use. As we have seen with recent applications, they can enable new forms of markets in the digital ecosystems such as non-fungible token (NFT)~\cite{sghaier2020capability}. The latter allows digital twins creators to sell their digital twins as unique assets by using the blockchain.

	\subsubsection{Biometric data}
	
	The metaverse uses data from the physical world (e.g., users' hand movements) to achieve an immersive user~\cite{duan2021metaverse}. For example, different sensors attached to users (e.g., gyroscope to track users' head movements) can control their avatar more realistically. Besides VR head-mounted displays, wearables, such as gloves and special suits, can enable new interaction approaches to provide more realistic and immersive user experiences in the metaverse. These devices can allow users to control their avatar using gestures (e.g., glove-based hand tracking devices) and render haptic feedback to display more natural interactions. The goal of capturing such biometric information is to integrate this mixed modality (input and output) to build a holistic user experience in the metaverse, including avatars' interactions with digital assets such as other avatars~\cite{duan2021metaverse}.
	
	However, all these biometric data can render more immersive experiences whilst opening new privacy threats to users. Moreover, as previously commented, digital twins used real-world data such as users' biometric data (e.g., health monitoring and sport activities) to simulate more realistic digital assets in the metaverse. Therefore, there exists a need to protect such information against attacks while still accessible for digital twins and other devices (e.g., wearables that track users' movements).

	\section{Trust and Accountability}\label{sec:trust}
	As the advancements in the Internet, Web technologies, and XR converge to make the idea of the metaverse technically feasible. And the eventual success would depend on how likely are users willing to adopt it, which further depends on the perceived trust and the accountability in the event of unintended consequences.
	
	\subsection{Trust and Information}
	Socrates did not want his words to go fatherless into the world, transcribed onto tablets or into books that could circulate without their author, to travel beyond the reach of discussion and questions, revision, and authentication. So, he talked and augured with others on the streets of Athens, but he wrote and published nothing. The problems to which Socrates pointed are acute in the age of recirculated ``news'', public relations, global gossip, and internet connection. How can rumours be distinguished from the report, fact from fiction, reliable source from disinformation, and trust-teller from deceiver? These problems have already been proven to be the limiting factor for the ubiquitous adoption of social networks and smart technologies, as evident from the migration of users in many parts of the world from supposedly less trustworthy platforms (i.e., WhatsApp) to supposedly higher trustworthy platforms (i.e., Signal)~\cite{WhatsApp}. For the same reason, in order for the convergence of XR, social networks, and the Internet to be truly evolved to the metaverse, one of the foremost challenges would to establish a verifiable trust mechanism. A metaverse universe also has the %has also 
	potential to solve many social problems, such as %such us 
	loneliness. For example, Because of the COVID-19 pandemic, or the lifestyle of the elderly in urban areas, the elderly people were forced to cancel the activities for their physical conditions or long distances. However, elderly people are almost most venerable to online scams/frauds, which makes coming up with solutions for the trust mechanism quite imperative.

	As in the metaverse universe, users are likely to devote %increasingly 
	more time to their journeys in immersive environments, and they would leave themselves vulnerable by exposing their actions to other (unknown) parties. %venerable to the action of other parties. 
	This can present another limiting factor. Some attempts have been to address this concern by exploiting the concept of ``presence'', i.e., giving users ``place illusion'' defined as the sensation of being there, and ``plausibility presence'' defined as the sensation that the events occurring in the immersive environment are actually occurring~\cite{10.1145/2970930.2970947}. However, it remains to be seen how effective this approach is on a large scale.
	
	%Novel interactions mechanisms in XR have proposed many hand gestures 
	Another direction towards building trust could be from the perspective of situational awareness. Research on trust in automation suggests that providing insight into how automation functions via situational awareness display improve trust~\cite{chang2019using}. XR can utilise the same approach of proving such information to the user's view in an unobtrusive manner in the immersive state.
	
	Dependability is also considered as an important aspect of trust. Users should be able to depend on XR technologies to handle their data in a way they expect to. Recent advances in trusted computing have paved a path for hardware/crypto-based trusted execution environments (TEEs) in mobile devices. These TEEs provide for secure and isolated
	code execution and data processing (cryptographically sealed memory/storage), as well as remote attestation (configuration assertions). The critical operations on user's data can be done through TEEs. However, the technology is yet to be fully developed to be deployed in XR devices while ensuring real-time experience.

	On the flip side, there is also a growing concern of over-trust. Users tend to trust products from big brands far too easily, and rightly so, since human users have often relied on using reputation as predominant metric to decide whether to trust a product/service from the given brand. However, in the current data-driven economy where user's information are a commodity, even big brands have been reported to engage in practices aimed to learn about the user as much as possible~\cite{10.1145/3432205}, i.e., Google giving access of users' emails to the third parties~\cite{u207s}. This concern is severe in XR, because XR embodies human-like interactions, and their misuses by the third parties can cause significant physiological trauma to users. The IEEE Global Initiative on Ethics of Autonomous and Intelligent Systems recommends that upon entering any virtual realm, users should be provided a ``hotkey'' tutorial on how to rapidly exit the virtual experience, and information about the nature of algorithmic tracking and mediation within any environment~\cite{ieee11}. Technology designers, standardisation bodies and regulatory bodies will also need to consider addressing these issues under consideration for a holistic solution.

	\subsection{Informed Consent}
	In the metaverse system, a great amount of potentially sensitive information is likely to leave the owner's sphere of control. As in the physical world during face-to-face communication, we place trust since we can check the information and undertakings others offer, similarly we will need to develop the informed consent mechanism which will allow avatars, i.e., the virtual embodiment of users, to place their trust. Such a consent mechanism should be allowed consent to be given or refused in the light of information that should be checkable. However, the challenges arise from the fact that avatars may not capture the dynamics of a user's facial expression in real-time, which are important clues to place trust in face-to-face communications.
	
	Another challenge that the metaverse universe will need to address is how to handle sensitive information of the minors since minors constitute a wide portion of increasingly sophisticated and tech-savvy XR users. They are traditionally less aware of the risks involved in the processing of their data. From a practical standpoint, it is often difficult to ascertain whether a user is a child and, for instance, valid parental consent has been given. Service providers in the metaverse should accordingly review the steps they are taking to protect children's data on a regular basis and consider whether they can implement more effective verification mechanisms, other than relying upon simple consent mechanisms. Developing a consent mechanism for metaverse can use general recommendations issued by the legal bodies, such as \textit{Age Appropriate Design Code} published by the UK Information Commissioner's Office.
	
	Designing a consent mechanism for users from venerable populations will also require additional consideration. Vulnerable populations are those whose members are not only more likely to be susceptible to privacy violations, but whose safety and well-being are disproportionately affected by such violations, are likely to suffer discrimination because of their physical/mental disorder, race, gender or sex, and class. Consent mechanisms should not force those users to provide sensitive information which upon disclosing may further harm users~\cite{10.1145/3415226}.
	
	Despite an informed consent mechanism already in place, it may not always lead to informed choice form presenting to users. %the user. 
	Consent forms contain technical and legal jargon and are often spread across many pages, which users rarely read. Oftentimes, users go ahead to the website contents with the default permission settings. An alternative way %potential way to go head 
	would be to rely on the data-driven consent mechanism that learns user's privacy preferences, change permission settings for data collection accordingly, and also considers that user's privacy preferences may change over time~\cite{10.1145/3172944.3172982,kumar2021theophany}.
	
	\subsection{Accountability}
	\begin{figure}[!t]
		\centering
		\includegraphics[width=\linewidth]{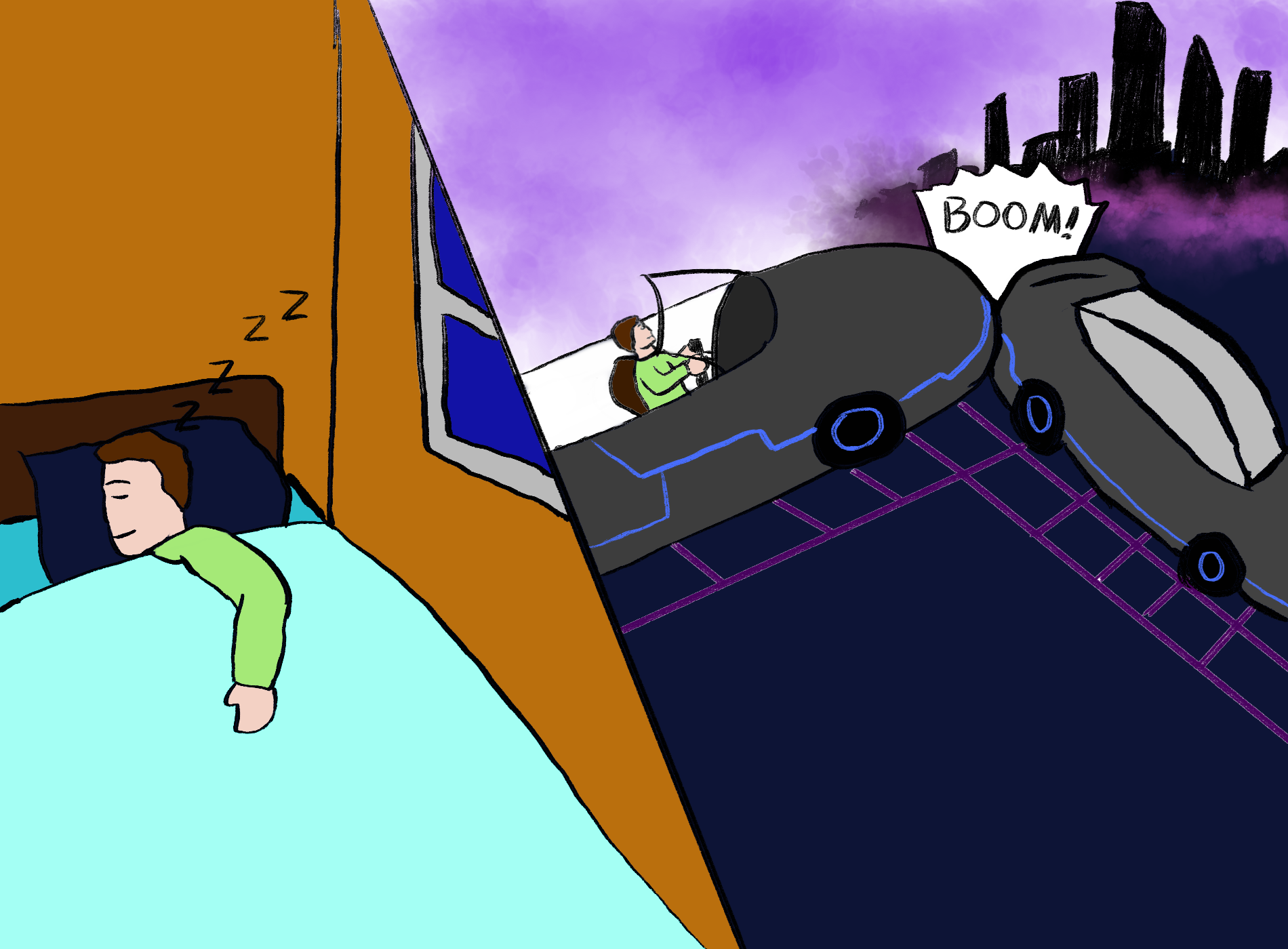}
		\caption{What is the user responsibilities with their digital copies as the avatars? For instance, the avatar's autonomous actions damage some properties in the metaverse. }
		\label{fig:uncontrol_avatar}
	\end{figure}
	Accountability is likely to be one of the major keys to realising the full potential of the metaverse ecosystem. Despite the technological advances making ubiquitous/pervasive computing a reality, many of the potential benefits will not be realised unless people are comfortable with and embrace the technologies, e.g., Figure~\ref{fig:uncontrol_avatar}. Accountability is crucial for trust, as it relates to the responsibilities, incentives, and means for recourse regarding those building, deploying, managing, and using XR systems and services.

	%\subsubsection{Online Harm Prevention}
	
	%\subsubsection{Content Moderation Policies}
	
	Content moderation policies that detail how platforms and services will treat user-generated content are often used in traditional social media to hold users accountable for the content which they generate. As outlined in Section~\ref{sec:avatar}, % (Avatar sec), 
	in the metaverse universe, users are likely to interact with each other through their avatars, which already obfuscates the user's identity to a certain extent. Moreover, recent advances in multi-modal machine learning can be used for machine-generated 3D avatars~\cite{nagano2018deep}. The metaverse content moderation will foremost need to distinguish where the given avatar embodies a human user, or is simply an auto-troll, since the human users are entitled to the freedom of expression, barring the cases of violent/extremist content, hate speech, or other unlawful content. In recent years, the content moderation of a popular Q \& A website, Quora, has received significant push-backs from users primarily based in the United States, since the U.S. based users are used to the freedom of expression in an absolute sense, and expect the same in the online world. One possible solution could be to utilise the constitutional rights extended to users in a given location to design the content moderation for that location. However, in the online world, users often cross over the physical boundary, thus making constitutional rights as the yardstick to design content moderation also challenging.

	Another aspect of accountability in the metaverse universe comes from the fact how users' data are handled since XR devices inherently collect more sensitive information like the user's locations and their surroundings than the traditional smart devices. Privacy protection regulations like GDPR rely on the user's consent, and 'Right to be forgotten', to address this problem. But, oftentime, users are not entirely aware of potential risks and invoke their ``Right to be forgotten'' mostly after some unintended consequences have already occurred. To tackle this issue, the metaverse universe should promote the principle of data minimisation, where only the minimum amount of user's data necessary for the basic function are collected, and the principle of zero-knowledge, where the systems retain the user's data only as long as it is needed~\cite{9156127}. Another direction worth exploring is utilising blockchain technology to operationalise the pipeline for data handling which always follows the fixed set of policies that have been already consented to. The users can always keep track of their data, i.e., keep track of decision provenance~\cite{9319656}.
	
	In traditional IT systems, auditing has often be used as a way to ensure the data controllers are accountable to their stakeholders~\cite{10.1007/978-3-319-68557-1_42}. Auditors are often certified third parties which do not have a conflict of interest with the data controllers. In theory, auditing can be used in the metaverse as well. However, it faces the challenge regarding how to audit secondary data which were created from the user's data, but it is difficult to establish the relationship between a given secondary data and the exact primary data, thus making it challenging for the auditor to verify whether the wishes of the users which withdrew their consent, later on, were indeed respected. The current data protection regulation like GDPR explicitly focuses on personally identifiable data and does not provide explicit clarity on the secondary data. This issue also relates the data ownership in the metaverse, which is still under debate.

	Apart from the data collection, stakes are even higher in the metaverse, since unintended consequences could cause not only psychological damage, but also physical harm~\cite{cloete2020call}. For example, the projection of digital overlays by the user's XR mobile headsets may contain critical information, such as manholes or the view ahead, which may cause life-threatening accidents. Regulatory bodies are still debating how to set up liabilities for incidents that are triggered by machines taking away user's full attention. In 2018, a self-Driving Uber Car which had a human driver killed a pedestrian in Arizona~\cite{uber11}. The accident could have been avoided if the human operator's full attention was on the driving. However, mandating full human attention all the time also diminishes the role of these assistive technologies. Regulatory bodies will need to consider broader contexts in the metaverse to decide whether the liability in such scenarios belong to the user, the device manufacturer, or any other third parties.
	
	%here?

	\section{research agenda and grand challenges}\label{sec:grand}
	
	\begin{figure*}[t]
		\begin{center}
			\includegraphics[width=\textwidth]{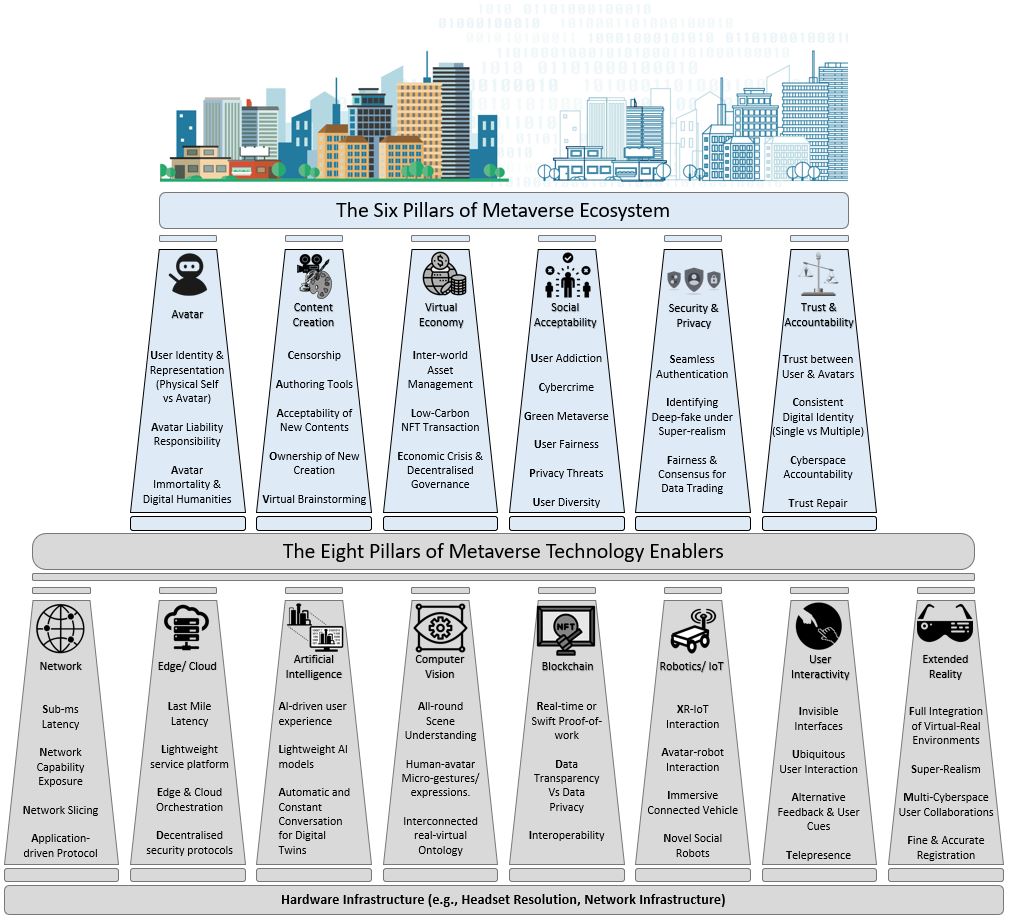}
		\end{center}
		\caption{The figure depicts a future roadmap for metaverse three-stage developments towards the surreality, i.e., the concept of \textbf{duality} and the final stage of \textit{co-existence of physical-virtual reality}. The technology enablers and ecosystem drivers help achieve self-sustaining, autonomous, perpetual, unified and endless cyberspace. }
		\label{fig:roadroad}
	\end{figure*}
	
	%These issues require human beings to constantly think and formulate corresponding guidelines.
	
	%\subsection{Grand Challenges}\label{ssec:grand-cha}
	
	We have come a long way since the days of read-only online content on desktop computers. The boundary between virtual and physical environments has become more blurred than ever before. As such, we are currently in the midst of the most significant wave of digital transformation, where the advent of emerging technology could flawlessly bind the physical and digital twins together and eventually reach the Internet featured with immersive and virtual environments.
	
	As mentioned in Section~\ref{sec:introduction}, the migration towards such an integration of physical and virtual consists of three stages: digital twins, digital natives, and the metaverse. As such, our immersive future with the metaverse requires both efforts to technology development and ecosystem establishment. 
	The metaverse should own \textbf{perpetual, shared, concurrent, and 3D virtual spaces that are concatenated into a perceived virtual universe}. We expect the endless and permanent virtual-physical merged cyberspace to accommodate an unlimited number of users, not only on earth, but eventual immigrants living on other planets (e.g., the moon and the mars), would inter-planetary travel and communication develop~\cite{alhilal2021roadmap}. Technology enablers and their technical requirements are therefore unprecedentedly demanding.
	The metaverse also emphasises the collection of virtual worlds and the rigorous activities in such collective virtual environments where human users will spend their time significantly. 
	Thus, a complete set of economic and social systems will be formed in the meta-cyberspace, resulting in new currency markets, capital markets, commodity markets, cultures, norms, regulations, and other societal factors. 
	
	Figure~\ref{fig:roadroad} illustrates a vision of building and upgrading cyberspace towards the metaverse in the coming decade(s).
	It is worthwhile to mention that the fourteen focused areas pinpointed in this survey are interrelated, e.g., ~\cite{edgexar} leverages IoT, CV, Edge, Network, XR, and user interactivity in its application design. Researchers and practitioners should view all the areas in a holistic view. For instance, the metaverse needs to combine the virtual world with the real world, and even the virtual world is more realistic than the real world. It has to rely on XR-driven immersive technologies to integrate with one or more technologies, such as edge and cloud (e.g., super-realism and zero-latency virtual environments at scale), avatar and user interactivity (e.g., motion capture and gesture recognition seamlessly with XR), artificial intelligence and computer vision for scene understanding between MR and the metaverse and the creation of creating digital twins at scale, Edge and AI (Edge AI) together for privacy-preserving AI applications in the metaverse, and to name but a few.

	In the remaining of this section, 
	we highlight the high-level requirements of the eight focused technologies for actualising the metaverse. 
	Accordingly, we pinpoint the six ecosystem aspects that could lead to the long-term success of the metaverse.

	%technology
	\textbf{Extended Reality.} 
	The metaverse moves from concept to reality, and VR/AR/MR is a necessary intermediate stage. To a certain extent, virtual environments are the technical foundation of the metaverse. The metaverse is a shared virtual space that allows individuals to interact with each other in the digital environment. Users exist in such a space as concrete virtual images, just like living in a world parallel to the real world. Such immersive technologies will shape the new form of immersive internet. VR  will allow users to obtain a more realistic and specific experience in the virtual networked world, making the virtual world operation more similar to the real world. Meanwhile, AR/MR  can transform the physical world. As a result, the future of our physical world is more closely integrated with the metaverse.
	
	More design and technical considerations should address the scenarios when digital entities have moved from sole virtual (VR) to physical (MR) environments. Ideally, MR and the metaverse advocate \textit{full integration} of virtual entities with the physical world. Hence, \textit{super-realistic} virtual entities merging with our physical surroundings will be presented \textit{everywhere and anytime} through large displays, mobile headsets, or holography. Metaverse users with digital entities can \textit{interplay} and \textit{inter-operate} with real-life objects. 
	As such, XR serves as a window to enable users to access various technologies, e.g., AI, computer vision, IoT sensors and other five focused technologies, as below discussed.
	
	\textbf{User Interactivity.} 
	\textit{Mobile techniques for user interaction} enable users to interact with digital overlays through the lens of XR. Designing mobile techniques in body-centric, miniature-sized and subtle manners can achieve \textit{invisible computing interfaces} for ubiquitous user interaction with virtual environments in the metaverse.
	Additionally, \textit{multi-modal feedback cues} and especially \textit{haptic feedback} on mobile techniques allow users to sense the virtual entities with improved senses of presence and realism with the metaverse, as well as to work collaboratively with IoT devices and service robots.
	
	On the other hand, virtual environments (VR/AR/MR) are enriched and somehow complex, which can only give people a surreal experience of partial senses, but cannot realise the sharing and interaction of all senses. Brain-Computer Interface (BCI) technology, therefore, stands out. Brain-computer interface technology refers to establishing a direct signal channel between the human brain and other electronic devices, thereby bypassing language and limbs to interact with electronic devices. Since all human senses are ultimately formed by transmitting signals to the brain, if brain-computer interface technology is used, in principle, it will be able to fully simulate all sensory experiences by stimulating the corresponding areas of the brain. Compared with the existing VR/AR headsets, a brain-computer interface directly connected to the human cerebral cortex (e.g., Neuralink\footnote{\url{https://neuralink.com/}}) is more likely to become the best device for interaction between players and the virtual world in the future meta-universe era.
	
	\textbf{IoT and Robotics.} 
	IoT devices, autonomous vehicles and Robots leverage XR systems to visualise their operations and invite human users to \textit{co-participate} in data management and decision-making. Therefore, presenting the data flow in comfortable and easy-to-view manners are necessary for the interplay with IoTs and robots. Meanwhile, appropriate designs of XR interfaces would fundamentally serve as a medium enabling \textit{human-in-the-loop} decision making. 
	To the best of our knowledge, the \textit{user-centric design} of immersive and virtual environments, such as \textit{design space} of user interfaces with various types of robotics, \textit{dark patterns} of IoT and robotics, \textit{subtle controls} of new robotic systems and so on, are in their nascent stage. Therefore, more research studies can be dedicated to facilitating the metaverse-driven interaction with IoT and robots. 
	
	\textbf{Artificial Intelligence.} 
	The application of artificial intelligence, especially deep learning, makes great progress in automation for operators and designers in the metaverse, and achieves higher performance than conventional approaches. However, applying artificial intelligence to facilitate users' operation and improve the immerse experience is lacking. Existing artificial intelligence models are usually very deep and require massive computation capabilities, which is unfriendly for resource-constrained mobile devices. Hence, designing light but efficient artificial intelligence models is necessary.
	
	\textbf{Blockchain.}
	%Dianlei
	Blockchain adopts the proof of work as the consensus mechanism, which requires participants to spend effort on puzzles to guarantee data security. However, the verification process for encrypted data is not as fast as conventional approaches. Hence, faster proof of work to accelerate the data accessing speed and scalability is a challenge to be solved. In addition, in public blockchains, their data is available to all users, which may lead to privacy issues. Hence, privacy protection mechanisms could be investigated in public blockchains.

	\textbf{Computer Vision.} 
	Computer vision 
	% lays a foundation for achieving the metaverse. It
	allows computing devices to understand the visual information of the user's activities and their surroundings. To build more a reliable and accurate 3D virtual world in the metaverse, computer vision algorithms need to tackle the following challenges. Firstly, in the metaverse, an interaction system needs to understand more complex environments, in particular, the integration of virtual objects and physical world. Therefore, we expect more precise and computationally effective spatial and scene understanding algorithms to use soon for the metaverse. 
	
	Furthermore, more reliable and efficient body and pose tracking algorithms are needed as metaverse is closed connected with the physical world and people. Lastly, in the metaverse, colour correction, texture restoration, blur estimation and super-resolution also play important roles in ensuring a realistic 3D environment and correct interaction with human avatars. However, it is worth exploring more adaptive yet effective restoration methods to deal with the gap between real and virtual contents and the correlation with avatars in the metaverse.

	\textbf{Edge and Cloud.} 
	The \textit{last mile latency} especially for mobile users (wireless connected) is still the primary latency bottleneck, for both Wi-Fi and cellular networks, thus the further latency reduction of edge service relies on the improvement of the last mile transmission, e.g., 1 ms promised by 5G, for seamless user experience with the metaverse.

	Also, \textit{MEC} involves multiple parties such as vendors, service providers, and third-parties. Thus, multiple adversaries may be able to access the MEC data and  steal or tamper the sensitive information.
	Regarding \textit{security}, in the distributed edge environment at different layers, even a small portion compromised edge devices could lead to harmful results for the whole edge ecosystem and hence the metaverse services, e.g., feature inference attack in federated learning by compromising one of the clients.
	
	\textbf{Network.} The major challenges related to the network itself are highly related to the typical performance indicators of mobile networks, namely latency and throughput, as well as the jitter, critical in ensuring a smooth user experience. User mobility and embodied sensing will further complicate this task. Contrary to the traditional layered approach to networks, where minimal communication happens between layers, addressing the strict requirements of user experience in the metaverse will require two-way communication between layers. 
	5G and its successors will enable the gNB to communicate network measurements to the connected user equipment, that can be forwarded to the entire protocol stack up to the application to adapt the transmission of content. Similarly, the transport layer, where congestion control happens, may signal congestion to the application layer. Upon reception of such information, the application may thus reduce the amount of data to transmit to meet the throughput, bandwidth, and latency requirements. Similarly, QoE measurements at the application layer may be forwarded to the lower layers to adapt the transmission of content and improve the user experience. 
	%Tristan
	
\begin{figure}[t]
		\begin{center}
			\includegraphics[width=\columnwidth]{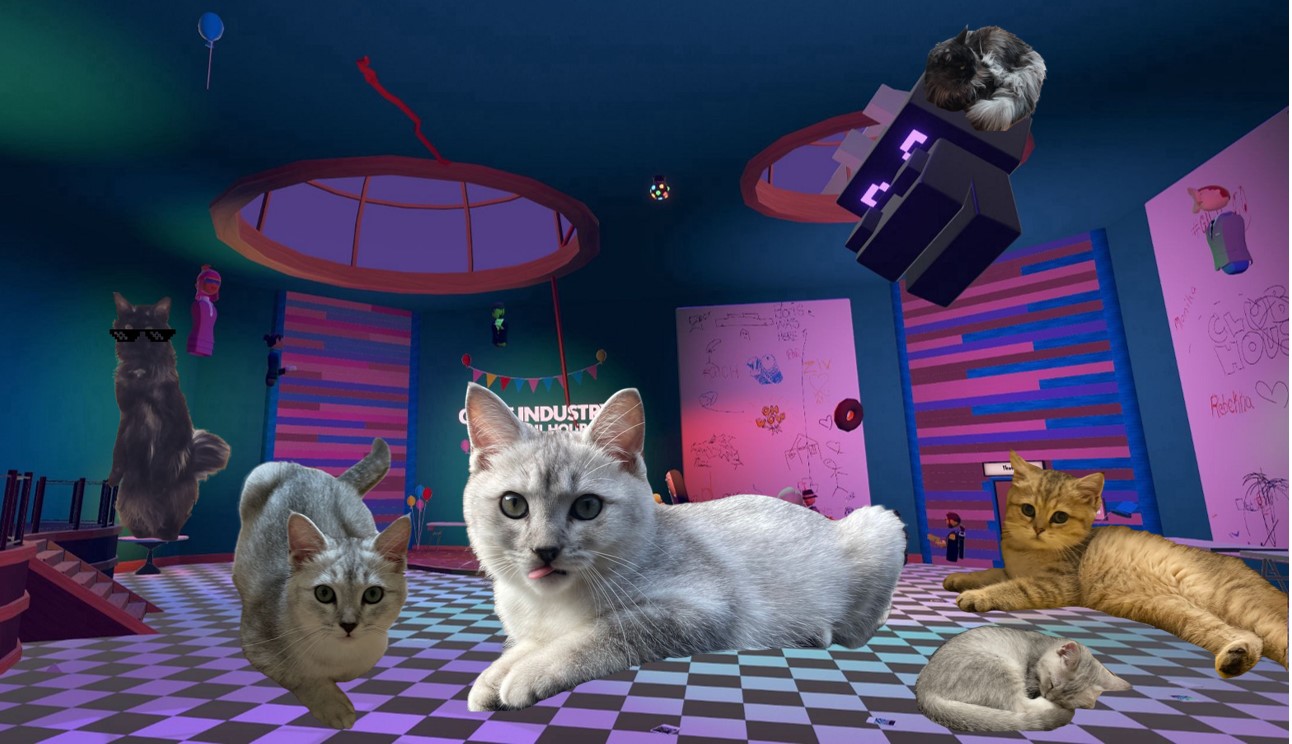}
		\end{center}
		\caption{Two perspectives: 1) ``In the Metaverse, nobody knows that you are a cat." is analogue to ``On the Internet, no one knows that you are a dog". 2) Metaverse can become a new horizon in Animal-Computer Interaction (ACI), e.g., a virtual environment as `\textit{Kittiverse}'. How to prepare the metaverse going beyond human users (or human avatars)?}
		\label{fig:cats}
	\end{figure}

	%Ecosystem
	\textbf{Avatar.} 
	Avatars serve as our digital representatives in the metaverse. Users would rely on the avatars to express themselves in virtual environments. Although the existing technology can capture the features of our physical appearance and automatically generate an avatar, the \textit{ubiquitous and real-time controls} of avatars with mobile sensors are still not ready for mobilising our avatars in the metaverse. Additional research efforts are required to enhance the \textit{Micro-expression} and \textit{non-verbal expression} of the avatars.
	Moreover, current gaps in understanding the design space of avatars, its influences to user perception (e.g., super-realism and \textit{alternated body ownership}), and how the avatars interact with \textit{vastly diversified smart devices} (IoTs, intelligent vehicles, Robots), should be further addressed. 
	The avatar design could also go farther than only human avatars. We should consider the following situations (Figure~\ref{fig:cats}): either human users employ their pets as avatars in the metaverse, or when human users and their pets (or other animals) co-exist in the metaverse and hence enjoy their metaverse journey together.
	
	Meanwhile, the ethical design of avatars and their corresponding behaviours/representation in cyberspace would also be a complicated issue. The metaverse could create a grey area for propagating offensive messages, e.g., race and could raise debate and prompt a new perspective to our identity. 
	An avatar creates a new identity of oneself in the metaverse, potentially raises a debate, and prompts new thinking of human life. That is, the \textit{digital clone of humanity} in the metaverse will live forever. Thus, even if the physical body, in reality, is annihilated, you in the digital world will continue to live in the meta-universe, retaining your personality, behavioural logic, and even memories in the real world.
	If this is the case, the metaverse avatars bring technical and design issues and ethical issues of the digital self. Is the long-lasting avatar able to fulfil human rights and obligations? Can it inherit my property? Is it still the husband of the father and wife of my child in the real world?
	
	\textbf{Content Creation.} 
	Content Creation should not be limited to professional designers -- it is everyone's right in the metaverse. Considering various co-design processes, such as \textit{Participatory design}, would encourage all stakeholders in the metaverse to create the digital world together. Investigating the \textit{Motivations} and \textit{Incentives} would enable the participatory design to push the progress of content creation in the metaverse. 
	More importantly, the design and implementation of \textit{automatic and decentralised governance} of censorship are unknown. Also, we should consider the establishment of creator cultures with cultural diversity, cross-generational contents, and preservation of phase-out contents (i.e., digital heritage). 
	
	\textbf{Virtual Economy.} 
	When it comes to the currency for the metaverse, the uncertainty revolves around the extent to which cryptocurrency can be trusted to function as money, as well as the innovation required to tailor it for the virtual world. Moreover, as the virtual world users will also be residents of the real world, the twin virtual and real economies will inevitably be \textit{intertwined} and should not be treated as two mutually exclusive entities. Therefore, a \textit{holistic} perspective should be adopted when examining what virtual economy truly means for the metaverse ecosystem. 
	
	Areas to be considered holistically include individual agent's \textit{consumption behaviours} in the virtual and real world as well as how aggregate economic activities in the two worlds can affect each other. In addition, a virtual world that is highly similar to the real world can potentially be used as a \textit{virtual evaluation sandbox} to test out new economic policies before we implement them in real life. Hence, to harness such merit, we need a conversion mechanism that optimally sets up the computer-mediated sandbox to properly simulate the reality with an accurate representation of economic agents' incentives.
	%Jerry Lin

	\textbf{Social Acceptability.} 
	Social acceptability is the reflection of metaverse users' behaviours, representing \textit{collective judgements and opinions} of actions and policies. The factors of social acceptability, such as privacy threats, user diversity, fairness, and user addiction, would determine the sustainability of the metaverse. 
	Furthermore, as the metaverse would impact both physical and virtual worlds, complementary rules and norms should be enforced across both worlds.
	
	On the other hand, we presume the existing factors of social acceptability can be applied to the metaverse. However, manual matching of such factors to the enormous metaverse cyberspace will be tedious and not affordable. And examining such factors case by case is also tedious. \textit{Automatic adoption of rules and norms} and subsequently the evaluation with social acceptability, to understand the collective opinions, would rely on many autonomous agents in the metaverse. Therefore, designing such agents at scale in the metaverse wold become an urgent issue. 
	
	More importantly, as the metaverse will be integrated into every aspect of our life, everyone will be impacted by this emerging cyberspace. Designing strategies and technologies for fighting cybercrime and reporting abuse would be crucial to improving the enormous cyberspace's social acceptability.

	\textbf{Security and Privacy.} 
	As for security, the highly digitised physical world will require users frequently to authenticate their identities when accessing certain applications and services in the metaverse, and \textit{XR-mediated IoTs and mechanised everyday objects}. 
	Additionally, protecting digital assets is the key to secure the metaverse civilisations at scale.
	In such contexts, asking textual passwords for frequent metaverse applications would be a huge hurdle to streamline the authentications with innumerable objects. The security researchers would consider new mechanisms to enable application authentications with \textit{alternative modalities}, such as biometric authentication, which is driven by muscle movements, body gestures, eye gazes, etc. As such, seamless authentication could happen with our digitised journey in various physical contexts -- as convenient as opening a door. However, such authentication system still requirements improvements in multitudinous dimensions, especially security levels, detection accuracy and speed, as well as the device acceptability.
	
	On the other hand, countless records of user activities and user interaction traces will remain in the metaverse. Accordingly, the accumulated records and traces would cause privacy leakages in the long term. The existing consent forms for accessing every website in 2D UIs would make users overwhelmed. Users with virtual 3D worlds cannot afford such frequent and recurring consent forms. Instead, it is necessary to design \textit{privacy-preserving machine learning} to automate the recognition of user privacy preference for dynamic yet diversified contexts in the metaverse.
	
	The creation and management of our digital assets such as avatars and digital twins can also have great challenges when protecting users against digital copies. These copies can be created to modify users' behaviour in the metaverse and for example share more personal information with `deep-fake' avatars.

	\textbf{Trust and Accountability.} 
	%abhishek your shine
	The metaverse, i.e., convergence of XR and the Internet, expands the definition of personal data, including biometrically-inferred data, which is prevalent in XR data pipelines. Privacy regulations alone can not be the basis of the definition of \emph{personal data}, since they can not cope up with the pace of innovation. One of the major grand challenges would be to design a principled framework that can define personal data while keeping up with potential innovations.
	
	As human civilisation has progressed from the past towards the future, it has accommodated the rights of \emph{minorities}, though after many sacrifices. It is analogous to how the \emph{socio-technical systems} on the world wide web have evolved, wherein the beginning, norms dictated acceptable or unacceptable actions, and these norms were decided by the democratic majority. As the metaverse ecosystem evolves, it must consider the rights of minorities and vulnerable communities from the beginning, because unlike in traditional socio-technical systems, potential mistreatment would have far more disastrous consequences, i.e., the victims might feel being mistreated as if they were in the real world.

	\section{Concluding Notes}
	On a final note, technology giants such as Apple and Google have ambitious plans for materialising the metaverse. 
	With the engagement of emerging technologies and the progressive development and refinement of the ecosystem, 
	our virtual worlds (or digital twins) will look radically different in the upcoming years.
	Now, our digitised future is going to be more interactive, more alive, more embodied and more multimedia, due to the existence of powerful computing devices and intelligent wearables. However, there exist still many challenges to be overcome before the metaverse become integrated into the physical world and our everyday life. 
	
	We call for a holistic approach to build the metaverse, as we consider the metaverse will occur as another enormous entity in parallel to our physical reality. 
	By surveying the most recent works across various technologies and ecosystems, we hope to have provided a wider discussion within the metaverse community. Through reflecting on the key topics we discussed, we identify the fundamental challenges and research agenda to shape the future of metaverse in the next decades(s).

	% \begin{thebibliography}{}

	% \end{thebibliography}
	
	\bibliographystyle{unsrt}
	\bibliography{references.bib}
	
	% , ref2.bib, lit-search.bib}

	\begin{IEEEbiography}[{\includegraphics[width=1in,height=1.25in,clip,keepaspectratio]{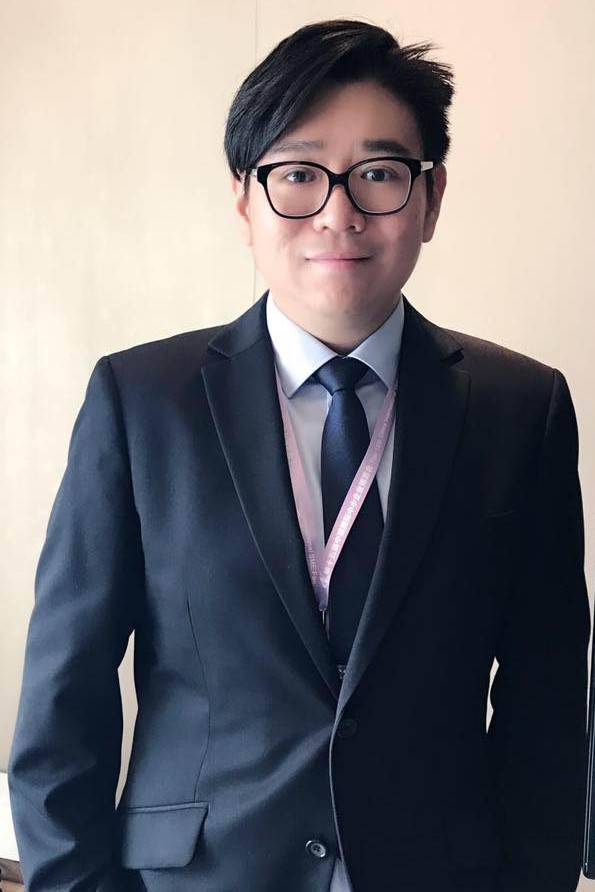}}]{Lik-Hang Lee} received the Ph.D. degree from SyMLab, Hong Kong University of Science and Technology, and the bachelor's and M.Phil. degrees from the University of Hong Kong. He is an assistant professor (tenure-track) with Korea Advanced Institute of Science and Technology (KAIST), South Korea. He is also the Director of the Augmented Reality and Media Laboratory, KAIST. He has built and designed various human-centric computing specialising in augmented and virtual realities (AR/VR). He is also a founder of an AR startup company, promoting AR-driven education and serving over 100 Hong Kong and Macau schools.
	\end{IEEEbiography}
	
	\begin{IEEEbiography}[{\includegraphics[width=1in,height=1.25in,clip,keepaspectratio]{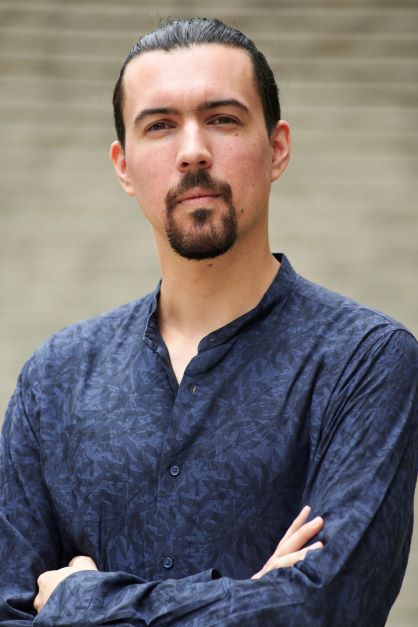}}]{Tristan Braud} is an assistant professor at the Hong Kong University of Science and Technology, within the Systems and Media Laboratory (SymLab). He received a Masters of Engineering degree from both Grenoble INP, Grenoble, France and Politecnico di Torino, Turin, Italy, and a PhD degree from Universit Grenoble-Alpes, Grenoble, France. His major research interests include pervasive and mobile computing, cloud and edge computing, human centered system designs and augmented reality with a specific focus on human-centred systems. With his research, Dr. Braud aims at bridging the gap between designing novel systems, and the human factor inherent to every new technology.
	\end{IEEEbiography}
	
	\begin{IEEEbiography}[{\includegraphics[width=1in,height=1.25in,clip,keepaspectratio]{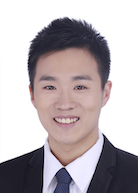}}]{Pengyuan Zhou} received his PhD from the University of Helsinki. He was a Europe Union Marie-Curie ITN Early Stage Researcher from 2015 to 2018. He is currently a research associate professor at the School of Cyberspace Science and Technology, University of Science and Technology of China (USTC). He is also a faculty member of the Data Space Lab, USTC. His research focuses on distributed networking AI systems, mixed reality development, and vehicular networks.
	\end{IEEEbiography}

	\begin{IEEEbiography}[{\includegraphics[width=1in,height=1.25in,clip,keepaspectratio]{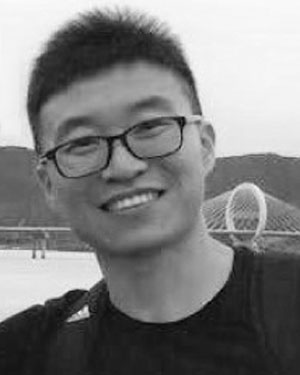}}]{Lin Wang} is a Postdoc researcher in Visual Intelligence Lab., Dept. of Mechanical Engineering, Korea Advanced Institute of Science and Technology (KAIST). His research interests include neuromorphic camera-based vision, low-level vision (especially image super-solution, HDR imaging, and image restoration), deep learning (especially adversarial learning, transfer learning, semi-/self-unsupervised learning) and computer vision-supported AR/MR for intelligent systems.
	\end{IEEEbiography}

	\begin{IEEEbiography}[{\includegraphics[width=1in,height=1.25in,clip,keepaspectratio]{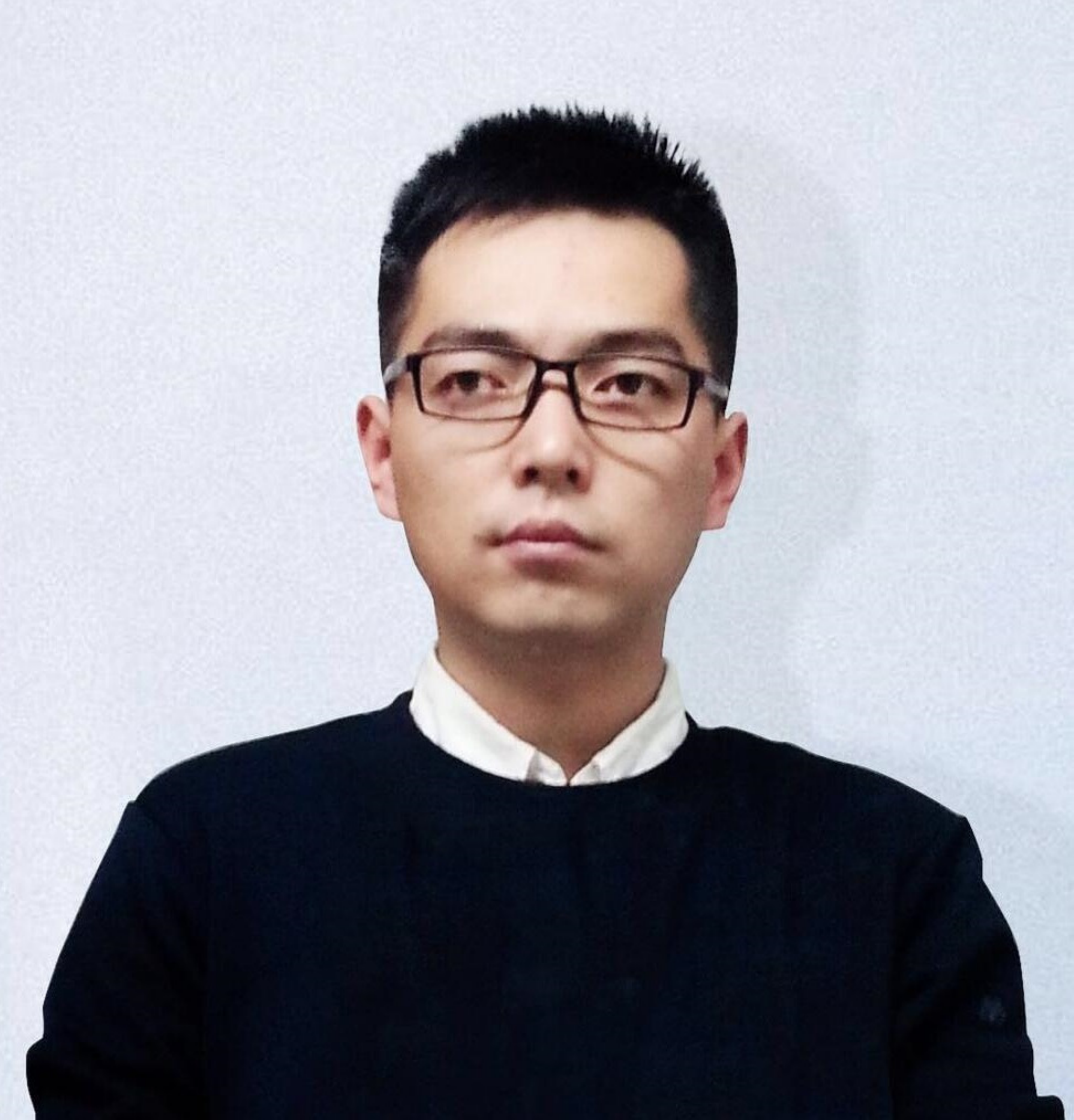}}]{Dianlei Xu}
		is a joint doctoral student in the Department of Computer Science, Helsinki, Finland and Beijing National Research Center for Information Science and Technology (BNRist), Department of Electronic Engineering, Tsinghua University, Beijing, China. His research interests include edge/fog computing and edge intelligence.
	\end{IEEEbiography}

	\begin{IEEEbiography}[{\includegraphics[width=1in,height=1.25in,clip,keepaspectratio]{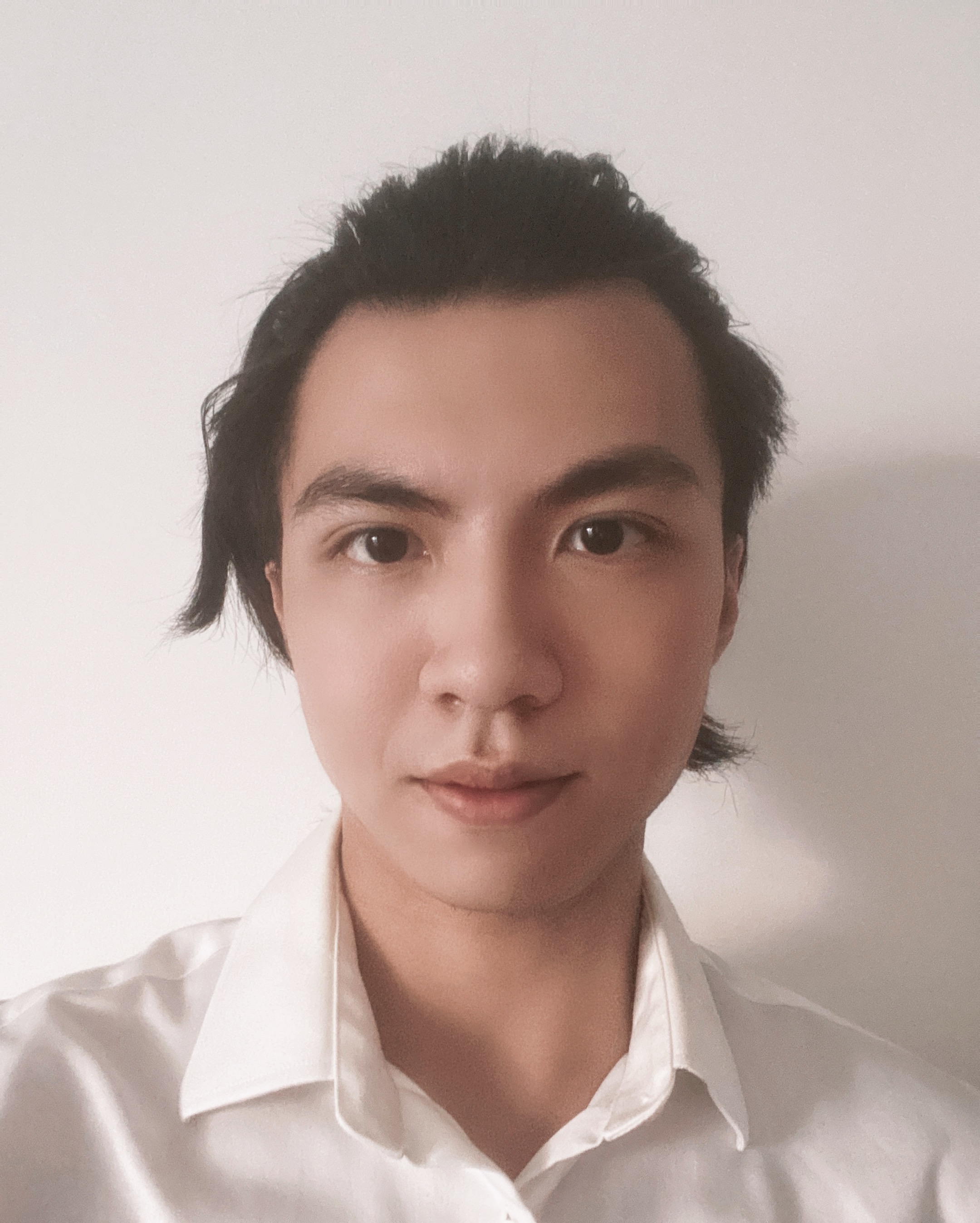}}]{Zijun Lin} is an undergraduate student at University College London (UCL) and a research intern at the Augmented Reality and Media Lab at Korea Advanced Institute of Science and Technology (KAIST). His research interests are the application of behavioural economics in human-computer interaction and the broader role of economics in computer science.
	\end{IEEEbiography}
	
	\begin{IEEEbiography}[{\includegraphics[width=1in,height=1.25in,clip,keepaspectratio]{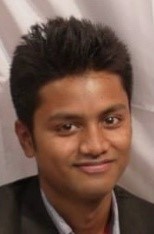}}]{Abhishek Kumar} is a PhD student at the Systems and Media Lab in the Department of Computer Science at the University of Helsinki. He has a MSc in Industrial and Systems Engineering, and a BSc in Computer Science and Engineering. His research focus is primarily in areas of Multimedia Computing, Multimodal Computing and Interaction with special focus on privacy.
	\end{IEEEbiography}

	\begin{IEEEbiography}[{\includegraphics[width=1in,height=1.25in,clip,keepaspectratio]{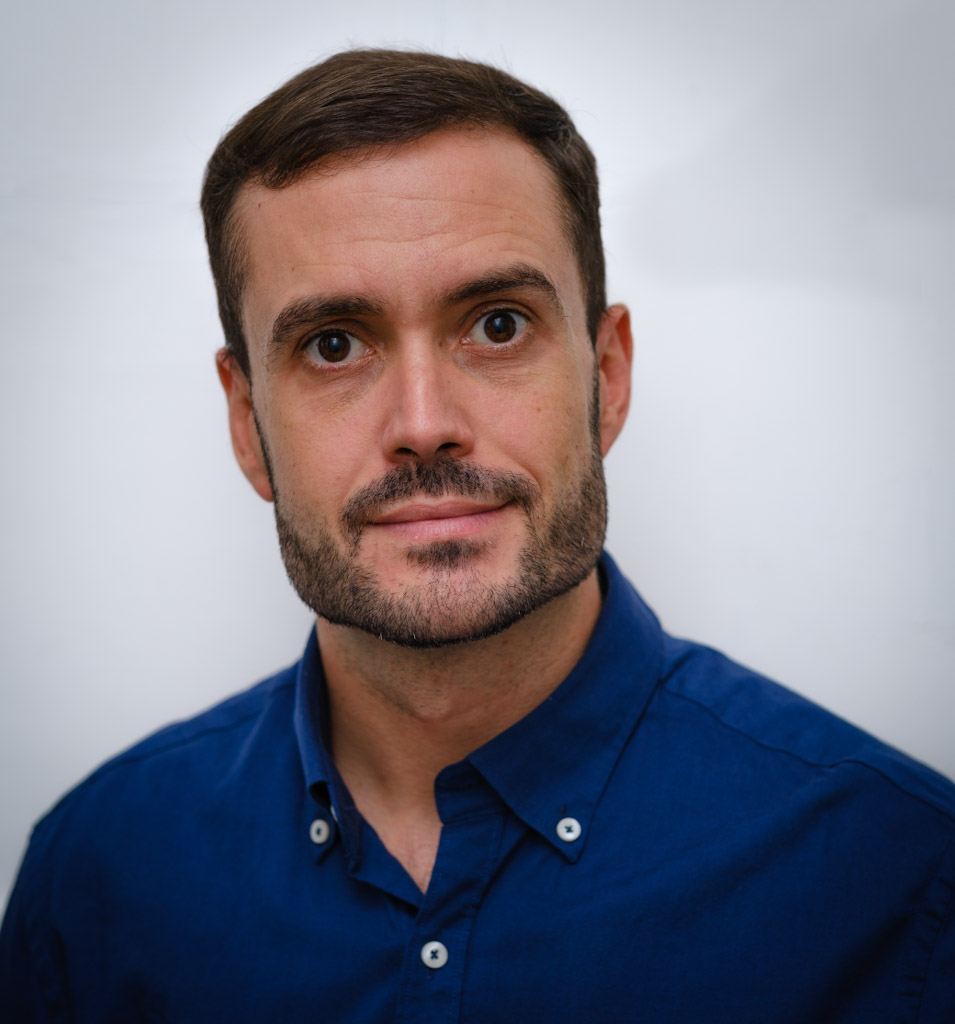}}]{Carlos Bermejo Fernandez} received his Ph.D. from the Hong Kong University of Science and Technology (HKUST). His research interests include human-computer interaction, privacy, and augmented reality. He is currently a Postdoc researcher at the SyMLab in the Department of Computer Science at HKUST.
	\end{IEEEbiography}

	\begin{IEEEbiography}[{\includegraphics[width=1in,height=1.25in,clip,keepaspectratio]{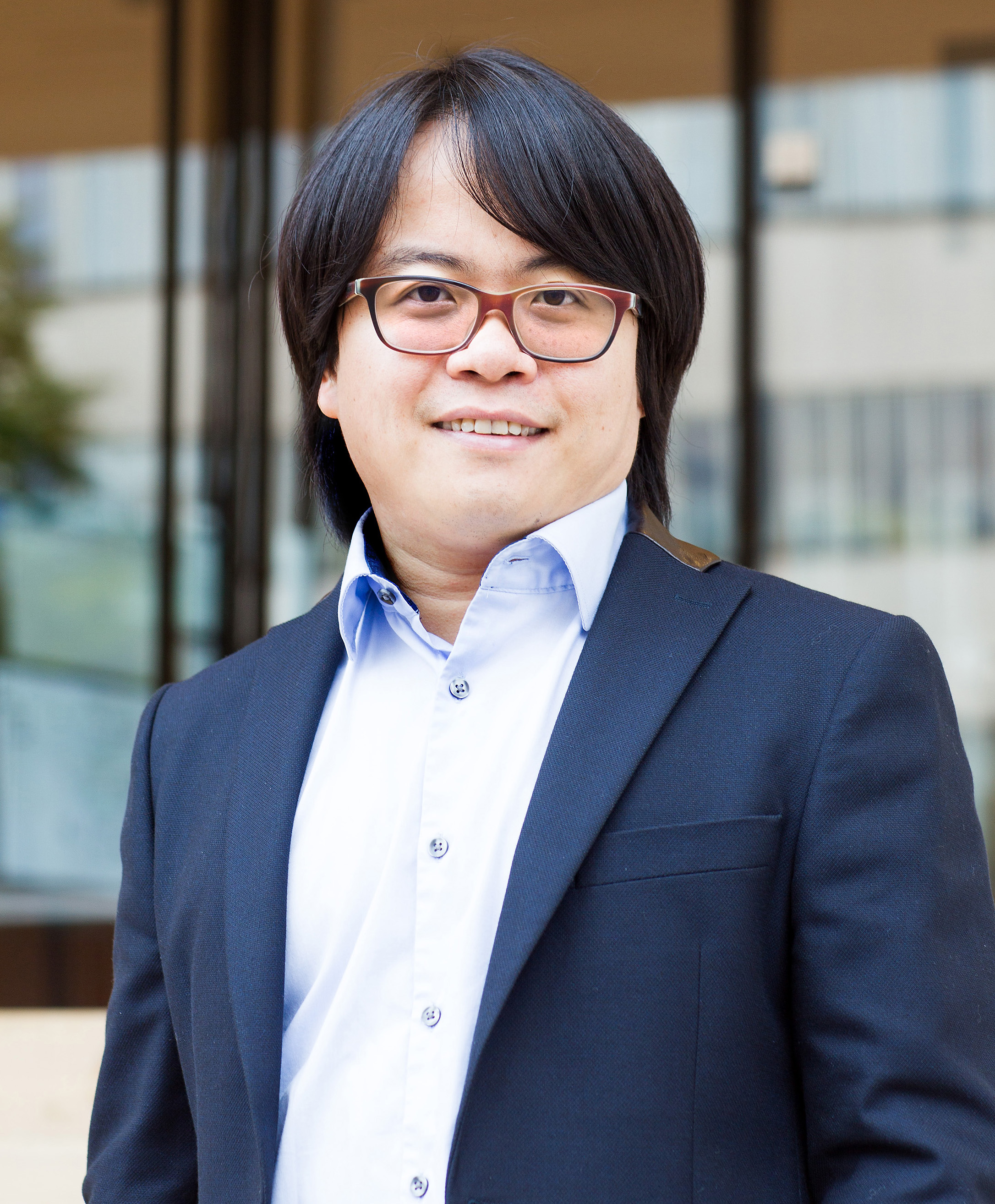}}]{Pan Hui} received the Ph.D. degree from Computer Laboratory, University of Cambridge, and the bachelor and M.Phil. degrees from the University of Hong Kong. He is a Professor of Computational Media and Arts and Director of the HKUST-DT System and Media Laboratory at the Hong Kong University of Science and Technology,  and the Nokia Chair of Data Science at  the University of Helsinki. He has published around 400 research papers and with over 21,000 citations. He has 32 granted and filed European and U.S. patents in the areas of augmented reality, data science, and mobile computing. He is an ACM Distinguished Scientist, a Member of the Academy of Europe, a Fellow of the IEEE, and an International Fellow of the Royal Academy of Engineering.
	\end{IEEEbiography}
	
	% \EOD
	
\end{document}